\documentclass[prd,preprintnumbers,superscriptaddress,amsmath,amssymb,nofootinbib,twocolumn,floatfix]{revtex4-2}

\usepackage{graphicx}
\usepackage{dcolumn}
\usepackage{bm}
\usepackage{setspace}
\usepackage{graphicx}
\usepackage{hyperref}

\usepackage{units}
\usepackage{nameref, color, xcolor}
\usepackage{pifont, amsmath, amssymb, mathtools}
\usepackage{tikz}
\usepackage[normalem]{ulem}

\usepackage{booktabs, acronym}

\newcommand{\Icomm}[1]{\textcolor{purple}{{#1}}}

\newcommand{\code}[1]{\texttt{#1}}

\begin{document}


\title{\textbf{Diffusing the assumptions: towards  realistic LISA stochastic component separation with physics-aware simulation-based inference} 
}%

\author{Iuliu Cuceu}
 \email{icuceu@oca.eu}
 \affiliation{
 Université Côte d’Azur, Observatoire de la Côte d’Azur, CNRS, Laboratoire Artemis, 06300 Nice, France}
\author{Nelson Christensen}
 \affiliation{
 Université Côte d’Azur, Observatoire de la Côte d’Azur, CNRS, Laboratoire Artemis, 06300 Nice, France}
\author{Astrid Lamberts}
 \affiliation{
 Université Côte d’Azur, Observatoire de la Côte d’Azur, CNRS, Laboratoire Artemis, 06300 Nice, France}
 \affiliation{
 Université Côte d’Azur, Observatoire de la Côte d’Azur, CNRS, Laboratoire Lagrange, 06300 Nice, France}
 \author{Quentin Baghi}
 \affiliation{Université Paris Cité, CNRS, Astroparticule et Cosmologie, F-75013 Paris, France}
 \author{Coen Rondeel}\affiliation{
 Université Côte d’Azur, Observatoire de la Côte d’Azur, CNRS, Laboratoire Artemis, 06300 Nice, France}
 \author{Natalia Korsakova}
 \affiliation{
 Université Côte d’Azur, Observatoire de la Côte d’Azur, CNRS, Laboratoire Artemis, 06300 Nice, France}

\date{\today}

\begin{abstract}
Simultaneous global-fit inference of the numerous sources observed by the \ac{LISA} depends critically on accurate modeling of its stochastic components: instrumental noise, the Galactic foreground, and astrophysical or \ac{SGWB}. Realistic features such as non-Gaussian transients and data gaps challenge traditional likelihood-based pipelines, creating an opportunity for \ac{SBI} to augment traditional analyses. Here, we present an \ac{SBI} framework for \ac{LISA} stochastic component separation built on the Simformer, an all-conditional transformer trained with diffusion or flow matching that ingests the full \ac{TDI} cross-spectral matrix via a physics-aware embedding. We progressively relax standard simplifying assumptions on simulated 11.6-day datasets containing a loud Galactic foreground and a faint power-law \ac{SGWB}. With unequal arms and per-link noise freedom, the networks match 17-dimensional \ac{MCMC} posteriors after importance reweighting ($20\%$ efficiency); full cross-spectra constrain each link's \ac{OMS} noise amplitude to $1\%$. When marginalizing over glitches or massive black hole binary transients, estimation errors remain flat up to transient \ac{SNR} of several hundred, whereas ignoring them severely biases likelihood fits. Under time-varying spline noise levels, realistic glitch rates, and gaps up to one day long, posterior coverage remains nominal or conservative across $\sim 10^4$ simulations. In addition, we test coverage in the local neighborhood of a given observation, by introducing a novel exact importance-weighted localized simulation-based calibration test. We find flow matching slightly better calibrated than diffusion in simpler regimes, although both are viable training objectives for \ac{GW} data, and observe modest gains from physics-aware template inputs. Finally, we outline how these networks can initialize the noise and stochastic blocks of a \ac{LISA} global fit. 
\end{abstract}

\maketitle

\section{Introduction}

\ac{LISA} will observe a millihertz \ac{GW} band dominated by signals rather than noise~\cite{colpi2024}: tens of millions of \ac{GBs}~\cite{cornish2017}, thousands of them individually resolvable~\cite{lamberts2019}, while others form a stochastic confusion foreground; \ac{MBHBs}~\cite{klein2016}, \ac{EMRIs}~\cite{babak2017}, \ac{SOBBHs}~\cite{sesana2016}, potential short duration burst-like signals~\cite{rubbo2006event,auclair2023cosmic}, and possibly stochastic \ac{GW} backgrounds of astrophysical and/or cosmological origin~\cite{amaroseoane2023, auclair2023}. Therefore, the statistical inference of all sources, as well as the instrumental noise, has to be performed jointly, an analysis paradigm referred to as the global fit~\cite{cornish2005, littenberg2023, katz2025, deng2025}. Within it, the stochastic components, namely the instrumental noise, the confusion foreground of unresolved \ac{GBs} and any other \ac{SGWB} present in the data, together set the noise covariance of every other source's likelihood, and must be estimated from the data~\cite{adams2014, boileau2021, baghi2023jcap, hartwig2023, karnesis2026}.

Separating these stochastic contributions serves the mission's science objectives~\cite{colpi2024}, as well as its operational performance. The former, through the discovery and characterization of a possible cosmological-in-orgin \ac{SGWB} and the ensemble characterization of the Milky Way's population of \ac{GBs}. In addition, though signal-dominated, \ac{LISA} detector characterization will be a critical part of its expected four-year duty cycle~\cite{colpi2024}. An active estimation of the instrument itself, including each interferometric measurement's noise levels, their time drift, and instrumental artifacts (glitches), informs whether it performs as designed~\cite{armano2016, muratore2023}.


Stochastic component separation analyses of \ac{LISA} data usually assume Gaussian, stationary noise, equal arm length and the diagonal $AET$ channel combinations of \ac{TDI}, a two-parameter noise model shared by all the raw interferometer measurements (called links, later combined to make the \ac{TDI} channels), and no transients. A primary goal is then to isolate each stochastic component's contribution without these assumptions. This implies giving the noise model more freedom: too little freedom can bias the stochastic signals, too much can absorb them~\cite{kume2025, karnesis2026}; the balance depends on understanding, for each noise model, just how capable will \ac{LISA} data be in constraining each model's parameters within a Bayesian \ac{PE} analysis. Likelihood-based analyses achieve this by exploring one data set at a time, either on the stochastic components alone, with everything else assumed removed, or inside a global fit. Global-fit prototypes have so far modeled stationary, Gaussian noise in the diagonal channels, with the two-parameter model~\cite{katz2025, deng2025} or phenomenological splines~\cite{littenberg2023}. There, the effect of each noise-model choice is entangled with the state of every other block and \ac{GBs} are the only stochastic signal source present in the data. 

In this work, we aim to relax each of the aforementioned assumptions: we allow freedom in the spectral shape of the noise using splines~\cite{baghi2023jcap, santini2025, karnesis2026, aimen2026, vajpeyi2026}; the noise amplitude itself is allowed to vary in time~\cite{edwards2020, alvey2024} and between links~\cite{hartwig2023}; we abandon the $AET$ channels because their utility is eliminated by more realistic data considerations, with unequal arm lengths and differences in the secondary noises in each interferometer link correlating them~\cite{hartwig2023, kume2025}; glitches should occur roughly twice or thrice per day in \ac{LISA}~\cite{armano2022, baghi2022lpf} and must be modeled or mitigated~\cite{spadaro2023, muratore2025}; and gaps interrupt the data~\cite{baghi2019, burke2025}. Left unmitigated, such artifacts substantially corrupt parameter estimation~\cite{castelli2025}.



\ac{SBI}~\cite{cranmer2020} is particularly suited for handling such data and performing Bayesian \ac{PE} on the stochastic elements: trained on simulated datasets, a neural estimator learns the target posterior directly and marginalizes implicitly over everything else the simulator can draw. This means omitting in the process of posterior estimation all included but undesirable elements (for example, noise transients), without requiring an analytic likelihood that explicitly models and marginalizes over the extra elements. \ac{SBI} frameworks are amortized over the prior, meaning a faster inference time relative to likelihood-based alternatives. This allows noise models to be prototyped on realistic data: retrained once per model, the network analyzes thousands of simulated data sets at the cost of sampling alone, with the other components of the data present and one effect changed at a time, a true ablation study. 

For ground-based detectors, neural posterior estimation already matches classical samplers on real events~\cite{dax2021, dax2023is} and now runs in real time~\cite{dax2025}. For the stochastic components of \ac{LISA} data, \ac{SBI} analyses have so far assumed a two-parameter noise model shared by all links (varying between segments in~\cite{alvey2024saqqara}), equal arms, and a single channel or the diagonal $AET$ channels~\cite{dimitriou2024, alvey2024, lyu2026}. To our knowledge, no \ac{SBI} analysis of \ac{LISA} data has yet used the cross-spectra between \ac{TDI} channels, per-link noise, or non-stationary noise with glitches and gaps. Therefore, on the path towards realism, the neural architecture has to be flexible enough to accommodate this gradual enrichment of the data. It needs to account for the variable parameter dimension from the different noise models used as well as from the changing number of sources included. Simultaneously, variable data content and the dependence structure have to be considered. The Simformer~\cite{gloeckler2024} -- an all-conditional transformer architecture, fed by a physics-aware embedding of the \ac{TDI} spectral matrix and trained with \ac{DSM} or \ac{FM}, is a solution to this complex problem.

The most important aspect in evaluating the performance and credibility of such architectures is the construction of reliable evaluation metrics. Where a likelihood exists, network posteriors can be compared with reference posteriors and the comparison needs full-dimensional distances with an interpretable scale. To this end, we utilize three likelihoods, two of which are exact, for the different data under consideration, and whenever it is computationally feasible, we provide such reference and quantify the discrepancy between network and \ac{MCMC} posterior distributions.

Where no reference exists, only posterior calibration tests built from the simulator remain~\cite{talts2018, hermans2021, lemos2023}. These need the power to resolve the small effect sizes of the posterior distributions, marginal by marginal and jointly, over the prior (population-level) and near the observation being analyzed. We combine both kinds throughout. Among the latter is a localized \ac{SBC}: its neighborhood of the observation is built as in Ref.~\cite{picchini2026}, which generalizes posterior \ac{SBC}~\cite{sailynoja2026}, and we introduce a novel contribution: the exact importance weighted ranks of a restricted region, so that it tests the network trained over the prior near an observation, without any refitting, retraining, or surrogate distribution required.

We thus move towards stochastic component separation in realistic \ac{LISA} data by sequentially removing the assumptions listed above, ending with data whose likelihood analysis would itself require a small global fit. We aim to answer four questions, two concerning the physics, and two the method:
\begin{itemize}
    \item (i) When including spectral and per-link freedom, nonstationarities, transients and gaps, how well can the Galactic foreground and an extragalactic \ac{SGWB} be separated from the instrumental noise?
    \item (ii) Considering the entire \ac{TDI} cross-spectral matrix, which parameters of the instrumental noise does the data constrain?
    \item (iii) Can one neural architecture follow required data changes by being reconfigured and retrained rather than redesigned, and how much does its physics-aware embedding or the choice between \ac{DSM} and \ac{FM} impact the posterior accuracy?
    \item (iv) To what extent can the \ac{SBI} posteriors be trusted, when likelihood-based comparisons are available or when they are computationally prohibitive or analytically intractable?  

\end{itemize}

Section~\ref{sec:preliminaries} introduces the shared elements: the simulator and its noise model (Sec.~\ref{sec:simulator}) for (i) and (ii), the neural estimator (Sec.~\ref{sec:sbi}) for (iii) and the checks (Sec.~\ref{sec:checks_intro}) for (iv). Sections~\ref{sec:stationary} to~\ref{sec:nonstationary} each open with their analysis setup, and report results on one simulated observation of $11.6$ days as well as consider population-level posterior calibration: Section~\ref{sec:stationary} keeps the data stationary, Gaussian and free of transients, but with unequal arms and per-link noise, and compares a single channel and the diagonal $AET$ channels with the full \ac{TDI} spectral matrix. Section~\ref{sec:gaussianity} frees the spectral shape of the noise, adds one glitch or one \acs{MBHB} for the networks to marginalize, and tests the physics-aware inputs of the embedding. Section~\ref{sec:nonstationary} lets the noise amplitude vary in time, adds glitches at a realistic rate and data gaps, and measures the time variation, the information it carries about the stochastic signals, and the information lost to glitches and gaps. Section~\ref{sec:discussion} discusses what the framework could become with changes of increasing size (a full parameter estimation of the stochastic components, a search ahead of the global fit, a monitor of the instrument) and what each would require of the simulator and the checks; Sec.~\ref{sec:conclusion} concludes with answers to the above questions as well as a summary of all novel elements of this work.

\section{Preliminaries}\label{sec:preliminaries}

Sections~\ref{sec:stationary} to~\ref{sec:nonstationary} will analyze increasingly realistic data with one framework, which we describe here once, in its most general form and with appropriate background. We infer the parameters $\theta$ of the stochastic components (the instrumental noise, the Galactic foreground and the \ac{SGWB}) from data $x$ built from the spectral matrix of the three \ac{TDI} channels $XYZ$, i.e.\ their auto- and cross-spectra over frequency. In \ac{SBI} the likelihood needs only be sampled from, not evaluated: a simulator draws $\theta \sim p(\theta)$ and $x \sim p(x \mid \theta)$, and on such pairs a neural estimator $q_{\phi}(\theta \mid x)$, with trainable weights $\phi$, learns to approximate the posterior $p(\theta \mid x) \propto p(x \mid \theta)\, p(\theta)$. The framework has three parts: the simulator (Sec.~\ref{sec:simulator}), which fixes what $\theta$ and $x$ contain; the inference network (Sec.~\ref{sec:sbi}), which sets how much of the information in $x$ reaches the posterior; and the checks (Sec.~\ref{sec:checks_intro}), which decide what can be claimed about that posterior\footnote{The framework is implemented in \code{stochformer}, a JAX package that will be made
publicly available~\cite{stochformer}.}
. 

\begin{figure*}
\includegraphics[width=\textwidth]{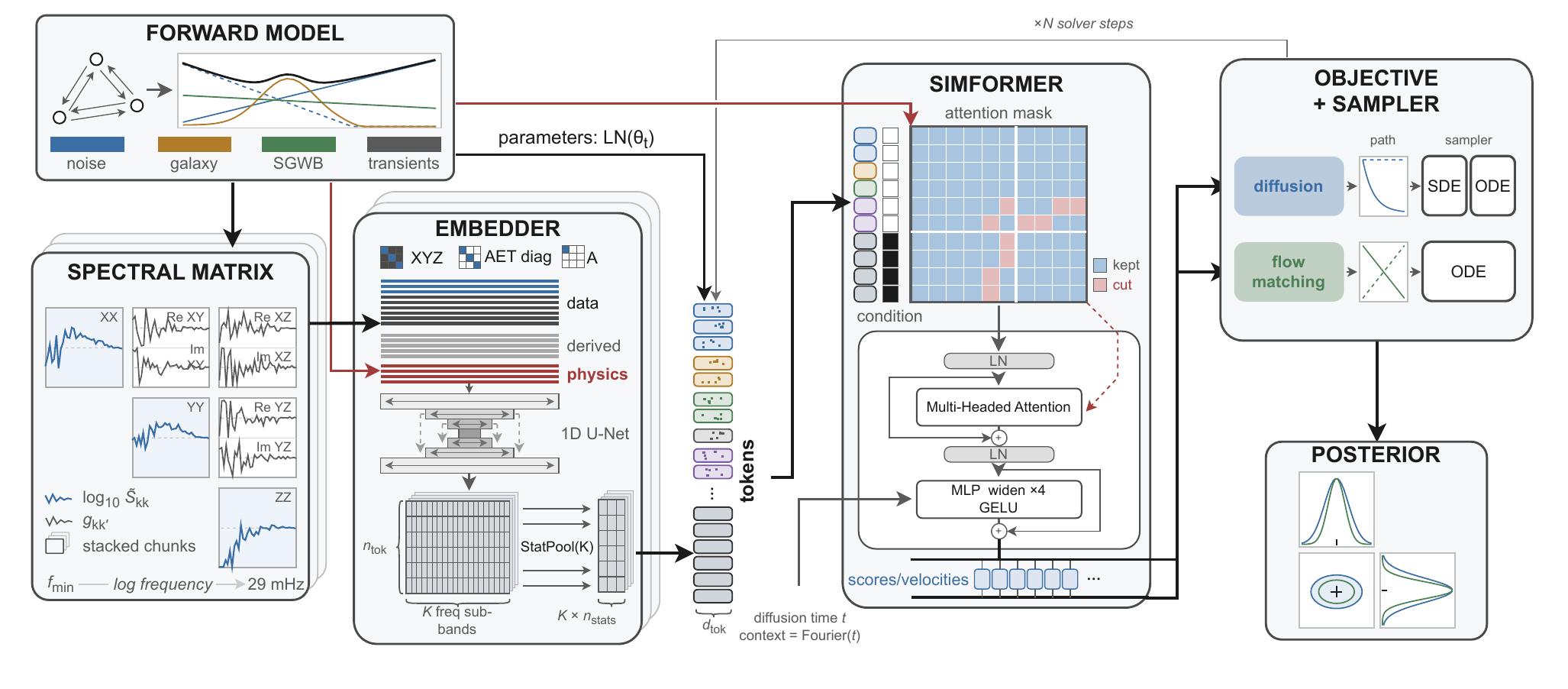}
\caption{From simulation to posterior. The forward model draws noise, Galactic foreground, \ac{SGWB} and transients and forms the spectral matrix of each record, or of each chunk of a segmented record: the log auto-spectra and the real and imaginary coherences over the coarse grid. The embedder reads the $XYZ$ matrix (or its $AET$-diagonal or TDI channel $A$), together with the derived and physics channels of Sec.~\ref{sec:embedding}, compresses them with a one-dimensional U-Net and pools $K$ frequency sub-bands into $n_{\rm tok}$ data tokens of width $d_{\rm tok}$. The Simformer attends over the parameter and data tokens, with a condition flag per token and an attention mask whose cut entries encode the dependence graph of the simulator, and returns the scores or velocities that the sampler integrates over $N$ solver steps, by diffusion or flow matching. Red arrows mark where knowledge of the simulator enters: the physics channels and the attention mask.}
\label{fig:pipeline}
\end{figure*}

Figure~\ref{fig:pipeline} is a snapshot of the entire framework, following one simulation through to a posterior. The forward model of stochastic signals (Sec.~\ref{sec:models}), instrumental noise (Sec.~\ref{sec:noise}) and transients (Sec.~\ref{sec:transients}) creates a cross spectral matrix as input data (Sec.~\ref{sec:datagen}). Then, an embedder (Sec.~\ref{sec:embedding}) compresses $x$ into $n_{\rm tok}$ data tokens, vectors in $\mathbb{R}^{d_{\rm tok}}$, and each parameter becomes a token of the same width. A transformer (Sec.~\ref{sec:network}), the Simformer, exchanges information among all tokens through attention and outputs a vector field (a score under \ac{DSM}, a velocity under \ac{FM}, Sec.~\ref{sec:sbi}) that a sampler integrates from a Gaussian draw to a posterior draw. The posteriors are then tested with classical, likelihood-based, comparisons, as well as \ac{SBI} calibration over the prior and the target observation (Sec.~\ref{sec:checks_intro}). Knowledge of the simulator enters at the red arrows: extra embedder inputs (physics channels) that project the spectral matrix onto templates of the modeled components, and an attention mask that cuts dependencies the simulator cannot produce. Dropping an assumption enlarges $x$ (the full $XYZ$ matrix rather than its $AET$ diagonal) or $\theta$ (a noise level per link, a free spectral shape, a level per time block), or adds transients (marginalized by the network) and gaps to the record. For the network, each change just means more tokens or embedder inputs.

\subsection{The forward model}\label{sec:simulator}

\ac{LISA} delivers six instrumental one-way link measurements, which are a realized superposition of projected \ac{GW} signals on the detector arm together with instrumental noise~\cite{colpi2024}. These can then be combined by \ac{TDI} into the three second-generation Michelson variables $XYZ$~\cite{tinto2021}. The signals further subdivide into short (or long duration) transients, such as \ac{MBHB} inspiral-coalescence-ringdown events, and long-lasting stochastic \ac{GW} signals, composed of a superposition of quiet sources of various astrophysical and cosmological origins. For stationary and Gaussian stochastic components, we can summarize their contribution entirely as a one-sided \ac{PSD} $S(f)$ per source, carried into the $3\times3$ \ac{TDI} cross-spectral matrix by a response kernel, assuming mutual independence:
\begin{align}\label{eq:signal_cov}
C_{\rm signal}(f\mid \theta) \;=\; &S_{\rm gal}(f\mid \theta_{\rm gal})\, \mathcal{R}_{\rm gal}(f) + \\ 
&S_{h}(f\mid \theta_{\rm sgwb})\, \mathcal{R}_{\rm sgwb}(f) \nonumber
\end{align}
with each component's sky-integrated response $\mathcal{R}$, whose time-dependence we do not write. It is anisotropic for the Galactic foreground, which has a response-driven amplitude modulation throughout the year, due to \ac{LISA}'s cartwheel motion around the sun~\cite{boileau2021}. This can be used to break degeneracies with isotropic components such as an \ac{SGWB}. The kernels are fixed by the constellation geometry (and the distribution of binaries in the sky for the galaxy) and by the choice of \ac{TDI} combination, and are given in Appendix~\ref{App:LISAmeasurement}. 
This section will briefly describe, from a modeling perspective, the stochastic signal components, Sec.~\ref{sec:models} as well as the \ac{OMS} and \ac{TM} noise contributions, Sec.~\ref{sec:noise}. Transients are not stochastic and do not enter Eq.~\eqref{eq:signal_cov}, being added coherently to the realisation instead. The two transient signal classes considered, instrumental glitches and \ac{MBHBs}, are also described below, Sec.~\ref{sec:transients} and lastly, the overall composition of our simulator.

\subsubsection{Foregrounds and backgrounds}
\label{sec:models}

We consider two different stochastic contributions: an unresolved superposition of millions of compact Galactic binaries, the Galactic foreground, as well as a generic \ac{SGWB} component. Both are assumed isotropic and stationary, such that the sky integral of the antenna response can be evaluated at a single epoch, and thus the response matrix, $\mathcal{R}(f)$ is not a function of time for either component.\footnote{The absence of a time-dependent response is not a limitation of our simulator, nor of the inference framework; in fact, as Section~\ref{sec:nonstationary} demonstrates with a model of noise variation in time, neither the network nor the embedding needs to change to accommodate a covariance that varies from segment to segment. Our treatment of the response and orbit is described in Appendix ~\ref{App:LISAmeasurement}.} For the $\le11.6$ day observation intervals considered in this work, this approximation is valid even for the foreground~\cite{digman2022}.

The Galactic foreground is modeled by a standard broken power-law empirical fit from~\cite{karnesis2021}:
\begin{align}\label{eq:galaxy}
S_{\rm gal}(f) \;=\; &A_{\rm gal}\, f^{-7/3} \exp\!\left[-\left(\frac{f}{f_{r1}}\right)^{\alpha}\right]\\ 
&\times\frac{1}{2}\left[1 + \tanh\!\left(\frac{f_{rk} - f}{f_{r2}}\right)\right], \nonumber
\end{align}
whose functional form is that of~\cite{cornish2017} and coefficients are taken from~\cite{karnesis2021}. We do not simulate individual Galactic binaries. This fit is a function of the \ac{SNR} threshold above which the bright (detectable) binaries are assumed to have been subtracted, which we fix through the parameters $\alpha=1.56$ and $f_{r2}=6.7\times 10^{-4}\rm Hz$. 

We parametrize an \ac{SGWB} signature as a power law in the fractional \ac{GW} energy density of the Universe to capture the large variety of possible spectral shapes assumed by various cosmological and astrophysical sources~\cite{Christensen:2018iqi}:

\begin{equation}
\Omega_{\rm gw}(f) \;=\; \Omega \left(\frac{f}{f_{\rm ref}}\right)^{n_t} ,
\qquad
S_h(f) \;=\; \frac{3 H_0^2}{4 \pi^2 f^3} \, \Omega_{\rm gw}(f) ,
\label{eq:sgwb}
\end{equation}
where $\Omega$ subsumes a factor $h^2$ that we do not write, and we take a reference frequency
$f_{\rm ref} = 1$~mHz. 

These stochastic signal parametrizations and their priors, Appendix~\ref{app:noise_models}, are chosen to provide a stringent model inference task (due to the low parameter count): one signal is loud with an expected \ac{SNR} $\ge100$ depending on observation duration for the foreground, source population and cosmology, while the \ac{SGWB} is expected to be comparatively faint $\le10$ and would in principle require mission-length observation intervals to gather sufficient \ac{SNR}. These two signals provide the parameter estimation baseline, whose constraint quality, relative to a likelihood-based classical analysis where possible, will be in the spotlight as we add more and more realistic \ac{LISA} noise and transient signals.

\subsubsection{Modeling instrumental noise} \label{sec:noise}

In contrast to the strongly parameterized stochastic signals, we treat the \ac{LISA} instrumental noise as flexibly as possible. Our aim for noise modeling is twofold. Firstly, in a generic noise-estimation scenario, freely absorbing unexplained power within the noise, as opposed to the known signals, is a desirable outcome; this is particularly important when the stochastic signals, by contrast, are very rigidly parametrized. Secondly, we want to explicitly include a large asymmetry in the number of parameters used to model the individual components of the overall \ac{PSD}. This is to stress-test how parameter dimensionality and likelihood information, or lack thereof, are absorbed by our transformer-based inference and thus showcase the advantage of attention-based inference frameworks. 

The principal component of noise in \ac{LISA} is laser frequency noise, expected to sit at more than $8$ orders of magnitude above any signal or other noise sources. In a generic unequal and time-varying arm length \ac{LISA} configuration, this is mitigated in postprocessing through second-generation \ac{TDI}~\cite{tinto2021}, see Appendix~\ref{app:tdi}. Other dominant noise contributions are tilt-to-length coupling and clock noise~\cite{hartwig2021clock}. We assume these to have been suppressed~\cite{Hartig_2025}, which leaves us with the \ac{TM} and \ac{OMS} contributions. Their overall covariance, assuming independence, thus reads as: 
\begin{equation} \label{eq:noise_cov}
    C_{\rm noise}(f\mid\theta_{\rm noise}) = \sum_{k} \sum_{n} S^{(n)}_k(f\mid \theta_{\rm noise})\, \mathcal{T}^{(n)}_{k}(f) ,
\end{equation}
with $n \in \{{\rm TM}, {\rm OMS}\}$ and $k$ running over the six links, $k \in \{12, 23, 31, 13, 21, 32\}$. Link by link independence gives each of the twelve single-link spectra its own \ac{TDI} transfer function $\mathcal{T}^{(n)}_{k}(f)$, a $3 \times 3$ matrix fixed by the constellation geometry and the \ac{TDI} combination, see Appendix~\ref{app:noise_tf}. The single-link spectra are generally parametrized as~\cite{babak2021}:

\begin{align}
S^{\rm TM}(f) = &A^2 \left[1 + \left(\frac{0.4\,{\rm mHz}}{f}\right)^{2}\right]
\left[1 + \left(\frac{f}{8\,{\rm mHz}}\right)^{4}\right] \nonumber \\
&\times \left(\frac{1}{2 \pi f c}\right)^{2} ,
\label{eq:analytic_noise_tm}
\end{align}

\begin{equation}
S^{\rm OMS}(f) = P^2 \left[1 + \left(\frac{2\,{\rm mHz}}{f}\right)^{4}\right]
\left(\frac{2 \pi f}{c}\right)^{2} ,
\label{eq:analytic_noise_oms}
\end{equation}
expressed in fractional-frequency units, with $A$ the noise of a single test mass, of which each link carries two, and $P$ is the effective total displacement noise of a one-way single-link test-mass-to-test-mass
measurement~\cite{LISA:2017pwj}. At the mission requirements  values $\bar{A} = 3 \times 10^{-15}\,{\rm m\,s^{-2}\,Hz^{-1/2}}$ and $\bar{P} = 15 \times 10^{-12}\,{\rm m\,Hz^{-1/2}}$~\cite{colpi2024, babak2021, hartwig2023}, they define the reference spectra $\bar{S}^{(n)}(f)$ around which all of our noise models are built. This literature-standard $2$-parameter noise model rests on three assumptions: all six links share one spectrum, its shape is known up to a level, and it does not change over the observation. This means that the $6 \times 6$ link matrix is described by only two numbers, and the \ac{TDI} covariance inherits an exact structure from this geometry alone. Succinctly, in the commonly utilized linear rotation of the Michelson \ac{TDI} basis, the $AET$ channels gain their orthogonal property only when considering equal-arm orbits and shared-link noise; see Appendix~\ref{app:noise_tf} for details and~\cite{hartwig2023} for a proof and in-depth discussion on \ac{TDI} basis orthogonality and the existence of ``null" channels.

Throughout, we use unequal-arm, time-varying orbits (Keplerian) and progressively drop the three assumptions. Our flexible noise model, allowing for spline-driven shape freedom with per-link amplitude deviations and time-varying amplitude deviations reads:
\begin{equation}
\ln S^{(n)}_{k, b}(f) \;=\; \ln \bar{S}^{(n)}(f) \;+\; \epsilon^{(n)}_{k} \;+\; \delta^{(n)}(\ln f)  \;+\; d^{(n)}_{b} ,
\label{eq:noise_model}
\end{equation}
where each deviation from the reference relaxes one of the three assumptions, which is recovered when $\epsilon^{(n)}_{k}$ is common to all links, $\delta^{(n)}$ is constant in frequency, or $d^{(n)}_{b}$ is common to all blocks, $b$. 

The per-link level $\epsilon^{(n)}_{k}$ lets the twelve single-link spectra differ by a frequency-independent factor. At fixed shape, and written as $\epsilon^{(n)}_{k} = 2\ln(a^{(n)}_{k}/\bar{a}^{(n)})$ in terms of per-link amplitudes about the specification values $\bar{a}^{(n)} \in \{\bar{A}, \bar{P}\}$, it gives the model of~\cite{hartwig2023}, which Sec.~\ref{sec:stationary} utilises to showcase diffusion-based \ac{SBI}. 

The shape term $\delta^{(n)}$ relaxes the rigid spectra of Eqs.~\eqref{eq:analytic_noise_tm} and~\eqref{eq:analytic_noise_oms} by modelling the single-link log-\ac{PSD} with a spline~\cite{baghi2023jcap}, specifically the Akima interpolant~\cite{akima1970} in $\ln f$ adopted for \ac{LISA} noise in~\cite{karnesis2026}. The Akima interpolant is $\delta^{(n)}(\ln f_i)=c^{(n)}_{i}$, through $n_s = 6$ control points $(\ln f_i, c^{(n)}_{i})$, where $\ln f_i$ are fixed knot log-frequencies \footnote{We fix the knots per component, placed by warping a logarithmic spacing towards low (or high) frequency; both the number and the positions are fixed before training by a fisher analysis and are thus never sampled, in contrast to the trans-dimensional treatment of~\cite{karnesis2026}.}. Applied identically to all six links it gives spectral flexibility but not per-link freedom, which our noise model restores by adding the $\epsilon^{(n)}_{k}$ about the now free-in-shape baseline. It is the model of Sec.~\ref{sec:gaussianity}, with more detail found in Appendix~\ref{app:noise_models} about the parametrization and priors used to minimize degeneracy. 

Lastly, $d^{(n)}_{b}$ relaxes stationarity: the observation is split into $n_b$ contiguous blocks, each carrying one level per component, all drawn independently so that blocks are defined relative to the shared baseline. The noise model of Sec.~\ref{sec:nonstationary}, combines the block levels with the shared spline and per-link freedom links. Each is written out in Appendix~\ref{app:noise_models}, and Table~\ref{tab:priors} includes the priors of all stochastic components.

We conclude this foray into noise modeling by mentioning why we restrict ourselves to amplitude offsets (per-link or per-segment) rather than to the number and position of the knots. Within a likelihood-based Bayesian analysis, the flexibility is a model choice that should be decided by the evidence~\cite{santini2025}, implying a data-driven transdimensional scheme, which is not an easy-ask for most \ac{SBI} architectures. \footnote{In the penalized-spline treatment of ~\cite{aimen2026}, they find that adaptive knot placement removes the need for transdimensional treatment and can be performed once at initialization only, as we do here.} Beyond the Bayesian perspective, there is one specific to \ac{SBI}: a neural estimator does not react to a change of noise model the way a likelihood analysis does. Changing the noise parameterization does not only deform the likelihood surface; it changes the dimension and the meaning of the vector being inferred, the statistics of the training distribution, the interaction with the data embedding, and the learning dynamics of the network itself. Disentangling those effects is far beyond the objective here. What we do set out to show is that the framework tolerates the parameter count and the kind of freedom that a realistic noise model requires.

\subsubsection{Transient perturbations} \label{sec:transients}
Two classes of non-Gaussian transients are considered; a short-lived instrumental test mass glitch and a comparatively long-lived \ac{MBHB} transient. We preface this by mentioning that we do not aim to infer the parameters of either transient; they are primarily meant to be marginalized over implicitly by the network; more on this in Sec.~\ref{sec:gaussianity}. 

\ac{LISA} Pathfinder measured transient acceleration events at a rate of roughly one per day, falling into fast readout events and slower true force events \cite{armano2022}. We use an analytic shapelet~\cite{baghi2022lpf,Berg__2019,muratore2025}, whose frequency-domain form for a single event is
\begin{equation}
\Delta \tilde{\nu}(f) \;=\; \frac{-2 \, \Delta \dot{v} \, \beta}{i\, c\, 2 \pi f} \;
\frac{e^{-i \tau 2 \pi f}}{\left(-i + \beta\, 2 \pi f\right)^{2}} ,
\label{eq:shapelet}
\end{equation}
with an amplitude $\Delta \dot{v}$, a decay scale $\beta$ and a trigger time $\tau$. The glitch is injected on a test mass and propagated to $X$, $Y$, $Z$ through the same single-link transfer that carries the \ac{TM} noise of that link, $t^{TM}_{k}(f)$, Eq.~\ref{eq:t_vectors}~\footnote{In general, glitches need not appear in \ac{TM} interferometers only; this is merely for the convenience of the simulation.}. That construction carries an obvious and exploitable structure: each \ac{MOSA} feeds exactly two of the three Michelson variables and leaves the third identically zero, see Appendix~\ref{App:LISAmeasurement}. The number of events per segment is Poisson distributed, and the population is drawn over amplitude and decay scale.

The second transient is an \ac{MBHB}. We generate the polarizations with a time-domain phenomenological waveform for quasi-circular, non-precessing binary~\cite{estelles2021}, keeping only its dominant $(2,2)$ mode, project them onto the six one-way link responses in the time domain, and then perform the \ac{TDI} combination in the frequency domain. Just as for the noise and the stochastic signals, the \ac{TDI} transfer uses the arm lengths at the start of the record (they change by at most $3$~ms over it), while the link responses follow the motion of the constellation. Eleven parameters describe the source and we use the \textsc{jax} implementation of the waveform, \textsc{Phentax}~\cite{santini2026phentaxjaximplementationimrphenomthm,phentax}. Contrasting with the glitch which affects one link and has no sky dependence, an \ac{MBHB} touches all six links and its channel content depends on sky position, inclination and polarization.

\subsubsection{Data generation and preprocessing}
\label{sec:datagen}

A single data realization's construction is shared between all experiments showcased below (the only noteworthy difference being the nonstationary noise case, where this is repeated multiple times over shorter timescales) and is produced as follows: 
\begin{itemize}
    \item The model covariance $C(f\mid \theta)$ is evaluated on the Fourier grid given by $T_{\rm obs}$, by constructing its components: $S_{\rm gal},S_{\rm h}$, response $\mathcal{R}$ (isotropic and shared, precomputed once at a single orbital epoch), $S_{\rm k}^{(n)}$ and their transfer functions $\mathcal{T}_{\rm k}^{(n)}$.
    \item Cholesky factor $L(f)$, with $L L^{\dagger} = C$, is formed; a standard complex normal vector $\varepsilon(f)$ is drawn and mapped to $z(f) = L(f) \varepsilon(f)$.
    \item If transients are included, they are added coherently: $z(f) \;\longrightarrow\; z(f) + \sum_{g} \delta z_{g}(f)$, with $\delta z_g$ the TDI-projected glitch (or glitches in Sec.~\ref{sec:nonstationary}) or an \ac{MBHB} waveform.
    \item The raw cross-periodogram, $z(f) z^{\dagger}(f)$, is then coarse-grained on a custom hybrid grid (single Fourier bin below a transition frequency, log-spaced above, with the band partitioned into $n_{\rm cg}$ disjoint bins): $S_j \;=\; \frac{1}{N_j} \sum_{f \in j} z(f) \, z^{\dagger}(f)$, where $N_j$ is the number of Fourier bins falling in coarse bin $j$.
    \item The last preprocessing step is prewhitening by a fixed reference covariance (precomputed on the same coarse grid) to form the coarse-grained sample covariance matrix: $\tilde{S}_j \;=\; L_{{\rm ref}, j}^{-1} \, S_j \, L_{{\rm ref}, j}^{-\dagger}$, with reference Cholesky factor $L_{\rm ref}$, an invertible linear map. This is for purely numerical reasons (to restrict the dynamic range). 
    \item Finally, the data artifact per coarse bin consists of $9$ real numbers: $\log_{10} {\rm diag}(\tilde{S}_j)$ which are real and positive; in addition, the real and imaginary parts of the three off-diagonal elements as coherences: $g_{k k'} \;=\; \frac{\tilde{S}_{k k'}}{\sqrt{\tilde{S}_{k k} \tilde{S}_{k' k'}}}$, where the coherence is bounded by Cauchy-Schwarz, $|g_{k k'}| \le 1 $.
\end{itemize}

A few remarks on the above construction. First, the following implicit assumptions are made: the covariance varies slowly enough to be treated as constant within a coarse bin, the Fourier bins are uncorrelated, and the averaging kernel is a boxcar, so that the equivalent noise bandwidth is unity and the degrees of freedom are the bin count, and not an effective count that would be required when starting with time-domain data. We refer the reader to~\cite{baghi2026} for a quantitative treatment of the error incurred by neglecting correlations between averaged bins. Second, the analysis band is $[3 \times 10^{-5}, 2.9 \times 10^{-2}]$~Hz. The low end is chosen to minimize the numerical issues associated with the rank-1 bins present en masse at low frequency (third and second eigenvalues of $\tilde{S}$ significantly below the primary eigenvalue numerically). Additionally, the requirements for \ac{LISA} do not apply below $0.1$~mHz. At the high end, the Michelson transfer functions have nulls at harmonics of the arm length. Near a null the response varies rapidly with frequency, so to preserve the validity of the above construction the high-frequancy cut-off is just before the first null. Both of these are chosen for convenience, but it does mean discarding information; for an example of a construction that utilizes very low/high-frequency regions, see~\cite{Caprini:2019pxz}. Lastly, the above coarse-graining scheme is not appropriate in general, as it discards coherently added information, the transients. In a realistic setting, no fixed analytic reduction is sufficient, and the burden of summarizing the data has to be carried by the embedding instead. That is why we construct and benchmark the physics-informed embedder of Sec.~\ref{sec:embedding}.

\subsection{Simulation-based inference with diffusion}
\label{sec:sbi}

In \ac{SBI}, the conditional estimator $q_\phi(\theta \mid x)$ targets the posterior distribution of parameters $\theta$, given the observation $x$, which we denote as the entire input data stream. Rather than restricting $x$ to the raw spectral bins alone, we define the data object as a multi-channel stack over the coarse frequency grid,
\begin{equation}
x \;=\; \Big\{ \big[ x_{j}^{\rm spec} ,\; x_{j}^{\rm aux} \big] \Big\}_{j=1}^{n_{\rm cg}} \;\in\; \mathbb{R}^{C \times n_{\rm cg}} ,
\label{eq:data_object}
\end{equation}
with the primary spectral channels,
\begin{equation}
x_{j}^{\rm spec} \;=\; \left\{ \log_{10} \tilde{S}_{j,kk} ,\;\; \mathrm{Re}\, g_{j,kk'} ,\;\; \mathrm{Im}\, g_{j,kk'} \right\} ,
\label{eq:spec_channels}
\end{equation}
where $1 \le k < k' \le 3$. Coarse-graining alone is already helpful in reducing the size of our data. However, we further employ a learned embedder for both additional compression and to assist our inference network by giving it access to a learned data representation, which can include auxiliary information. Our embedder is then ``physics-aware'' in the sense that our input channels do not only constitute the usual summary statistic for stochastic sources (the spectral matrix), but also simulator-dependent summaries. The companion channels $x_{j}^{\rm aux}$ are of three types: deterministic, model-free and physics-informed summaries; these can include frequency band coordinates, matrix basis invariants of $\tilde{S}_j$, and projected template banks; the exact forms and roles are detailed in Sec.~\ref{sec:embedding}. 

The auxiliary channels provide the convolutional trunk with explicit physical inductive biases, such as scale-free matched-filter contractions and non-linear matrix invariants, which are challenging for a linear-filtering front end to synthesize under a finite simulation budget. Throughout, $\theta$ denotes the inferred physical parameters of Sec.~\ref{sec:models} and Sec.~\ref{sec:noise}. We describe in turn what we ask a network to learn from the pair $\hat{x}=(\theta, x)$, the transformer architecture, and the custom tokenization scheme.

\subsubsection{Neural posterior estimation and generative objectives}
\label{sec:objectives}

Traditional \ac{NPE}~\cite{papamakarios2016fast, greenberg2019} fits a conditional density $q_{\phi}(\theta \mid x)$ directly, with network parameters $\phi$, to pairs drawn from the simulator's joint distribution, by minimizing the loss $-\, \mathbb{E}_{p(\theta)\, p(x \mid \theta)} \left[ \log q_{\phi}(\theta \,\mid\, x) \right]$. The expectation runs over the entire joint distribution $p(\hat{x})=p(\theta)\, p(x \mid \theta)$, so any quantity the simulator draws and $\theta$ omits is integrated out, ``marginalized over''. Training is amortized over the whole prior. This holds only where the joint distribution has support. The direct route to representing $q_{\phi}$ is a normalizing flow, which parametrizes an invertible map with a tractable Jacobian and optimizes the aforementioned loss~\cite{papamakarios2021normalizing}.

The route we take, under the unified \ac{SI} framework for simplicity~\cite{albergo2025}, specifies a continuous path between a tractable reference distribution and the posterior, and learns a vector field that transports along it. Both our objectives share that construction, for a path time $t \in [0,1]$:
\begin{equation}
\hat{x}_{t} \;=\; \mu(t)\, \hat{x} \;+\; \sigma(t)\, \xi , \qquad \xi \sim \mathcal{N}(0, \mathbb{I}) ,
\label{eq:path}
\end{equation}
with $(\mu, \sigma)$ and time convention such that $t = 1$ recovers $\hat{x}$ and $t = 0$ is a distribution we can sample, usually Gaussian, which is the flow matching literature's convention~\cite{lipman2023, dax2023}. Please note, the Simformer formulation and the diffusion literature reverse the time-flow direction~\cite{song2021, gloeckler2024}. As a consequence, all diffusion t-functionals are reparametrized, so our $t=1-t_{\rm{SDE}}$, with $t_{\rm{SDE}}$ the native Simformer formulation. The two choices used here are
\begin{align}
\mu(t) &= 1 , & \sigma(t) &= \sigma_{\min} \left( \sigma_{\max} / \sigma_{\min} \right)^{1-t} , \label{eq:path_ve} \\
\mu(t) &= t , & \sigma(t) &= \max \! \left( 1 - t,\, \gamma_{\min} \right) , \label{eq:path_ot}
\end{align}
the first, a variance-exploding schedule that leaves $\hat{x}$ in place and inflates the noise geometrically, the second being the optimal-transport path standard in flow matching~\cite{lipman2023}. $\gamma_{\min}$ is a small numerical regularizer that we omit in writing from now on. The corresponding objectives are
\begin{align}
\mathcal{L}_{\rm DSM} &= \mathbb{E} \left[ \lambda(t) \left\| s_{\phi}(\hat{x}_t, t) - \nabla_{\hat{x}_t}\log p_t(\hat{x}_t\mid\hat{x}) \right\|^{2} \right] , \label{eq:dsm} \\
\mathcal{L}_{\rm FM} &= \mathbb{E} \left[ \left\| v_{\phi}(\hat{x}_t, t) - u_t(\hat{x}_t\mid\hat{x}) \right\|^{2} \right] , \label{eq:fm}
\end{align}
where $p_t(\hat{x}_t\mid\hat{x})=\mathcal{N}\!\left(\hat{x}_t;\mu(t)\hat{x},\sigma^2(t)\mathbb{I}\right)$ is the Gaussian conditional path density and expectations are over $t,\hat{x},\xi$. Equation~\eqref{eq:dsm} is \ac{DSM}~\cite{vincent2011, song2021}: $\lambda(t) = -{\rm d}\sigma^{2}/{\rm d}t$ on the variance-exploding path, needed because that target diverges as $\sigma \to 0$~\cite{song2021}. Equation~\eqref{eq:fm} is the sample-conditional \ac{FM} objective~\cite{lipman2023, dax2023}. Each loss is minimized by the conditional expectation of its target~\cite{albergo2025}, so the networks learn the score $s(\hat{x}_t, t)$ of the noised density or the velocity $v(\hat{x}_t, t)$ that carries the reference at $t = 0$ to the posterior at $t = 1$\footnote{For a fixed conditioning mask~\cite{gloeckler2024}, see also Sec.~\ref{sec:network}.}.

To obtain a posterior, we sample to integrate the learned field across the reverse-time dynamics. For score-based networks, this means:
\begin{equation}
{\rm d}\hat{x}_t \;=\; \tfrac{1}{2} \left( 1 + \kappa^{2} \right) g^{2}(t)\, s_{\phi}(\hat{x}_t, t)\, {\rm d}t \;+\; \kappa\, g(t)\, {\rm d}\bar{W} ,
\label{eq:reverse}
\end{equation}
with $g^{2}(t) =\lambda(t)$ for the variance-exploding schedule and $\bar{W}$ is a Wiener process. $\kappa = 1$ reproduces the reverse \ac{SDE} and $\kappa = 0$ the deterministic probability-flow \ac{ODE}, which shares its marginals~\cite{song2021}; for flow-matching we integrate ${\rm d}\hat{x}_t = v_{\phi}(\hat{x}_t,t)\, {\rm d}t$ over the same interval. \footnote{Throughout this work, \ac{DSM} objectives are sampled with the reverse \ac{SDE} unless stated otherwise, and \ac{FM} objectives with a Heun integrator on the velocity, both at $200$ steps. In addition, both allow for a probability-flow \ac{ODE} to be used to evaluate $\log q_{\phi}(\theta\mid x)$ exactly up to \ac{ODE} discretization as a continuous normalizing flow, which is needed for importance sampling, Sec.~\ref{sec:checks_intro}~\cite{dax2023is}.}

The choice of $\kappa$ (\ac{ODE} vs \ac{SDE}) has a main consequence: a deterministic map integrates whatever error the learned field carries coherently, so a systematic mistake in $s_{\phi}$ or $v_\phi$ becomes a fixed transport error that lands in the same place on every draw, whereas a stochastic sampler re-injects noise at each step and is repeatedly pulled back onto the model's own marginals.

\subsubsection{A transformer-based inference network}
\label{sec:network}

Either $s_{\phi}$ or $v_{\phi}$ are realised by the same network, a transformer~\cite{vaswani2017} over a sequence of tokens, following the all-conditional construction of Gloeckler et al.~\cite{gloeckler2024}; Fig.~\ref{fig:pipeline} shows the path from the stored data object to the posterior. A forward model first creates the spectral matrix as the sufficient data summary for stochastic processes. Each scalar component of $\theta$ is mapped to a token of width $d_{\rm tok}$ by a learned linear layer, and the data contributes $n_{\rm tok}$ tokens produced by the embedder of Sec.~\ref{sec:embedding}. The hand-crafted ``Frequency series embedder'' then creates the learned data summary and tokenization process, including additional information (derived and physics channels) and allowing for a flexible input channel number as well as token number and depth output.

To each token we concatenate a learned vector scaled by a binary flag recording whether it is conditioned on or sampled~\cite{gloeckler2024}, which are then passed to a transformer-based inference network~\cite{peebles2023}. Self-attention~\cite{vaswani2017} runs over the full sequence of $N_{\rm tok} = \dim\theta + n_{\rm tok}$ tokens, and a per-token linear head reads one component of the score or velocity from each parameter token. Because conditioning is a per-token flag instead of a fixed division between inputs and outputs, one trained network can in principle express the posterior, the likelihood, and any conditional in between~\cite{gloeckler2024}. We train in posterior mode throughout this work: the conditioning mask holds the data components $x$ fixed, so the path, objectives and sampling dynamics act only on the parameter components $\theta$; only $\theta$ is noised, so we learn the score/velocity fields corresponding to the posterior $p(\theta\mid x)$, rather than the joint $p(\hat{x})$. The number of inferred parameters is then merely a sequence length choice. Across this paper $\dim\theta$ runs from $17$ to $37$ and $N_{\rm tok}$ from $18$ to $72$ with nothing architecturally altered but the token counts. This permits our noise model to grow from two parameters to a per-link spline without a new network or a re-tuned architecture.  The embedder and inference network are trained jointly under a single loss. Finally, the resulting scores/velocities are used to integrate reverse time (through an \ac{SDE} or \ac{ODE}) to obtain a posterior.

Two distinct masks act on the token sequence: the conditioning flag above is one per token and selects which conditional is learned; the attention mask is a matrix over token pairs and selects which token may attend to which. This is where knowledge of the forward model enters the inference network directly, since a pairwise dependence the simulator cannot generate need not be representable. The conditional independences of the forward model constrain the score exactly only at the data end of the path. The Gaussian smoothing of Eq.~\eqref{eq:path} does not preserve the sparsity of the attention matrix, so away from $t = 1$ a restricted mask is an inductive bias. Every configuration with a stationary data object uses the unrestricted mask; the segmented configuration uses one derived from the dependence graph implied by Eq.~\eqref{eq:noise_model}: per-block noise levels are independent, see Sec.~\ref{sec:nonstationary}.

The cost of this generality is as follows: first, the number of simulations needed to train, which dominates the whole procedure is addressed in Appendix.~\ref{app:configs}. The second is the cost of a posterior: a normalizing flow samples in one forward pass, whereas both objectives integrate a solver, so what is lost relative to a flow is the amortization of sampling and not that of training, and the cost is then per observation, rather than per likelihood evaluation. The third is the shape of the network, with token count and token width: attention cost grows with the number of tokens while the work per token grows with its width, so the tokenization scheme is what primarily determines the cost of the inference network.

\subsubsection{Physics-aware embedding and tokenization}
\label{sec:embedding}

The data tokens fed to the transformer are at once the bottleneck and the freedom of this architecture. A bottleneck, because whatever the transformer can use about $x$ must be carried by $n_{\rm tok}$ vectors of width $d_{\rm tok}$; with $\dim\theta$ of order tens, a token per coarse bin would be impractical, since it would bury the parameters in a sequence dominated by the data. On the flip side, a token need not be a slice of the data: it can carry any deterministic function of it. It is the point at which the physics we know can be handed to the network instead of having to be learned, and it is what we mean by a physics-aware embedding.

The embedder reads a stack of channels over the coarse grid: Firstly, Eq.~\eqref{eq:spec_channels}, or a subset of it (the same stored object serves a single-channel, an autos-only and a full-matrix analysis). The second is model-free (the derived part of the embedder in Fig.~\ref{fig:pipeline}): two coordinate rows, $\log_{10} f_{j}$ rescaled to $[-1, 1]$ across the band and $N_{j}^{-1/2}$, and the rotation invariants (under a unitary change of channel basis):
\begin{equation}
\log_{10} \mathrm{tr}\, \tilde{S}_{j} , \qquad
\frac{I_{2}(\tilde{S}_{j})}{\left( \mathrm{tr}\, \tilde{S}_{j} \right)^{2}} , \qquad
\frac{\det \tilde{S}_{j}}{\left( \mathrm{tr}\, \tilde{S}_{j} \right)^{3}} ,
\label{eq:invariants}
\end{equation}
with $I_{2} = \tfrac{1}{2} [ ( \mathrm{tr}\, \tilde{S} )^{2} - \mathrm{tr}( \tilde{S}^{2} ) ]$. The two ratios are bounded by $1/3$ and $1/27$ by construction. The trunk below these is a convolution with equal padding, hence translation equivariant along the frequency axis, so without them, the only absolute-frequency information reaching a token would be which sub-band it came from. On the hybrid partition of Sec.~\ref{sec:datagen} the degrees of freedom $N_{j}$ span more than two orders of magnitude across the band, so the uncertainty of a coarse bin is information that helps separate the bin-uncertainty from parameter-uncertainty.

The third tier is physics-informed, in the sense that it commits to a model: a bank of fixed Hermitian templates $\mathsf{B}_{e}$ contracted against the data, one channel per template,
\begin{equation}
\frac{\mathrm{Re}\, \mathrm{tr} \! \left( \mathsf{B}_{e}^{\dagger}\, \tilde{S}_{j} \right)}
{\left\| \mathsf{B}_{e} \right\|_{F}\; \mathrm{tr}\, \tilde{S}_{j}} .
\label{eq:contraction}
\end{equation}
The normalization makes the channel scale-free, so any frequency envelope in $\mathsf{B}_{e}$ cancels and only its direction in the nine real dimensions of Hermitian $3 \times 3$ matrices remains. Eq.~\eqref{eq:contraction} is effectively the per bin integrand of an incoherent matched filter. Our three template kinds are then: the directions in which each modelled component pushes the covariance, the per-link directions along which a glitch enters through the transfer functions of Sec.~\ref{sec:transients}, and the least-sensitive direction of a reference noise covariance. Which bank is present is a per-configuration choice: none in Sec.~\ref{sec:stationary}, where the data is purely stochastic and coarse-graining is already sufficient, and the banks of Sec.~\ref{sec:gaussianity} and Sec.~\ref{sec:nonstationary} written out where they are used. Section~\ref{sec:gaussianity} also reports the same network with its bank switched on and off to showcase the contribution of these design choices.

The stack is then compressed. A one-dimensional convolutional trunk of U-Net form of~\cite{ronneberger2015}; its output is split into groups: each group is pooled over $K$ contiguous frequency sub-bands with more than one pooling statistic (called ``StatPool''), then mapped by a linear layer to the token width. Pooling over sub-bands rather than the whole band is an intentional choice: a token summarizing the entire band with one number per feature would rapidly discard frequency dependence\footnote{Throughout the experiments we ran in designing this embedder, we found that mean-pooling routinely washed out large gradients in frequency and systematically led to wider posteriors, even under different U-Net bottleneck sizes. We determined that 20 sub-band groups, coupled with three summary statistics for each: mean, logsumexp (temperature-scaled, has smoother gradients than a plain max) and the linear (first-order) coefficient of a weighted least-squares polynomial fit (essentially a gradient)—preserved the most amount of information.}. The trunk runs at full frequency resolution while the transformer only ever sees tens of tokens, so roughly $80$ to $90\%$ of the forward floating-point cost is in the embedder and not in the inference network. However, the data tokens are constant along a solver trajectory: they are computed once and reused for all sampling steps.

If an entire data object is reduced to a single token, the same construction applied to sub-records turns a record into a stream of tokens. This is how the non-stationary case is handled: the observation is cut into contiguous chunks, each reduced as above, and one shared trunk is applied to each, so the tokens index time, and the positional embedding of each is the epoch of its chunk. What attention then has to create from a sequence of per-chunk spectra is a time-frequency reconstruction. The details, including how a missing chunk is expressed, can be found in Sec.~\ref{sec:nonstationary}. The configurations used in the following sections, with their token layouts, network sizes, objectives and budgets, are collected in Appendix~\ref{app:configs}.

\subsection{Posterior checks}
\label{sec:checks_intro}

A reference posterior from classical methods exists only where the forward model is also the truth and the resulting likelihood is tractable. This is fully the case in Sec.~\ref{sec:stationary}, partially in Sec.~\ref{sec:gaussianity}—where a transient forces a non-central likelihood—and not at all in Sec.~\ref{sec:nonstationary}, where gaps and multiple transients would require a mini-global fit. Throughout this work, we use the following likelihood constructions: for data containing coarse-grained stationary stochastic signals only, the \ac{$L0$} construction is used in the figures below; when a single transient is added coherently, the resulting coarse-grained sufficient statistic is the \ac{$L1$}~\cite{ratnarajah2005}, which infers both the transient's parameters as well as the stochastic components'; lastly, we also consider a \ac{$L2$}, for the network agreement test in Sec.~\ref{sec:transients}. The exact forms, as well as a description of how likelihood information is canonically used in \ac{SBI}, are included in Appendix~\ref{app:likelihoods}, as they are standard tools.  

Alternatively, likelihood-free methods are commonly employed in the \ac{SBI} literature. Likelihood-free checks are those built from the simulator itself. Divided by scope, population-level tests certify the estimator over the whole prior but say nothing about the single observation; effects over the prior can easily blend into a correct-on-average metric. In between these sits a test over the neighborhood of an observation. 

At the population level we draw pairs from the joint, $p(\theta,x)$ and ask whether the truth sits appropriately in the posterior. \ac{SBC}~\cite{talts2018} does this marginally, ranking each true parameter among draws from its corresponding posterior: the ranks are uniform for a calibrated estimator. We display the ranks, normalized to $u \in [0, 1]$, through the difference between their \ac{ECDF} and the uniform one, together with the simultaneous confidence band of an exactly calibrated estimator~\cite{sailynoja2022}; we also summarize each test by a single number when using many parameters/networks: the Kullback--Leibler divergence, in nats, from the standard normal (the law of $\Phi^{-1}(u)$ for uniform ranks) of the Gaussian with the same mean and variance:
\begin{equation}
D_{u} \;=\; \tfrac{1}{2} \left( m^{2} + s^{2} - 1 - \ln s^{2} \right) ,
\label{eq:kl_rank}
\end{equation}
where $m$ and $s^{2}$ are the mean and variance, over the simulated datasets, of $\Phi^{-1}(u)$, with $\Phi$ the standard normal distribution function. $D_u$ vanishes for a calibrated estimator and, whatever the shape of the posteriors, it bounds from below the divergence between the true and the estimated posterior averaged over datasets (Appendix~\ref{app:sbc})\footnote{As a reference scale of $D_u$, a posterior displaced by $0.15$ standard deviations, or one that is $10\%$ too narrow, gives $D_{u} \approx 10^{-2}$ assuming infinitely many draws. We use this as our tolerance.}. Therefore, $D_u\coloneq \mathrm{KL}(\mathcal{N}(m, s^2)||\mathcal{N}(0,1))\leq1/N$, with N the number of ranks (simulated datasets used). A marginal test is blind to a misestimated correlation, so we pair it with expected coverage of a credible region, either with random reference points~\cite{lemos2023} or with the highest-density ordering~\cite{hermans2021}.

A population-level verdict need not transfer to a single observation. Posterior \ac{SBC}~\cite{sailynoja2026} therefore repeats the test with the posterior at the observation in the role of the prior. It requires posteriors conditioned on the observation together with each simulated replicate, which single-realization networks cannot provide. However, Bayes' theorem holds for any prior, and under a reweighted prior $p_{g}(\theta) \propto p(\theta)\, g(\theta)$, for any non-negative function $g$ in the support, the posterior is the original one similarly reweighted. An estimator that is correct under $p(\theta)$ is therefore correct under $p_{g}$ once its draws carry the weights $g$, see Appendix~\ref{app:sbc} for the proof. 

We call this ``localised \ac{SBC}''~\cite{picchini2026} and use it around an observation of interest $x_{\rm obs}$. We take $g\propto \mathcal{N}$, with its mean in Eq.~\ref{eq:tilt_gauss_mean} and covariance in Eq.~\ref{eq:tilt_gauss_cov}, draw parameters $\theta^{*} \sim p_{g}$, simulate a dataset $x$ for each, and rank $\theta^{*}$ among $K$ network draws $\theta_{k}$ with the weights $w_{k} = g(\theta_{k})$ and $w_{*} = g(\theta^{*})$:
\begin{equation}
u \;=\; \frac{\sum_{k=1}^{K} w_{k}\, \mathbf{1} \! \left[ f(\theta_{k}) < f(\theta^{*}) \right] \;+\; V\, w_{*}}{w_{*} + \sum_{k=1}^{K} w_{k}} ,
\label{eq:weighted_rank}
\end{equation}
with $\mathbf{1}[\cdot]$ the indicator, $V$ uniform on $[0, 1]$, and $f$ a test quantity\footnote{Following the philosophy in~\cite{modrak2023}, our test quantities are the parameter marginals themselves as is common in \ac{SBC}, but we also consider two additional test quantities: the two best-constrained directions of the posterior at $x_{\rm obs}$.}. For a correct estimator $u$ is exactly uniform, for any $g$ and any $K$, and $g = 1$ recovers the ordinary rank (Appendix~\ref{app:sbc}). The statement this supports concerns datasets like the observation, not the observation itself, but unlike the over-the-prior classical version of \ac{SBC}, the restrictive volume here is significantly less prone to posterior distortions cancelling each other out. And because these test quantities depend on the parameters alone, an estimator that uses only part of the information in the data can still pass~\cite{modrak2023}: the test therefore certifies calibration, not that the data have been used in full (which would require a likelihood, or some other data-derived test quantity, which we leave for future work). 

The checks can thus be ordered by their requirements: reference posteriors and importance sampling need the likelihood, population tests only the simulator, and localized SBC the simulator and one observation. The following sections progressively drop the first kind while keeping the other two, starting from stationary, Gaussian and transient-free data, where every check applies and the network can be compared to the exact posterior.

\begin{figure*}
    \centering
    \includegraphics[width=0.7\linewidth]{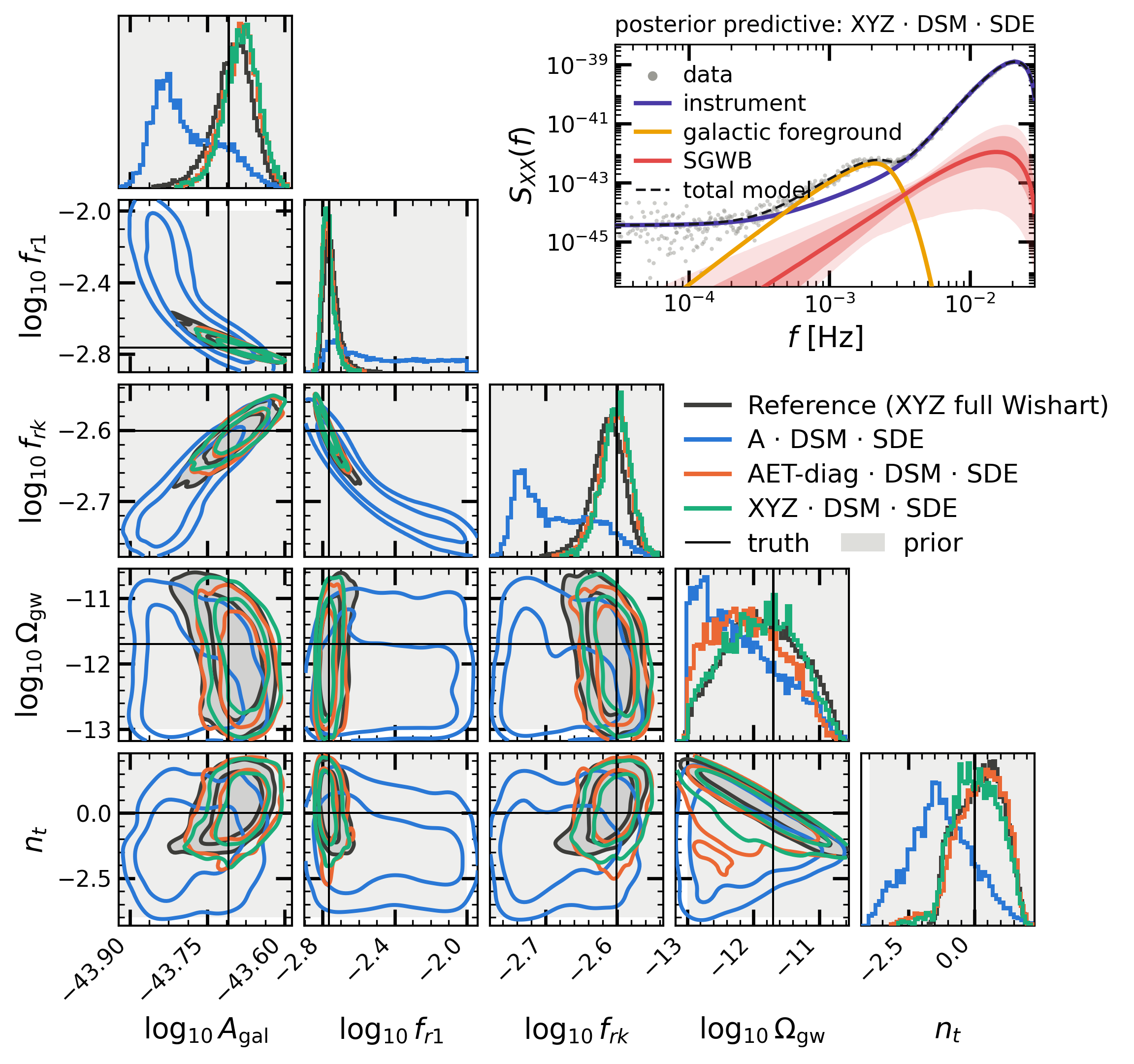}
    \caption{Joint posteriors for the three Galactic-foreground and two \ac{SGWB} parameters, from the \ac{DSM}/\ac{SDE} networks: $A$ channel-only (blue), $AET$-diagonal (orange), $XYZ$ (green); against the exact $XYZ$ Wishart reference (black-grey). Black lines mark the injected values; contours are the $68\%$ and $95\%$ highest-density regions; the grey background  is the prior. All four posteriors are conditioned on the same data, so the spread between them is information lost by considering a partial spectral-matrix. Inset: the measured $X$-channel coarse-grained periodogram (grey points) with the posterior-predictive instrument, Galactic-foreground and \ac{SGWB} spectra drawn from the $XYZ$ (\ac{DSM}, \ac{SDE}) network, each as its median and intervals; the dashed curve is their sum.}
    \label{fig:comp_sep_sci}
\end{figure*}

\section{Stochastic component separation with diffusion}
\label{sec:stationary}

To showcase our networks' performance in a controlled environment, we consider $11.6$ contiguous days of data populated only by stochastic contributions: per-link component instrumental noise, Eq.~\ref{eq:perlink}, Galactic foreground Eq.~\ref{eq:galaxy} and an \ac{SGWB} power-law contribution Eq.~\ref{eq:sgwb}. We utilize unequal arm-length orbits (Keplerian), second generation \ac{TDI}, and investigate the impact on parameter recovery of considering a single channel ($A$), the usual diagonalization ($AET$), and the full Michelson-combination spectral matrix ($XYZ$). With the data construction described in Sec.~\ref{sec:datagen} and only stochastic contributions, the reference is the central Wishart likelihood of Eq.~\ref{eq:wishart_like}.

\begin{figure*}
    \centering
    \includegraphics[width=0.95\linewidth]{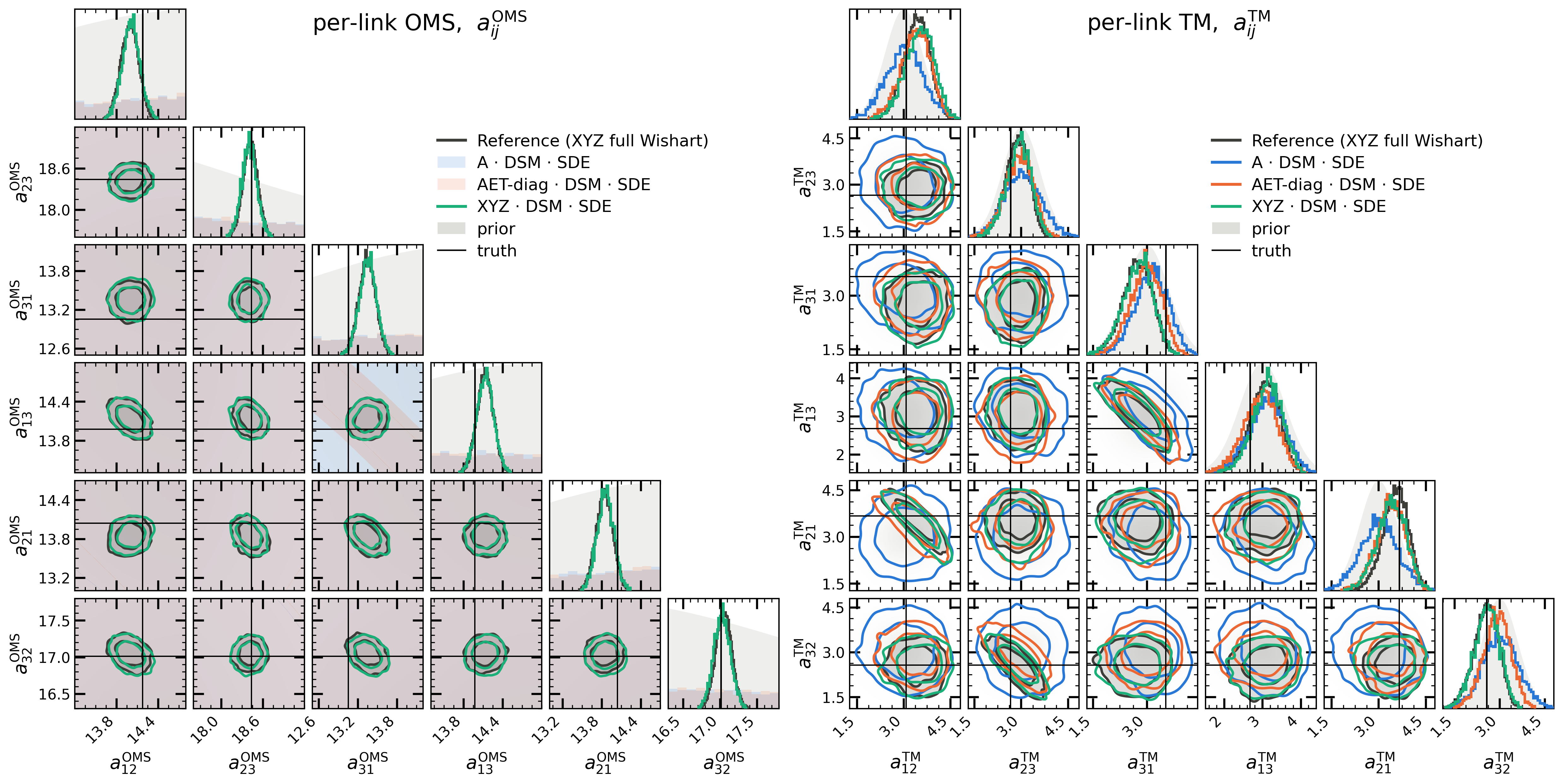}
    \caption{Per-link instrument amplitudes, split by noise component: the six \ac{OMS} amplitudes (left) and the six \ac{TM} amplitudes (right). Colours, reference and injected values as in Fig.~\ref{fig:comp_sep_sci}. Left: the axes are zoomed onto the reference and $XYZ$ posteriors, which have a marginal standard deviation of $0.12$ against a prior of $3.0$. $A$ and $AET$-diag are $~20\times$ wider (i.e. at the prior, the grey background) and are drawn as a filled wash instead of a full contour. Their one \ac{OMS} direction visible in the marginals is the $3-1$ arm sum $a_{31}+a_{13}$, which is why the wash in the $(a_{31}, a_{13})$ panel alone has a visible diagonal edge. Right: the \ac{TM} amplitudes are weakly measured by every data combination, except the arm-symmetric-link degeneracies, which are resolved by $AET$/$XYZ$.}
    \label{fig:comp_sep_noise}
\end{figure*}

In addition to the incomplete-data study, the performance is benchmarked (in terms of accuracy) by utilizing the Simformer native's objective (\ac{DSM}) versus the state of the art in generative modeling in \ac{SBI} (\ac{FM}) by an objective-only ablation study. Sec.~\ref{sec:comp_sep_res} discusses the effect of incomplete-data through a single observation, while Sec.~\ref{sec:comp_sep_check} includes network calibration and the objective ablation.

\begin{figure}
    \centering
    \includegraphics[width=0.8\linewidth]{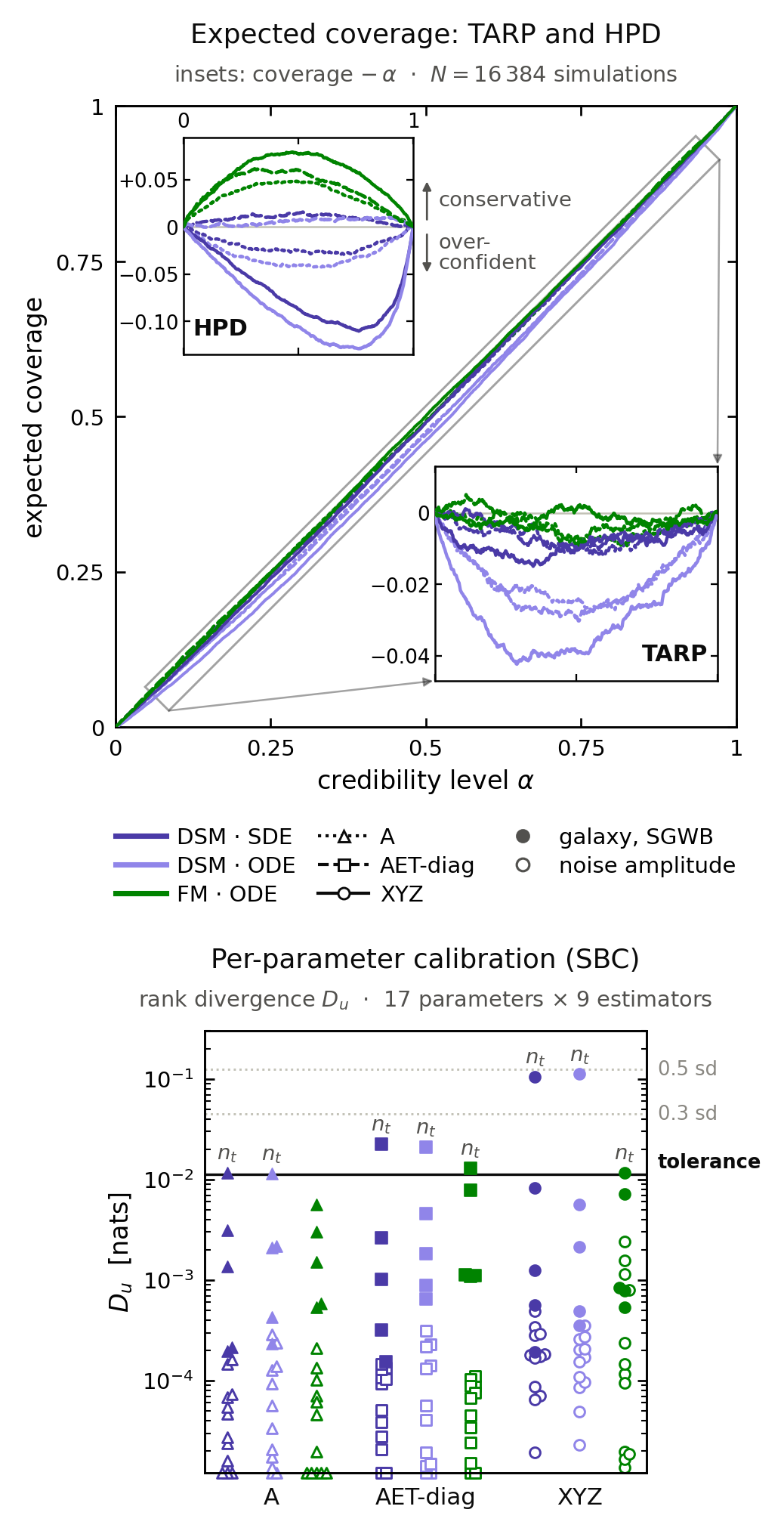}
    \caption{Population-level calibration of the nine network ($A$, $AET$-diag, $XYZ$ each paired with \ac{DSM} \ac{SDE}/\ac{ODE} and \ac{FM}) posteriors. Colours distinguish objective and sampler, line styles and markers the channel set. Top: expected coverage of credible regions built around random reference points (\ac{TARP}) against the credibility level $\alpha$. The lower-right inset magnifies the framed band along the diagonal as coverage $-\,\alpha$, and the upper-left inset shows the same quantity for \ac{HPD} regions of a Gaussian fit to each posterior. Bottom: \ac{SBC} of each of the $17$ parameters, one marker per parameter and posterior, summarized by the rank divergence; filled markers are the Galactic-foreground and \ac{SGWB} parameters, open ones the per-link noise amplitudes. For reference, we plot a predefined tolerance as the $D_u$ of a Gaussian posterior displaced by $0.15$ (solid) of its standard deviation (also $0.3$ and $0.5$ dotted).}
    \label{fig:comp_sep_checks}
\end{figure}

\subsection{Effect of $A$, diagonal $AET$ or $XYZ$} 
\label{sec:comp_sep_res}

Fig.~\ref{fig:comp_sep_sci} displays the ``science'' parameters, foreground and \ac{SGWB}, of the partial-channel information study, across three diffusion-based (\ac{SDE}) networks, as well as the ground truth Wishart likelihood for one observation. For the $12$-parameter, per-link noise model of Eq.~\ref{eq:perlink} (at a per-link draw with a $20\%$ scatter about the specification values, shared by the reference observation of every section), this foreground achieves an \ac{SNR} of $\sim250$ over the observation period, while the \ac{SGWB}'s \ac{SNR} is $\sim5$. The $XX$ component of the spectral matrix (grey dots) is represented in the Fig.~\ref{fig:comp_sep_sci} inset: instrumental noise (dark blue), foreground (yellow), \ac{SGWB} (red) and their sum (dashed black), as well as their $1\sigma$ and $2\sigma$ contours for the full-data \ac{DSM} (\ac{SDE}) network.

The conditioning data is shared across the networks, including the noise realization, with the only difference being the information content extracted from the full spectral matrix. In the corner plot of Fig.~\ref{fig:comp_sep_sci}, the blue posterior is a network trained on single-channel ($AET$ basis, the $A$ channel) data, orange is a diagonal-only-trained network, while green is the complete spectral matrix-trained network, with the reference Wishart in grey. The $A$/$AET$ conditioning data was obtained by rotating the $XYZ$-basis spectral matrix used for the $XYZ$ network and Wishart reference and selecting the $(1,1)$ or diagonal-only components, respectively.

In line with previous work~\cite{hartwig2023,vajpeyi2026}, we find that the $AET$-diagonalization is suitable in replicating the full $XYZ$ spectral matrix's sensitivity to a loud Galactic foreground as well as a faint \ac{SGWB}, even in an unequal arm-length, unequal per-link noise scenario for a short $11.6$ day observation window. Unsurprisingly, the approximately orthogonal basis, implies that considering only the $AA$-component  results in a strictly lower \ac{SNR} and worse  constraints. This result, however, is a consequence of the assumptions made (isotropic foreground and background as well as stationary noise) and only holds for the ``science'' parameters.

To see why, we showcase in Fig.~\ref{fig:comp_sep_noise} the per-link noise posterior for the same networks and data-conditioning scheme. The left-hand-side corner plot displays the per-link \ac{OMS} amplitude contributions. Beside the clear agreement of our $XYZ$-conditioned network (green) to the reference posterior, it is immediately noticeable that the $A$/$AET$-conditioned networks (blue and orange, filled) provide significantly wider posteriors. In fact, they recover the prior, with the only exception being the $13-31$ link-sum degeneracy being mildly constrained by the $AET$ network. The motivation for this difference between $AET$/$XYZ$ is that the \ac{OMS} parameter's \ac{SNR} sits predominantly in the off-diagonal terms\footnote{An observation also confirmed by a Fisher analysis and \ac{MCMC}.}. Conversely, the per-link \ac{TM} amplitude contributions (right-hand-side corner plot) are mostly unconstrained by the available data (true of both the reference and all three networks). As expected due to the symmetric-in-link \ac{TM} noise response, the best constrained are the symmetric $\{12-21, 13-31, 23-32\}$ link-sums, whose degeneracy structure is recovered from the prior by both $AET$/$XYZ$-conditioned networks.

Before we address the network's performance, it is perhaps instructive to make a remark on the above results: In the simplified setting of an $11.6$-day observation (and thus defensible to freeze the naturally time-varying response), we already notice clear ability to constrain the per-link noise parameters (at least of \ac{OMS} and symmetric \ac{TM} directions). As we diminish the assumptions (including transients, realistic orbits, a time-varying response, and flexible noise models) ignoring part of the our instrument's data is a losing strategy. Instead, we should aim to increase the complexity of both the data and our network's ability to extract information from said data. To this end, developing trust in the fidelity of a posterior estimator is critical, from both a likelihood-informative and a likelihood-starved perspective.

\subsection{Calibration and the choice of objective}\label{sec:comp_sep_check}

We present the population-level calibration checks with \ac{TARP}~\cite{lemos2023} and \ac{SBC} (with a rank divergence summary) for all the networks and data-inclusion levels in Fig.~\ref{fig:comp_sep_checks}. The upper panel shows deviations from expected coverage under \ac{HPD} and \ac{TARP} over $16384$ trials. Because of the much higher-than-common in the \ac{GW} literature trials used, we do not plot null bands here, since they can become arbitrarily small, instead focusing on effect sizes: although \ac{FM} (green) outperforms the \ac{DSM} objective under either sampler, all three, across all data-incompleteness settings, are well-calibrated over the full-dimensional parameter space (\ac{TARP} and \ac{HPD}). Investigating the marginals in the lower-plot \ac{SBC} summary, it is clear that consistently over the parameters, the highest deviations from uniformity (least calibrated) are the spectral tilt $n_t$, of the \ac{SGWB}. For the interpretability of $D_u$, Eq.~\ref{eq:kl_rank}, we plot the equivalent posterior deviations that would result in a similar $D_u$; the tolerance is chosen by a $0.15\sigma$ bias (alternatively a $10\%$ too narrow posterior) of a distorted Gaussian. In other words, the effect sizes here are also small, but nonetheless we identified the source: the networks' difficulty in inferring $n_t$ seems to mainly be driven by loud \ac{SGWB} (hundreds to thousands in \ac{SNR}), where the resulting hyper-narrow posterior (relative to the prior) becomes a hard inference target for both objectives.\footnote{Curriculuum learning in particular is a simple solution for this; in the case of more extreme prior-to-posterior contractions, both learning and objectives would need to be modified.}

\begin{figure}
    \centering
    \includegraphics[width=0.9\linewidth]{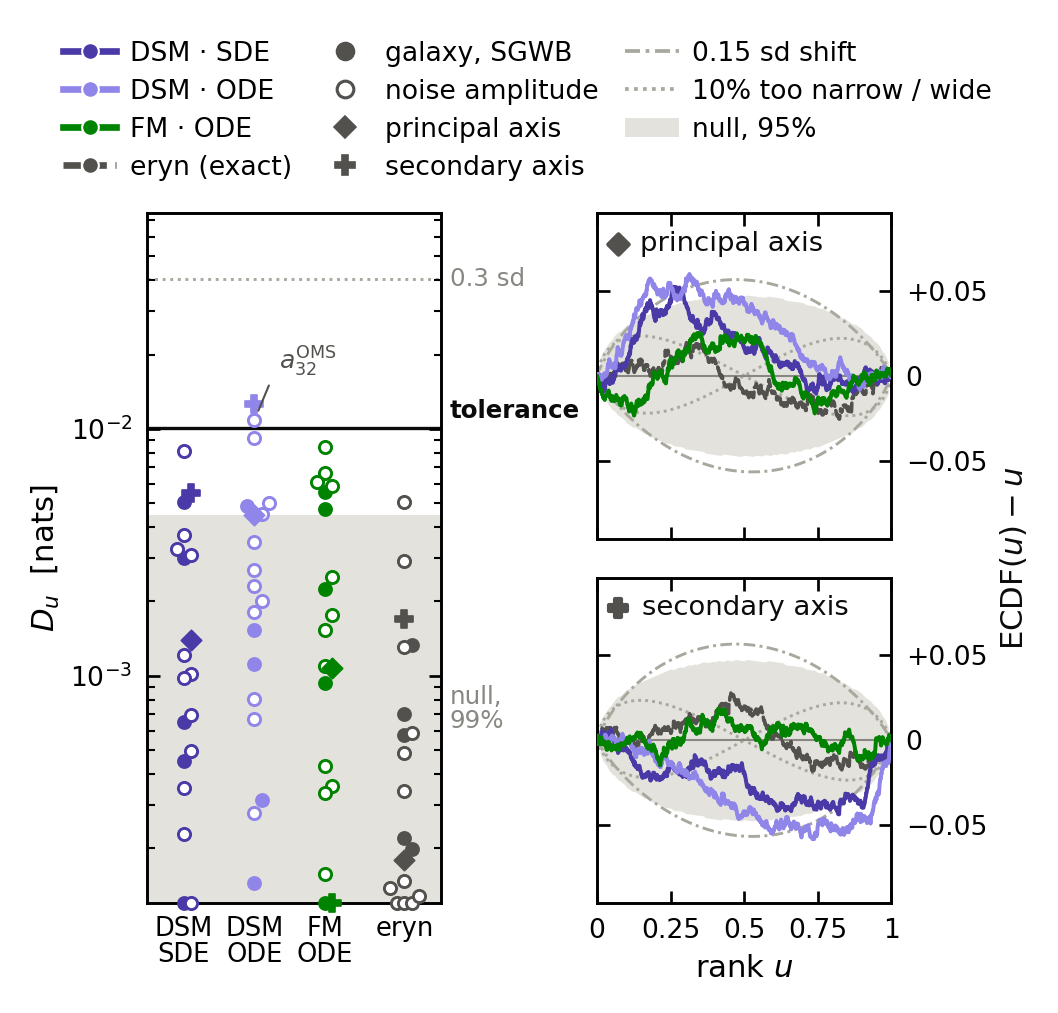}
    \caption{Localised \ac{SBC} of the three $XYZ$ posteriors around the reference observation $x_{\rm obs}$: $1024$ datasets simulated from parameters drawn under the tilt $g$ of Sec.~\ref{sec:checks_intro}, each ranked among $K = 256$ network draws with the weighted rank of Eq.~\eqref{eq:weighted_rank}. \ac{MCMC} reference (Eryn~\cite{karnesis2023eryn}) samples the exact likelihood and so shows what an exactly calibrated estimator reads. Left: rank divergence of the $19$ test quantities of each estimator, namely the $17$ marginals (filled: Galactic-foreground and \ac{SGWB} parameters; open: per-link noise amplitudes) and the principal (diamond) and secondary (plus) axes. Right: empirical distribution function of the ranks minus the uniform one, for the principal and secondary axes. These are the two best-constrained directions of the posterior at $x_{\rm obs}$. Dash-dotted and dotted curves are defined as in Fig.~\ref{fig:comp_sep_checks}. With $1024$ datasets the range of an exactly calibrated estimator is comparable to these scales, so we draw it in grey: the level that $99\%$ of single readings of $D_u$ stay below (left), and the $95\%$ simultaneous band~\cite{sailynoja2022} (right).}

    \label{fig:comp_loc_sbc}
\end{figure}

Fig.~\ref{fig:comp_loc_sbc} showcases the localized \ac{SBC} construction for the $XYZ$ posterior of the same observation as in Fig.~\ref{fig:comp_sep_sci} and Fig.~\ref{fig:comp_sep_noise}. We simulated $1024$ datasets in the posterior-region according to the protocol described in Sec.~\ref{sec:checks_intro} and computed the rank-divergence summary statistic as well as \ac{ECDF} curves. The left hand side of Fig.~\ref{fig:comp_loc_sbc} displays the calibration on posterior-like observations for the three $XYZ$ networks (\ac{DSM}-\ac{SDE}/\ac{ODE} and \ac{FM}) as well as reference \ac{MCMC} (Eryn~\cite{karnesis2023eryn}) for comparison. The right hand side is the \ac{ECDF} for the two best-constrained directions, in units of the prior width: the principal axis in prior-width units, is $\propto 0.67 \log_{10} f_{r1} + 0.62 \log_{10} f_{rk} + 0.40 \log_{10} A_{\rm gal}$ and the secondary axis $\propto 0.90 \log_{10} f_{rk} - 0.41 \log_{10} A_{\rm gal} + 0.14 \log_{10} f_{r1}$, with negligible other-parameter coefficients. 

We also quote \ac{IS} effective sample size and sampling efficiencies, since for the data presented here the likelihood is exact: The \ac{DSM}, \ac{ODE} network achieves a sampling efficiency $\sim19\%$ with an \ac{ESS}$=964$ while the \ac{FM} network has a sampling efficiency $\sim21\%$ with an \ac{ESS}$=1046$. In addition, the sliced 2-Wasserstein distance before \ac{IS} sits at $0.35$ and $0.26$ for \ac{DSM} and \ac{FM} respectively, while after \ac{IS}, we have $0.12$ and $0.12$, with a null~\cite{tran2026} of $\sim 0.09$, so effectively undistinguishable distributions, as expected.

Finally, before increasing the data complexity, a remark on the performance of the training objectives: a denoising objective (\ac{DSM}) or a velocity objective (\ac{FM}) both match calibration performance across the prior well, when using their native samplers. Both Fig.~\ref{fig:comp_sep_checks} and Fig.~\ref{fig:comp_loc_sbc} showcase the \ac{ODE} solver with a denoising objective being consistently the worst performer (still with minimal effect sizes, however).

\section{Breaking Gaussianity}
\label{sec:gaussianity}
\begin{figure*}
    \centering
    \includegraphics[width=\textwidth]{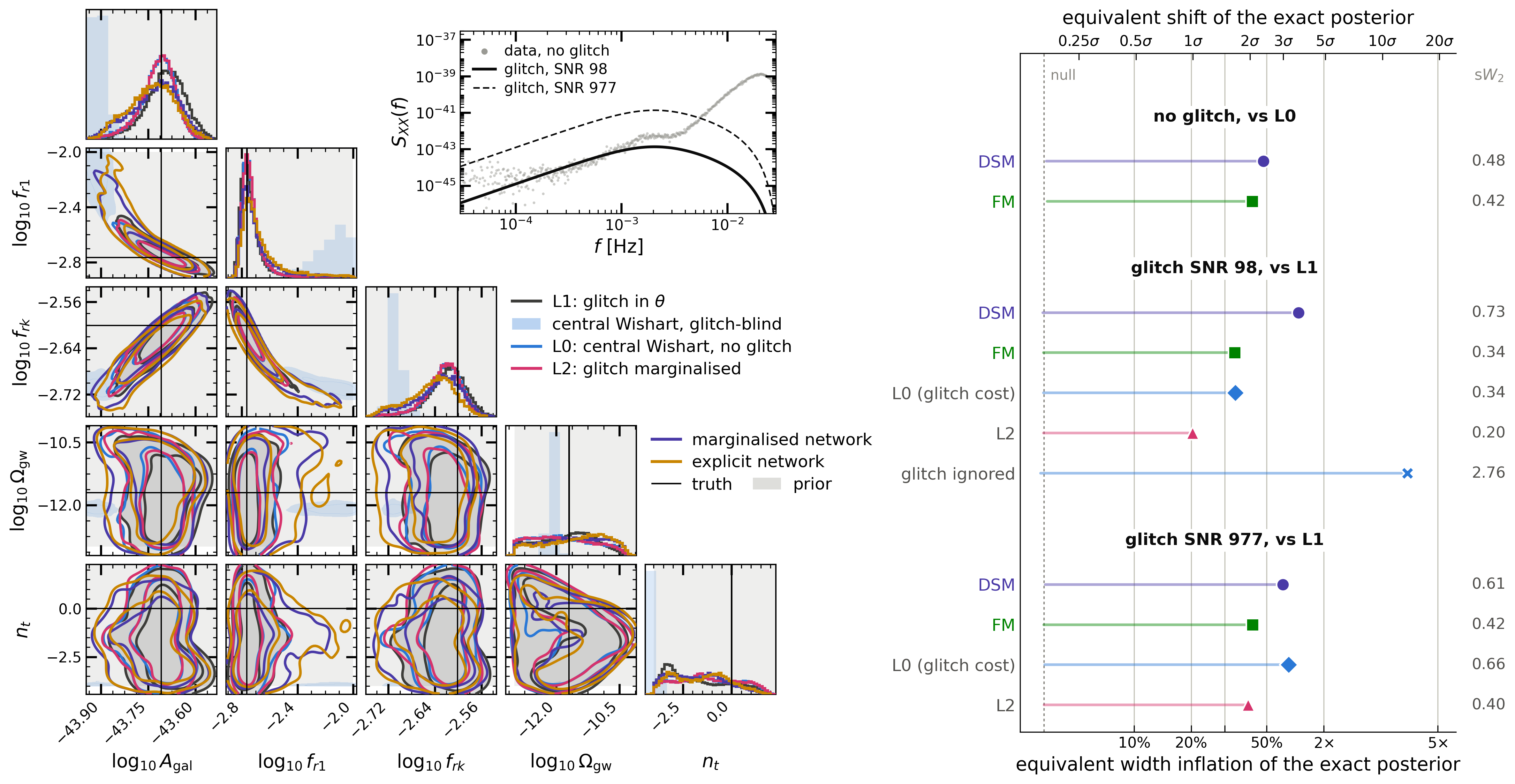}
    \caption{Left: joint
    posteriors of the three Galactic-foreground and two \ac{SGWB} parameters. The grey contours are the \ac{$L1$} likelihood, which carries the glitch in $\theta$ and serves as the reference; red is the \ac{$L2$} likelihood, which marginalizes the glitch; the blue line is \ac{$L0$}, the central Wishart likelihood on the same data without the glitch, and the pale blue histogram is the same likelihood on the glitched data (the glitch ignored). Purple and amber are the \ac{DSM} networks trained to marginalize the glitch and to infer it explicitly. Contours are the $68\%$ and $95\%$ highest-density regions, black lines mark the injected values and the grey background is the prior. Inset: the coarse-grained $X$-channel periodogram of the data and glitches. Right: the sliced 2-Wasserstein distance ($sW_2$, printed in the column) of each posterior from
    the exact one (\ac{$L0$} without a glitch, \ac{$L1$} with it). Each value is placed on two equivalent scales: the width inflated or rigidly shifted posterior's equivalent $sW_2$ distance. The dotted line is the null, $sW_2 = 0.03$
    between two independent draws of the same posterior. \ac{DSM} and \ac{FM} in purple and green, respectively.}
    \label{fig:agreement}
\end{figure*}

\ac{GW} data is not Gaussian, generally because of instrumental glitches appearing. If one can model said glitch, it can be treated like a signal transient, and thus the Gaussianity of the remaining instrumental noise is preserved. In \ac{LISA} data, with the real possibility of simultaneous and morphologically distinct and ambiguous glitches appearing, and the certainty of a signal-dominated regime, this is not as trivial. Therefore, ``breaking gaussianity'' here means that \ac{SBI} methods need not assume a transient morphology (in this work a glitch or an \ac{MBHB}) to be able to estimate stochastic parameters of interest, a relevant feature for both our considered transients: glitch morphology need only be simulated~\cite{dooney2024,dooney2026}, not analytically tractable, while waveform systematics for \ac{MBHBs} can be ignored\footnote{For example, by training with multiple waveform approximants, since waveform surrogate modelling is not as reliable as glitch simulators, where morphological overlap need not be as exact.}.

This section will present the performance of our architecture for stochastic component separation in the presence of transients. The same $11.6$-day in frequency-domain setup for the data, which includes: Galactic foreground and \ac{SGWB}, with the noise spectrum shape freedom through splines as well as per-link amplitude freedom, Eq.~\ref{eq:noise_model} without non-stationarities (the block amplitudes). In addition to these, we include a single transient in the training data with $50\%$ probability (or $100\%$ when asking the network to estimate the glitch parameters as well). 

Sec.~\ref{sec:agreement_test} presents stochastic component separation in the presence of a glitch on link $12$, under a marginalization agreement test; estimating the glitch parameters versus marginalizing over them, networks versus analytic, must all result in the same posterior distributions. Sec.~\ref{sec:MBHBs_physics_emb} then presents stochastic component separation in the presence of an \ac{MBHB} as well as quantifies the effect of our physics-aware embedding scheme with a network-ablation test. Finally Sec.~\ref{sec:pop_checks} discusses the calibration of the networks and the impact of the transient's \ac{SNR}.

\subsection{Marginalizing over glitches}\label{sec:agreement_test}

To certify the ability of our networks to marginalize out a glitch present in the training data, whose parameters are inferred, we compare four posterior distributions: a network trained with glitches present in the data, but without estimating their parameters; another network that is asked to estimate the glitch parameters jointly with the stochastics; and the two likelihoods of Eq.~\ref{eq:non_central_wishart} (joint parameter estimation of glitch and stochastic contributions) and Eq.~\ref{eq:ML_marg_glitch} (maximum-likelihood expansion and marginalization over the glitch parameters).

Fig.~\ref{fig:agreement} summarizes the agreement test with the science posterior on the left and the high-dimension ($24$-parameter) sliced 2-Wasserstein distance metric on the right. The corner plot includes two networks (\ac{DSM}, \ac{SDE}) in purple and amber, the former implicitly marginalizing the glitch, while the latter is asked to estimate its parameters (so we explicitly select only the science parameters of the full posterior). The blue curve and pale blue histogram correspond to the standard Wishart likelihood, the former applied to de-glitched data while the latter to the glitched data, like the $X$-channel frequency spectrum inset on the left-hand-side plot. Finally, the grey and red contours are the glitch-estimated and glitch-marginalized likelihoods, with the former representing the most accurate reference posterior here.

Although it is immediately clear that the science posteriors agree, we wish to quantify if this is true, including the noise parameters. To this end, the right-hand-side of Fig.~\ref{fig:agreement} displays the sliced 2-Wasserstein distance (insets on the right) using two calibrated rulers for interpretability: the width inflation of the exact posterior, every deviation from its mean scaled by $1+x$, that gives the same $sW_2$ (bottom, $sW_2 = x$), and the rigid shift of the exact posterior by $\delta$ of its standard deviations along the shift direction that gives the same $sW_2$ (top, $sW_2 = \delta/\sqrt{d}$). The first thing to note is the blue cross, the distance between the \ac{$L0$} and \ac{$L1$} posterior for a \ac{SNR} $100$ glitch, quantifying the effect of ignoring the glitch, i.e. a catastrophic bias on the posterior estimate, increasing with \ac{SNR} as expected, not plotted as it is far to the right. Another interesting reference point is the middle-rung's blue diamond, \ac{$L0$} on de-glitched data versus \ac{$L1$}, which is the information cost of having such a glitch in the data; considering the low \ac{SNR} and singular nature of the glitch and assuming correct modelling (which is the case here by construction) one does not expect the glitch to significantly disturb the parameter estimation of the stochastic components, which does indeed seem to be the case here. Most relevant for the networks' performance however, we see that both \ac{DSM} and \ac{FM} (purple circles and green squares) are comparable in effect-size, across the no-glitch, \ac{SNR} $\sim100$ and $\sim1000$ glitch cases to the reference correct likelihood \ac{$L1$}. In other words, the networks are about as close to the exact posterior as a perfectly glitch-subtracted residual analysis is.

\begin{figure*}
    \centering
    \includegraphics[width=0.7\linewidth]{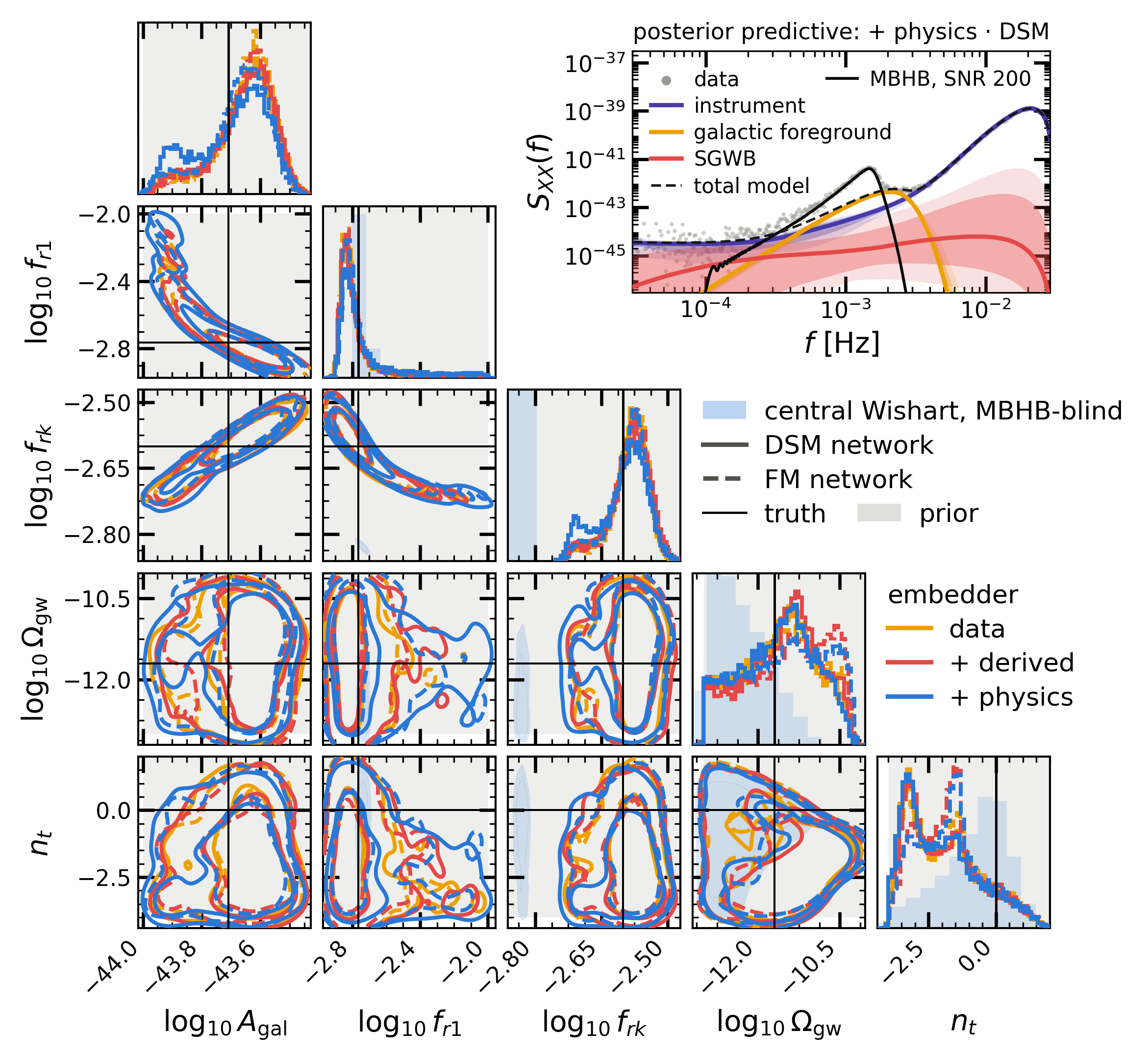}
    \caption{Joint posteriors of the Galactic-foreground and \ac{SGWB} parameters with an \ac{MBHB} in the data: the noise realization and parameters of Fig.~\ref{fig:agreement}, with an \ac{MBHB} at \ac{SNR} $200$ in place of the glitch. The six networks differ in objective, \ac{DSM} (solid) or \ac{FM} (dashed), and in what their embedder reads (Sec.~\ref{sec:embedding}): the spectral matrix alone (orange), with the derived channels added (red; band coordinates and the invariants of Eq.~\eqref{eq:invariants}), or with the physics channels added as well (blue; the template contractions of Eq.~\eqref{eq:contraction}). No exact reference due to computational cost on an \ac{MBHB} template for \ac{$L1$}; the pale blue histogram is the central Wishart likelihood on the same data, blind to the \ac{MBHB}, whose $\log_{10} A_{\rm gal} \simeq -42.6$ lies outside the plotted range. Contours, truth, prior and inset as in Fig.~\ref{fig:agreement}.}

    \label{fig:MBHB_sci}
\end{figure*}

\begin{figure}
    \centering
    \includegraphics[width=0.9\linewidth]{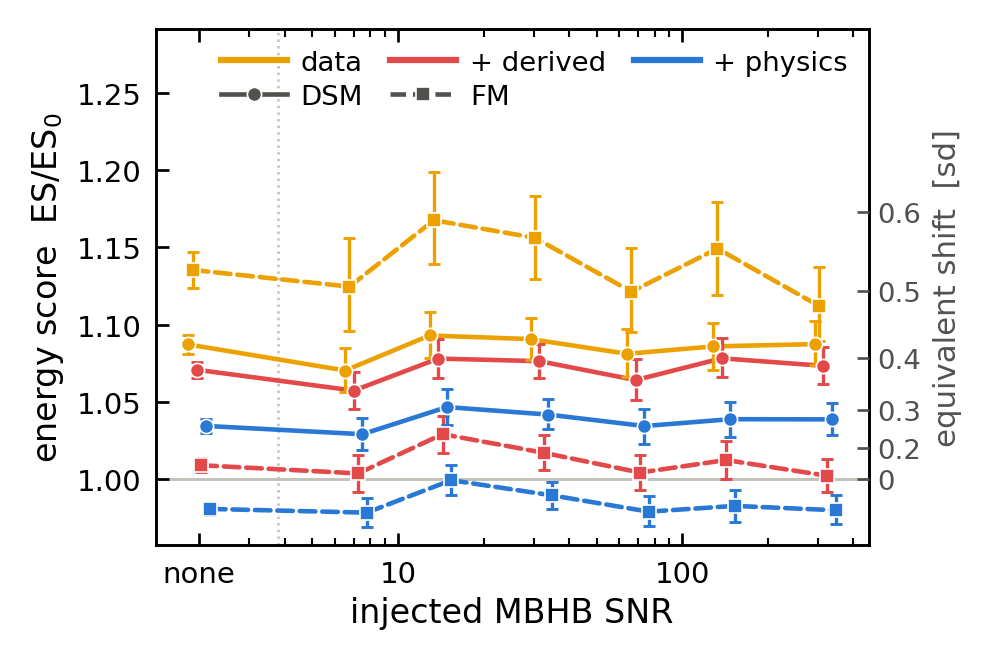}
    \caption{Energy score of the six \ac{MBHB} networks against the injected parameters, by \ac{MBHB} signal-to-noise ratio (``none'': the trials without an \ac{MBHB}). Colours and line styles as in Fig.~\ref{fig:MBHB_sci}. For each test simulation, the energy score distance proxy from the truth is computed in the Mahalanobis metric of the mean posterior covariance of the six networks on that simulation. It is shown relative to ${\rm ES}_0$, the score of an exact Gaussian posterior with that covariance, and the ideal in expectation over datasets is $1$. Points are means over the simulations in each bin, and bars are $95\%$ bootstrap intervals. The right ordinate translates the score into the displacement, in every direction and in units of the posterior standard deviation, that raises an exact Gaussian posterior from ${\rm ES}_0$ to that level.}

    \label{fig:MBHB_es}
\end{figure}

\subsection{MBHBs and physics-informed embedders}\label{sec:MBHBs_physics_emb}

The other expected-high-\ac{SNR} transient, an \ac{MBHB}, is a $11$-dimensional waveform, whose response is projected onto all six links; there is no insensitive channel, unlike a test-mass glitch. The maximum-likelihood expansion and marginalization of Eq.~\ref{eq:ML_marg_glitch} as well as the jointly-inferred likelihood of Eq.~\ref{eq:non_central_wishart} would not be as straightforward to apply; the former due to numerics, the latter due to computational cost. Therefore, we display in Fig.~\ref{fig:MBHB_sci} only network posteriors (\ac{DSM} continuous, \ac{FM} dashed), and the \ac{MBHB}-blind central Wishart likelihood as a pale blue histogram, biased away from the truth. In addition, to determine the effect of our physics-aware embedder, we plot the results of an ablation study on the input channels. Spectral matrix-only (the data) is in orange, data plus derived quantities (rotation invariants, coarse-graining information) in red, and data plus derived quantities plus physics (projections, template banks) in blue.

On the science-posterior-only of Fig.~\ref{fig:MBHB_sci} all objective and embedder combinations seem to perform similarly, so we complement it with a full-dimensional posterior-to-truth summary. Fig.~\ref{fig:MBHB_es} shows the energy score, ${\rm ES} = \mathbb{E}_q\|\theta - \theta^{*}\| - \tfrac{1}{2}\mathbb{E}_q\|\theta - \theta'\|$, with $\theta, \theta' \sim q_\phi(\theta\mid x)$ and $\theta^{*}$ the truth, computed over the $24$-dimensional space~\cite{gneiting2007}. With the same color scheme of Fig.~\ref{fig:MBHB_sci}, we see that for \ac{DSM} (continuous lines) each extra layer of information brings the networks slightly closer to the truth. \ac{FM} (dashed lines) shows the most improvement with the addition of derived quantities, with further investigation showing that specifying the bin error $N_j$ had the most impact. For interpretability, we also specify on the right-hand-side ordinate the equivalent Gaussian displacement required to achieve similar scores. Although clearly not major effect sizes (the intuition being that all the scores would get closer to one with longer training), for a fixed training duration we see a measurable effect on the posterior quality across all \ac{MBHB} \ac{SNR} ranges. Furthermore, it is noteworthy that, considering the bootstrap intervals, all networks are equally far away from the truth as a function of \ac{MBHB} \ac{SNR} over the $16384$ simulations considered. This is a good sign of the networks having correctly learned how to marginalize over the transient. In fact, we have observed the same energy score as a function of \ac{SNR} behavior in the glitch networks as well.

\subsection{Calibration with transients}\label{sec:pop_checks}

\begin{figure*}
    \centering
    \includegraphics[width=0.95\textwidth]{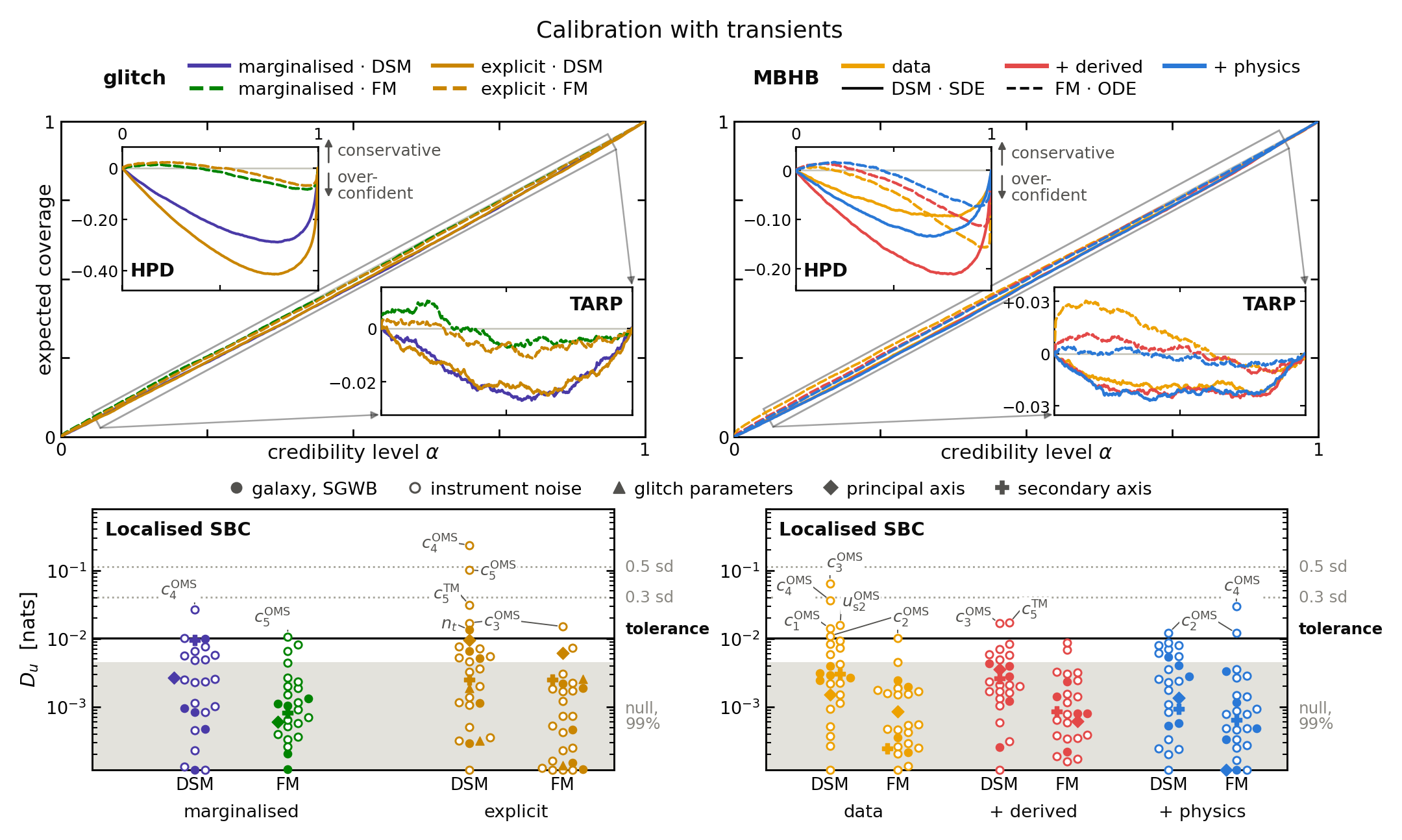}
    \caption{Population and localized network calibration, glitch (left) and \ac{MBHB} (right). Top: expected coverage of \ac{TARP}/\ac{HPD} regions,  as in Fig.~\ref{fig:comp_sep_checks}. Bottom: localized \ac{SBC} of the same networks, as in Fig.~\ref{fig:comp_loc_sbc}, with $K = 512$ draws per dataset, around the glitched observation of Fig.~\ref{fig:agreement} (left) and the \ac{MBHB} observation of Fig.~\ref{fig:MBHB_sci} (right). The principal and secondary axes are here the two best-constrained directions of the three Galactic-foreground parameters, since the best-constrained directions overall are single noise knots. Colors distinguish the treatment of the glitch and the embedder, line styles the objective, and each \ac{SBC} column is the network of the matching color.}

    \label{fig:trans_calib}
\end{figure*}

We conclude this section with the different networks' performance ($10$ in total) at the population level, as well as in the vicinity of the posterior. Fig.~\ref{fig:trans_calib} displays on the left-hand-side, the marginalised and explicit glitch networks (\ac{DSM} continuous line, \ac{FM} dashed; upper plot, \ac{TARP} and \ac{HPD}; lower plot localised \ac{SBC} for the \ac{SNR} $100$ glitch); the right-hand-side is the \ac{MBHB} embedder ablation study (data-only channels in orange, plus derived invariants in red and plus the physics template banks in blue).

From the perspective of \ac{TARP} and \ac{HPD}, it is clear again that the \ac{FM} training objective outperforms \ac{DSM}, once more with minor effect sizes, both being well-calibrated (on the \ac{TARP}-side). Unlike Sec.~\ref{sec:comp_sep_check} however, we can track some noticeable effects: \ac{TARP}/\ac{HPD} diagnostics are both necessary conditions for posterior calibration, and they provide complementary information. Due to our \ac{TARP} implementation being $\theta$-dependent only, it amounts to a test of locality first and foremost (i.e. deviations are due to bias), with significantly less sensitivity to width errors (due to the random reference points being prior-dominated draws). In contrast, the per-trial \ac{HPD} \footnote{The \ac{HPD} statistic per-trial ranks each trial's truth by its Mahalanobis distance in each posterior's covariance (splitting the number of draws, half for covariance estimation, half for ranks), so every direction counts in units of its posterior width, exact for any posterior shape (the regions are highest-density only for a Gaussian posterior). It amplifies width errors with dimension: in $24$ parameters as here, a uniform $5\%$ ($10\%$) under-dispersion already gives an \ac{HPD} deficit of $0.14$ ($0.28$) at $\alpha = 0.68$, as do two or three directions that are $40\%$ too narrow. } is mainly a width-sensitive metric and thus driven by under- or over-confident posteriors. In fact, the \ac{HPD} deficits of the \ac{DSM} networks are consistent with the few noise knots flagged by \ac{SBC} rather than with a global miscalibration. As seen in the bottom panels of Fig.~\ref{fig:trans_calib}, the high-frequency \ac{OMS} spline knot amplitudes in particular are the least calibrated ones, usually due to mild overconfidence. The \ac{MBHB} localized \ac{SBC} panel also displays a clear attenuation of this effect on the \ac{DSM} side, with fewer of the \ac{OMS} noise parameters away from the null, when going from data-only channels to the addition of derived or plus physics channels. This effect does not seem to translate to the population level, with \ac{DSM}'s \ac{TARP} being similar for all three embedders, while its \ac{HPD} actually prefers the data-only variant; on the flip side, both \ac{HPD} and \ac{TARP}, for the \ac{FM}-arm are closer to calibration with the additional embedder channels activated. The full noise corner plot companions to Fig.~\ref{fig:agreement} and Fig.~\ref{fig:MBHB_sci} can be found in App.~\ref{app:extras} and we note that all the noise parameters under consideration can be constrained by the data, including the components' spline shapes and the per-link amplitude deviations.

Every network so far analyzed one stationary $11.6$-day record with at most one transient, a setting in which the likelihood references of Sec.~\ref{sec:checks_intro} remain within reach. Section~\ref{sec:nonstationary} removes the remaining assumptions together: the noise changes level over the observation, glitches arrive at a realistic rate, and chunks of data go missing.

\section{Facing non-stationarity}
\label{sec:nonstationary}

The previous sections broke the usual assumptions one at a time; this one breaks them all at once. Over the same $11.6$ days,  we change the instrumental noise level every $2.33$ days (a conservative choice), glitches arrive at a realistic rate, and whole stretches of data go missing, while the stochastic signals stay as they were. The data presented here would require a mini-global fit (a stochastics block and a transient block for the glitches), so for the realistic case everything we claim about the posteriors has to come from simulator and network checks.

We ask four questions of the networks: whether they remain calibrated, over the prior and around an observation; whether they measure the non-stationarity itself, the noise level of each block; whether the time variation of the noise gets us information about the signals, as it should~\cite{alvey2024}; and how gracefully the posteriors degrade as glitches multiply and data goes missing.

\subsection{Segmented data and experimental design}
\label{sec:segmented_data}

\begin{figure*}
    \centering
    \includegraphics[width=0.9\textwidth]{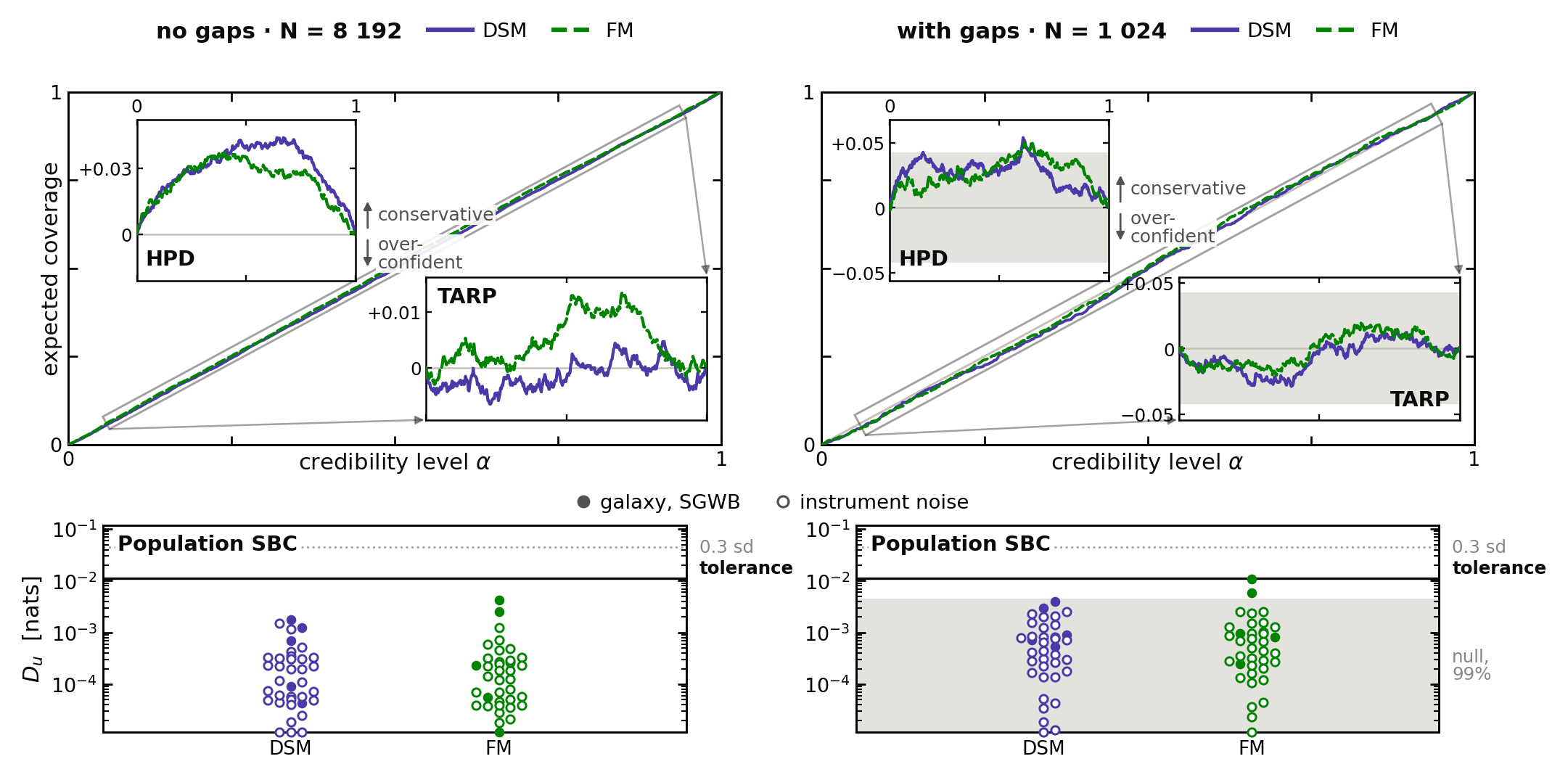}
    \caption{Population calibration of the two Sec.~\ref{sec:nonstationary} networks, \ac{DSM} (solid) and \ac{FM} (dashed).
Left: $8192$ test simulations drawn from the training process (non-stationary block levels, glitches at the training rate, every chunk present). Right: the first $1024$ of them, each with a gap of one to three consecutive chunks drawn from the training gap model. Top: \ac{TARP} expected coverage; the lower-right inset magnifies the framed band along the diagonal, and the upper-left inset shows the same quantity for \ac{HPD}. Bottom: \ac{SBC} rank divergence of each of the $37$ parameters, filled markers for the Galactic-foreground and \ac{SGWB} parameters, open ones for the instrumental noise. With $1024$ simulations the range of an exactly calibrated estimator is comparable to these scales, so it is drawn in grey: the $95\%$ simultaneous band of the coverage, and the level that $99\%$ of single readings of $D_u$ stay below. At $8192$ they are $\pm 0.015$ and $5.6 \times 10^{-4}$ so we do not plot them.}

    \label{fig:nonstat_calib}
\end{figure*}

The $11.6$ days are cut into $35$ contiguous chunks of $8$~h, $f_{\rm min}=6.9\times 10^{-5}$, and every seven consecutive chunks form a block of $2.33$ days with its own \ac{OMS} and \ac{TM} levels $d^{(n)}_{b}$ of Eq.~\eqref{eq:noise_model}, drawn independently with $\sigma_d = 0.2$, i.e. a $20\%$ scatter of the noise \ac{PSD} from block to block~\cite{edwards2020, alvey2024}. The rest of the noise model is that of Sec.~\ref{sec:gaussianity}, except that all five \ac{TM} contrasts are now inferred, so that $\theta$, of dimension $37$, fixes the covariance of every chunk exactly. Every chunk receives a Poisson number of glitches with mean $0.8$ (at most four), about $2.4$ per day over the constellation, so that a glitched record typically has $13$ to $26$ of its $35$ chunks affected, and $20\%$ of the training records are drawn without any glitch. We note that $20\%$ glitch-free simulations are not enough for our network to truly explore the stochastic-only regime in $37$ dimensions. Networks will thus not be perfectly calibrated in the glitch-free regime, it is merely a sanity check. We judged the training time to be better spent on the harder realistic glitch population and gapped, nonstationary data regime instead. Gaps~\cite{baghi2019} remove whole chunks: $30\%$ of the records lose a run of one to three consecutive chunks ($8$ to $24$~h), half of them placed at a block boundary, since the antenna repointing behind a scheduled gap is also a natural point for the noise level to change~\cite{edwards2020}.

Each chunk is reduced as in Sec.~\ref{sec:datagen}, on a $162$-bin hybrid grid of single Fourier bins below $2.67$~mHz, and stored with the nine spectral channels of Eq.~\eqref{eq:spec_channels} and thirteen derived ones: $\log_{10} \mathrm{tr}\, \tilde{S}_{j}$, the two ratios of Eq.~\eqref{eq:invariants}, and a bank of ten templates in Eq.~\eqref{eq:contraction}, namely the \ac{OMS}, \ac{TM} and signal directions, the six per-link glitch directions and the null direction of the reference noise covariance. A single trunk, shared by all chunks, turns each into one token, to which learned embeddings of its chunk and block indices are added: $35$ data tokens, so $72$ total tokens including the parameters. A missing chunk enters as a block of zeros, from which the embedder reads a presence channel. The attention mask follows the dependence graph of Eq.~\eqref{eq:noise_model}, so a block-level attends only to the data tokens of its block and not to the levels of the other blocks. As before, the \ac{DSM} and \ac{FM} networks differ only in their objective.

\subsection{Network calibration}

Because we have no exact reference for this data construction, short of a global fit, we instead begin with an \ac{SBI} network calibration analysis, as there is little point in looking at individual posteriors before certifying calibration over the prior. Fig.~\ref{fig:nonstat_calib} displays \ac{TARP}/\ac{HPD} expected coverage, and population \ac{SBC} for both gap-free and gapped datasets. Due to the increased simulation cost for the datasets presented here, we restricted the full population calibration to $8192$ trials (half of previous sections) and $1024$ for gapped datasets, so we also plot the null for the latter. Keeping in mind that (true) effect sizes are independent of the number of trials, we observe this to be the most calibrated \ac{DSM}-objective network in this work. \ac{HPD} coverage displays a minuscule underconfidence, while \ac{TARP} reports \ac{DSM} as least biased over the prior. All parameter marginals are also calibrated over the prior, and a few missing chunks do not modify these conclusions.

We thus conclude the ongoing \ac{DSM} versus \ac{FM} objective ablation by remarking that prior-level calibration consistently preferred \ac{FM}-trained networks over diffusion up until now: Fig.~\ref{fig:comp_sep_checks} only through small effect sizes, while Fig.~\ref{fig:trans_calib} showed more significant ones. \ac{SBC} also identified similar flaws in both, either because of loud \ac{SGWB} datasets poisoning $n_t$ calibration or high-frequency spline amplitudes for \ac{OMS}/\ac{TM}. What they have in common is a difficult inference target (tight posteriors), which for both objectives is alleviated with further training. Perhaps more pertinent is the fact that the various networks' weak points throughout this study affected both objectives similarly: data completeness, degenerate noise models, embedder input channel content, and glitch/gap frequencies, all displayed larger effect sizes on posterior quality than objectives alone. We thus consider both flow matching and denoising score matching to be viable training objectives for \ac{GW} data analysis.

\subsection{Nonstationary noise effects}

\begin{figure*}
    \centering
    \includegraphics[width=\textwidth]{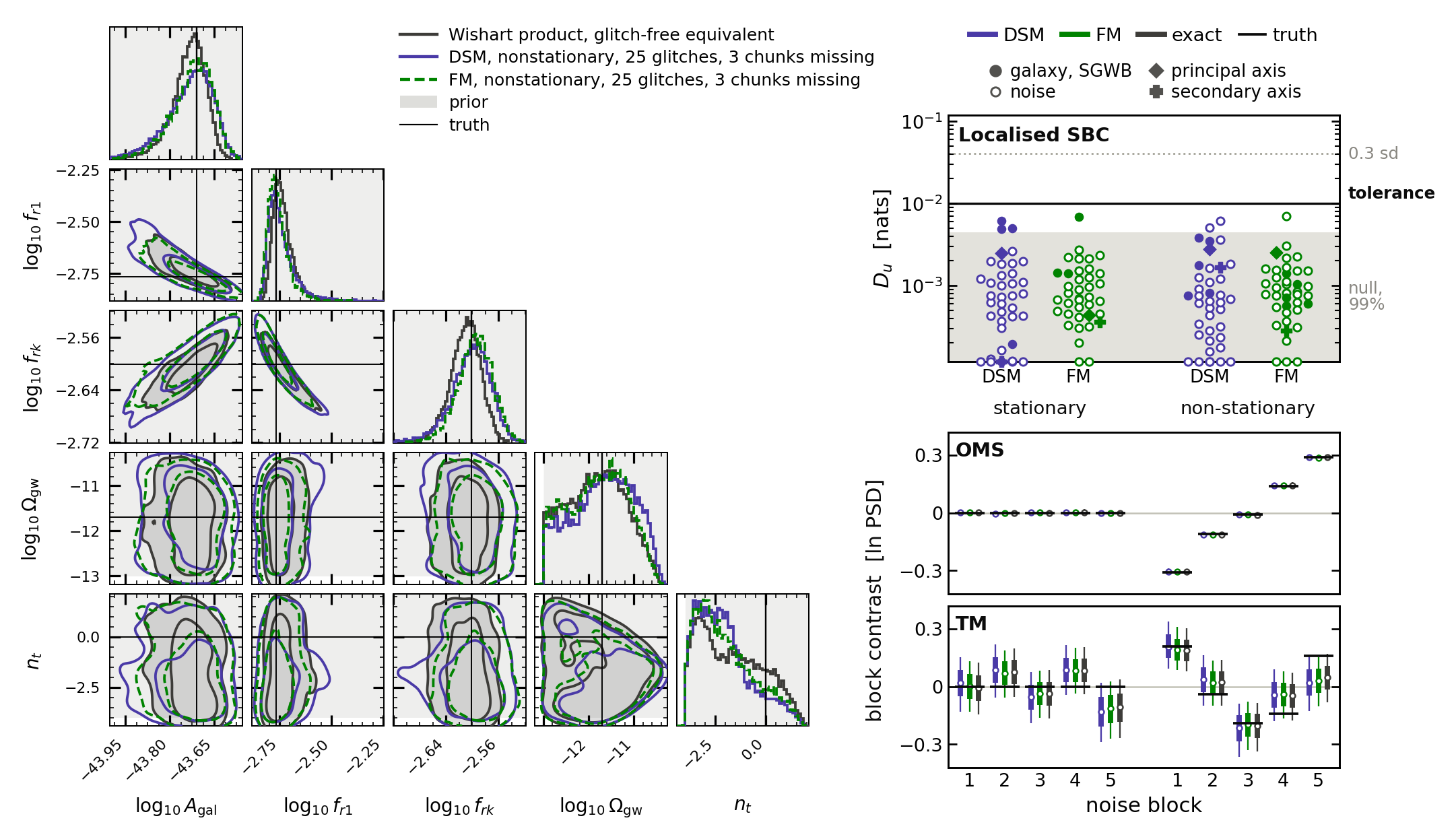}
\caption{Gliched, gapped and nonstationary stochastic component separation. Left: joint posteriors of the three Galactic-foreground and two \ac{SGWB} parameters, with non-stationary noise, $17$ glitched chunks and chunks $7$--$9$ missing, from the \ac{DSM} (solid) and \ac{FM} (dashed) networks; filled is the exact posterior of the same data without the glitches (the per-block central Wishart likelihood, gap-aware, the closest available reference). Contours are the $68\%$ and $95\%$ highest-density regions, black lines mark the injected values and the grey background is the prior. Top right: localised \ac{SBC} of both networks around the stationary observation and around its non-stationary counterpart (the same noise realization with the noise level of each block changed), as in Fig.~\ref{fig:comp_loc_sbc}, with $1024$ datasets and $K = 512$ draws. The datasets are simulated from the training process at the tilted parameters. Bottom right: the noise levels of the five blocks as contrasts, each level minus their mean (the mean itself is degenerate with a common shift of the spline knots), \ac{OMS} above \ac{TM}, on the stationary observation and on its non-stationary twin: median, $68\%$ (thick) and $95\%$ (thin) intervals of \ac{DSM}, \ac{FM} and the exact posterior; black bars are the injected contrasts.}

    \label{fig:money_plot}
\end{figure*}

Fig.~\ref{fig:money_plot} showcases the full flexibility of our architecture. $11.6$ days of data are analyzed as $35$ tokens, with a realistic glitch population (25 glitches in 17 chunks), and $3$ contiguous chunks are missing, in addition to the overall noise amplitude changing every 7 chunks. This is on top of the already explored spline-driven \ac{PSD} freedom as well as per-link variation, in a $37$-dimensional parameter space. 

We plot both \ac{DSM} (purple) and \ac{FM} (green) galaxy plus \ac{SGWB} parameters (left-hand-side of Fig.~\ref{fig:money_plot}) as well as an \ac{MCMC} posterior (black) on the equivalent (same noise realization) glitch-free dataset for comparison. The agreement is reinforced by our localized \ac{SBC} scheme, top right-hand-side of Fig.~\ref{fig:money_plot}. All $37$ parameters across \ac{DSM} and \ac{FM}, tested in the same posterior region as the above observation, as well as on its stationary counterpart, are well calibrated across $1024$ trials with $512$ draws each. In addition, we obtain constraints on the \ac{OMS} nonstationarity (upper part of the lower right hand side) and resolve the degeneracy structure (in a prior-dominated regime however) of the \ac{TM} nonstationarity (lower part). These are consistent between networks (on the realistic observation) and an \ac{MCMC} reference (in grey) on the de-glitched counterpart. The full noise parameter corner plot equivalent can be found in App.~\ref{app:extras}.

\subsection{Glitch multiplicity, noise variation, and gaps}

\begin{figure}
    \centering
    \includegraphics[width=0.9\linewidth]{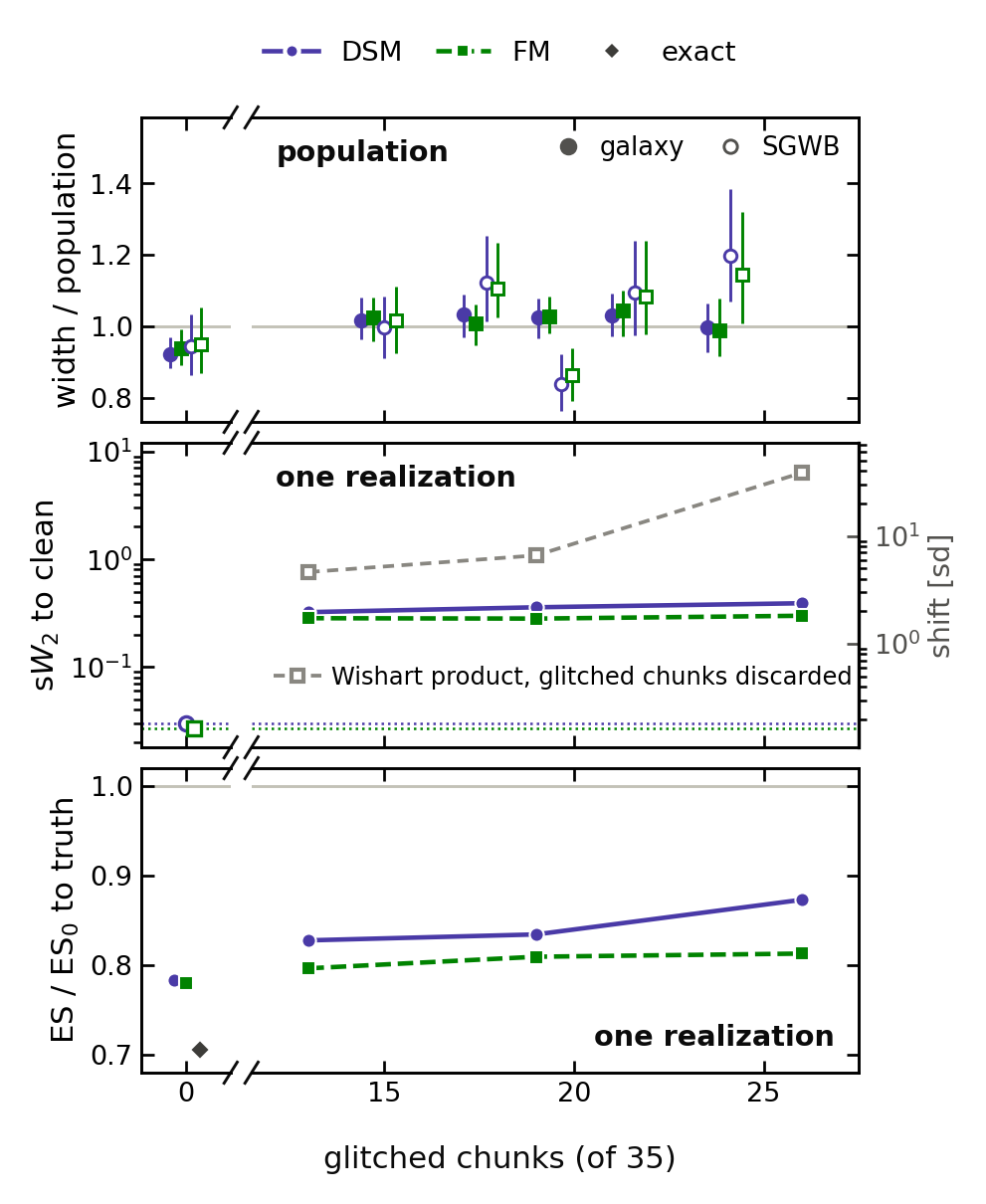}
\caption{Top: posterior width of the $8192$ test simulations grouped by their number of glitched chunks, as the median posterior standard deviation in the group over that of all simulations, geometric mean over the Galactic-foreground (filled) and \ac{SGWB} (open) parameters; bars are bootstrap $68\%$ intervals. Glitches are drawn independently of $\theta$, a glitched simulation has $13$ to $26$ glitched chunks in $98\%$ of cases, hence the break of the axis. Middle and bottom: one noise realization $13$, $19$ and $26$ glitched chunks, each set containing the previous one. Middle: sliced 2-Wasserstein distance of each posterior from the same network's posterior on the glitch-free data, on all $37$ parameters, both whitened by the latter; at zero, and dotted, the distance of an independent redraw of the glitch-free posterior (null). The right axis is the equivalent rigid shift along one direction, $\sqrt{37}\, sW_2$. Grey squares are the expected distance for an exact analysis that discards the glitched chunks instead; at $26$, an entire block is missing. Bottom: energy score to the injected parameters, in the Mahalanobis frame of the mean covariance, relative to ${\rm ES}_0$, the score of an exact Gaussian posterior with that covariance; the diamond is the exact posterior of the glitch-free data for reference. The equivalent shift is not plotted in the bottom panel, because with one realization and all values below 1, it means the truth is inside the distributions; values closer to unity for the energy score indicate wider posteriors.}

    \label{fig:nonstat_glitches}
\end{figure}

We now explore in more detail the impact of a realistic glitch population as well as excising entire chunks of data (either in reaction to a particularly loud glitch or due to antenna pattern repointing) and quantify the effect of noise nonstationarity.

Fig.~\ref{fig:nonstat_glitches} summarizes the effects on stochastic component separation posterior quality with our neural estimators. The shared abscissa is the number of chunks that include one or more glitches, while the ordinates report posterior width (upper), distance to the true posterior (middle), and distance to the data generating parameter point (bottom). The posterior width is the median over $8192$-prior draws, geometric mean over parameters; the $sW_2$ construction and plotting scheme is the same as in Fig.~\ref{fig:agreement}, and the energy score calculation matches Fig.~\ref{fig:MBHB_es}. \ac{DSM} and \ac{FM} (purple and green respectively) both show a clear ability to marginalize over the glitches, without any significant degradation with increasing glitch rates, in the resulting $37$-dimensional stochastic component posterior. The middle plot is relative to the glitch-free posterior, and we show in grey squares the effect of discarding the glitched chunks (in the extreme case, throwing away 26 out of 35 data segments).

\begin{figure}
    \centering
    \includegraphics[width=0.9\linewidth]{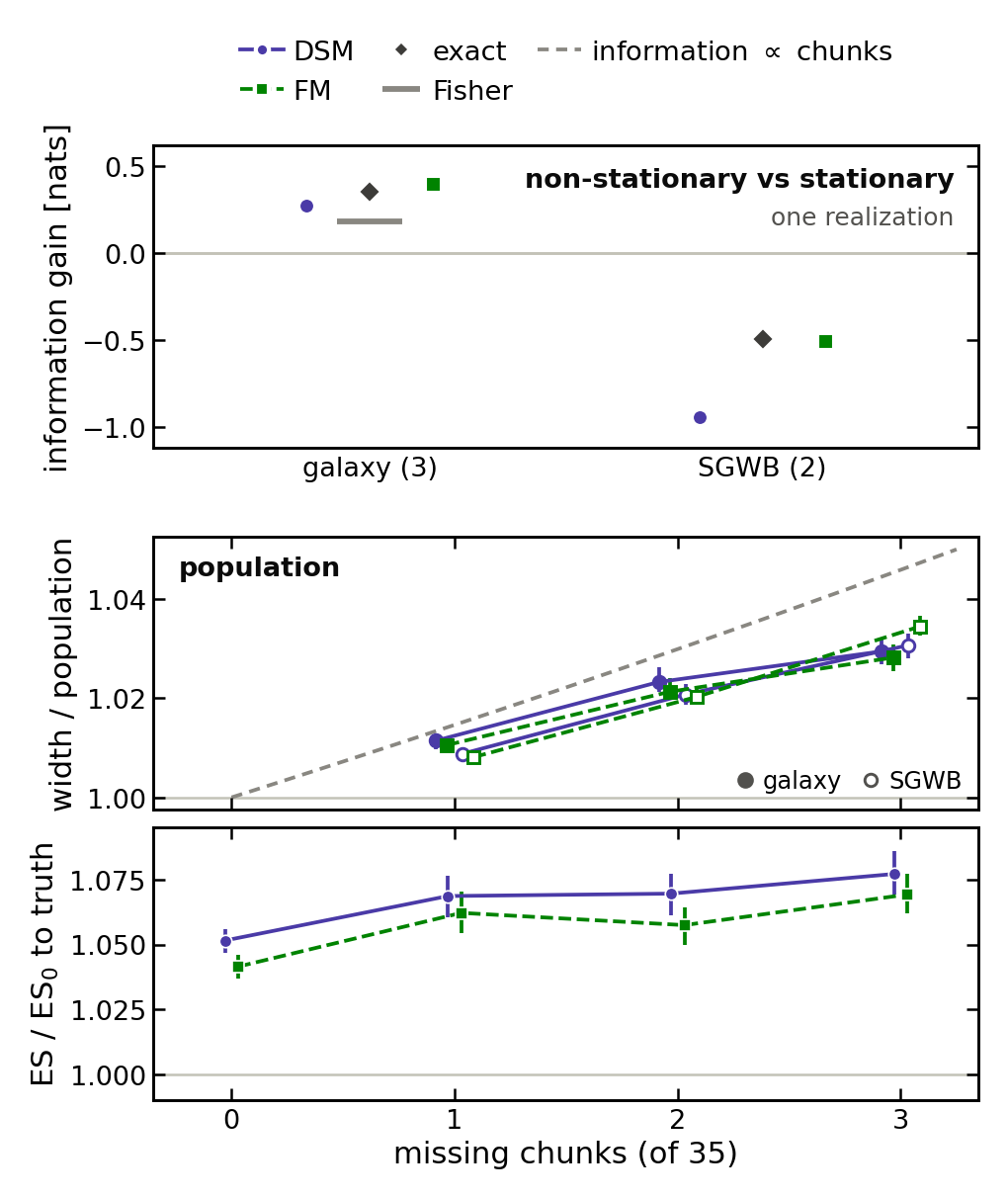}
\caption{Nonstationarity and gaps. Top: information gained by the foreground and \ac{SGWB} parameters when the noise level of each block changes: the difference of the Kullback--Leibler divergences of the posterior from the prior, between the non-stationary counterpart and the observation itself, for \ac{DSM}, \ac{FM} and the exact posterior; the grey bar is the expected gain of the exact model at the injected parameters, averaged over draws. The divergences are computed from $10^4$ draws by a nearest-neighbor entropy estimate. Middle and bottom: the $1024$ paired test simulations of Fig.~\ref{fig:nonstat_calib}, each with and without its gap, against the number of missing chunks. Middle: median over simulations of the ratio (with and without the gap) of posterior standard deviations, geometric mean over the foreground (filled) and \ac{SGWB} (open) parameters; we select only the simulations whose data constrain them (since fully prior-dominated regimes are redundant); the dashed curve is information proportional to the number of chunks, $[35/(35-k)]^{1/2}$; width is below because of partial prior domination. Bottom: energy score to the injected parameters; zero missing chunks are the same simulations without their gaps. Bars are bootstrap $68\%$ intervals.}

    \label{fig:nonstat_gaps}
\end{figure}

Beyond robustness, the time variation of the noise should itself carry information: the gain from low-noise blocks is expected to overcompensate the loss of \ac{SNR} in high-noise ones, as Ref.~\cite{alvey2024} showed for an \ac{SGWB} over a year of time-varying noise. We only consider $11.6$ days of data (but with a more aggressive amplitude noise variation, see Fig.~\ref{fig:money_plot}), and quantify the Bayesian information gain through the marginalized posterior to prior KL divergence in the upper part of Fig.~\ref{fig:nonstat_gaps}. The galaxy marginal shows a gain in information, which shows up as sharper breaking of the $A_{\rm gal},f_{{\rm r}1},f_{\rm rk}$ degeneracy, i.e. a better resolved Galactic shape. This is expected considering we started with a \ac{SNR} $\sim250$ galaxy. In contrast, the prior-dominated \ac{SNR} $\sim5$ \ac{SGWB} actually loses a bit of information from the presence of nonstationary noise. This effect is consistent between our two networks (\ac{DSM} in purple, \ac{FM} in green), the reference glitch-free \ac{MCMC} (in black-grey diamonds) and a Fisher analysis for the Galactic marginals (see~\cite{alvey2024} for the formalism). We do not show the Fisher line for the \ac{SGWB}, because the locally Gaussian approximation of the likelihood is subdominant to the prior; it registers a ``fake'' information gain that is not representative of the posterior. We cannot claim whether the \ac{SGWB} information loss from nonstationarity is due to similarly-treated (isotropic) stochastic signal components, the high flexibility of the noise model, or simply due to the low \ac{SNR}. What is certain however, is that over a longer observation period (with a corresponding anisotropic, galaxy-shaped response), nonstationary noise will provide us more information about any possible underlying \ac{SGWB}, naturally assuming a modest noise budget. We expect this to be true even with such noise model freedom, since an \ac{SGWB} ought to be the only isotropic and stationary component in the \ac{LISA} band.

The lower part of Fig.~\ref{fig:nonstat_gaps} shows the effect of gaps (shared abscissa as counts) over $1024$ prior draws; median width in the upper plot and energy score to the true generating parameters on the bottom. In addition to \ac{DSM} and \ac{FM} networks, we plot separately as filled circles the Galactic parameters and \ac{SGWB} as empty circles (geometric mean over them). The dashed grey line corresponds to the expected $\sqrt{\rm T_{\rm obs}}$ \ac{SNR} dependency (assuming Gaussian posteriors) and is plotted for reference. As expected, both metrics show a clear trend upwards, i.e. information is lost with increasing number of gaps.

In short, the networks of this section remain calibrated, measure the block-to-block variation of the noise, gain galaxy information from it as the exact posterior does, and lose little as glitched chunks multiply or chunks go missing.

\section{Discussion: simulation-based inference in LISA}\label{sec:discussion}

The utility of stochastic component separation is widespread, as already mentioned in the introduction: both the characterization of the Galactic foreground and a possible detection of cosmological-in-origin \ac{SGWB} are part of the science objectives of \ac{LISA}~\cite{colpi2024}. In addition, the current \ac{LISA} analysis paradigm, the global fit, is a very computationally intensive Bayesian framework that will require appropriate initialization, or a search phase. Lastly, detector characterization, through an active noise estimate will be critical to the operations of the mission. The \ac{SBI} framework introduced in this work (including the architecture, embedding scheme and accuracy checks) already constitutes a Bayesian parameter estimation pipeline for the stochastic components. This framework, could, with changes of increasing size, serve each of the aforementioned: as a full parameter estimation on \ac{LISA} data, a global fit search, or as a monitor of the instrument. Each of these roles implies a different extension and the nature of the task, and a different calibration and checking mentality. For the stochastic components, the cost of each role falls, as in Secs.~\ref{sec:stationary} to~\ref{sec:nonstationary}, on the simulator and the checks more than on the network. We describe below, in a bit more detail, how this work can be adapted to suit each of these roles.

\subsection{Parameter estimation}\label{sec:disc_pe}

Our networks already return the marginal posterior of the noise, the Galactic foreground and the \ac{SGWB}, which a global fit would obtain only by sampling every other source. If this were to constitute a full analysis of \ac{LISA} data (targeting only stochastics but including all sources in the data) it would unlock a large swath of pre-launch studies as well as a comparative result for traditional analyses. However, the \ac{SBI} framework would need a stricter accuracy and precision standard than the calibration our checks established.

Where a likelihood exists, importance sampling of the network's draws makes the result asymptotically exact~\cite{dax2023is}, here at efficiencies of about $20\%$ (Sec.~\ref{sec:comp_sep_check}), but this is not realistic when considering all sources in the data. In such cases, as in Sec.~\ref{sec:nonstationary}, simulation-based checks test calibration, not correctness, which needs data-dependent test quantities (Sec.~\ref{sec:checks_intro}) or references on nearby problems, such as the glitch-free records of Sec.~\ref{sec:nonstationary}. Correctness matters most over a mission: combining the posteriors of independent segments, as Ref.~\cite{alvey2024} does for a hundred segments of $11.5$ days, can compound slight per-segment errors into bias. The Galactic foreground also needs its annually modulated response~\cite{adams2014, boileau2021}, which requires no new architecture (Sec.~\ref{sec:models}), and a freer shape than our empirical fit, a freedom that can cost up to an order of magnitude in \ac{SGWB} uncertainty~\cite{kume2025}.

One key benefit of amortization, coupled with an ability to simulate realistic data, is that it allows use cases that would be impractical for a global fit: large-scale detectability and model prototyping and checking studies. Detectability statements are usually made, deliberately, in idealized settings: against a power-law sensitivity curve built from a known noise spectrum~\cite{thrane2013, auclair2023}, or on data containing the stochastic components alone, as if all else were perfectly subtracted~\cite{Caprini:2019pxz, flauger2021}. 

Trained on data as realistic as can be simulated, or deliberately more varied, a network can complement them over the full model prior:  given, e.g., a cosmic-string  template, is such a background~\cite{auclair2023} detectable (and at what $\Omega_{\rm GW}$), and if not, does observation time limit it or a degeneracy with a component that the idealized setting leaves out? Section~\ref{sec:nonstationary} did so on a small scale, measuring over $8192$ and $1024$ simulations what glitch multiplicity and gaps cost the stochastic posteriors. Such population statements only need calibration over the prior, which our checks established, rather than \Icomm{correctness} on one observation, since we'd implicitly be aiming for detectability statements, so upper limits. Such a study could be performed with the framework introduced here, with minimal changes.

\subsection{Searches ahead of the global fit}\label{sec:disc_search}

Ground-based search pipelines look for rare, short signals in noise-dominated data: they rank candidates by a detection statistic whose background is measured empirically, e.g.\ from time-shifted data of different detectors, assign each a false-alarm rate, and leave parameter estimation only for the significant ones~\cite{klimenko2016, usman2016, messick2017, aubin2021, gwtc4methods}. Signal-dominated \ac{LISA} data cannot be used like this: in general, whether a source is present can only be confirmed within a global fit against all other sources and the noise; a \ac{LISA} search would need to deliver an overcomplete set of candidates and parameters, not detections. It would then target one source class, marginalizing over the other sources rather than ignoring them. The natural interface is initialization, the first stage of the global fit~\cite{cornish2022}: the walkers start from the search's output, as the \ac{MBHB} walkers of Ref.~\cite{katz2025} start from the samples of its \ac{MBHB} search. Search results should not become priors, which would use the data twice, nor, in our view, proposals: proposals strongly contracted by a search would move the burden of population modeling and coverage onto the search, whose task is the opposite in our opinion, to provide an initial estimate, while ignoring everything it cannot handle. Proposal should be used instead to raise acceptance rates and explore astrophysical populations~\cite{korsakova2024, katz2025}.

Two requirements follow: speed where latency matters, as for \ac{MBHB} alerts~\cite{cornish2022}, and errors on the side of width rather than of bias or over-confidence, since overdispersed starting points are what convergence diagnostics presuppose~\cite{gelman1992}; a start that is too wide costs some burn-in, whereas one that is too narrow or displaced can strand the walkers in a secondary mode, or produce chains that agree with one another before any of them has fully explored the posterior, a failure that diagnostics built on the comparison of chains (i.e. standard \ac{MCMC} convergence criteria) cannot detect. Where the number of sources is uncertain, no candidate list is both complete and pure, and that uncertainty belongs in the spread of the walkers, some starting with more sources than others. A search should therefore be conservative (i.e. overcomplete in the case of source number uncertainty) by design, as balanced estimators tend to be~\cite{delaunoy2022, delaunoy2023}, with ensembles being better than single estimators~\cite{hermans2021}, and be checked against a one-sided criterion, expected coverage at or above the nominal level, over the prior and near the observation (localized \ac{SBC}), with bias tested for separately.

\ac{LISA} sources differ in the type of problem they cause a search. For \ac{MBHBs}, the main issue is speed: low-latency detection and sky localization~\cite{cornish2022}, ideally before the merger~\cite{houba2024mbhb}. \ac{GBs} are difficult due to their unknown (large) number, since a search must decide how many are resolvable, a number our foreground fit fixes implicitly, through its resolvability threshold (Sec.~\ref{sec:models}). A \ac{GBs} search ought to estimate how many binaries are resolvable and the amplitude, frequency, frequency derivative and sky position of each, marginalizing over the unresolved remainder, transdimensionally; \ac{SBI} has so far treated isolated \ac{GBs} of known count~\cite{delmond2026, mao2025}. In contrast, \ac{EMRIs} require precision and careful treatment of multimodality. Large prior-to-posterior contraction is a weak point of our networks too (the \ac{SGWB} tilt at high \ac{SNR}, the high-frequency noise knots; Secs.~\ref{sec:comp_sep_check} and~\ref{sec:pop_checks}), so \ac{GBs} and \ac{EMRIs} require developments in network architecture, embeddings and training schemes, such as simulations focused on the posterior region~\cite{cole2026, lyu2026}.

For such classes, checks serve the architectural design phase primarily, before they certify correctness: here they ranked channel sets, embedder inputs and objectives (Secs.~\ref{sec:stationary} to~\ref{sec:nonstationary}), and for \ac{GBs} and \ac{EMRIs} it is such ablations that ought to decide what architecture to build. Design should also anticipate scale. The Gaussian likelihood of a model with one binary extends to ten million, only costlier. An \ac{SBI} configuration (data representation, embedder, inference scheme) that works with noise and a single target, as in most \ac{SBI} studies of \ac{LISA} data~\cite{alvey2024saqqara, dimitriou2024, spadaro2026, delmond2026, cole2026}, or with a few sources in two-hour segments of stationary noise~\cite{houba2025}, carries no such guarantee. This point will be made explicit when considering stochastic searches in particular, a primary target of ours when designing this framework, which we discuss below.

\subsection{A stochastic search}\label{sec:disc_stoch}

The stochastic search is the closest to this work and the one that must run first: the stochastic components set the noise covariance of every other likelihood, so their search starts from data from which nothing has been removed, as does the first noise and foreground fit of Ref.~\cite{katz2025}. It should also assume as little as possible about the signals, marginalizing every transient it can simulate, as our networks did glitches and an \ac{MBHB}, and staying agnostic to their morphology. Whether the searches for other sources should in turn utilize the information from its noise estimate, is still unclear. In its minimal form, the search runs a network like that of Sec.~\ref{sec:nonstationary} on each segment and starts the noise and stochastic walkers of the global fit from its draws, as previously described. In a more involved scheme, the noise estimate provided by the stochastic search would inform the searches for other sources as well.

The only major missing component between the work here and a full stochastic search pipeline is a richer simulator. Here, the noise was drawn from the parametrized family that the network inferred, so every check was well specified by construction. Since a network marginalizes whatever the simulator draws and $\theta$ omits, but is reliable only on data that its training distribution covers~\cite{cannon2022, schmitt2023}, a search should simulate more than it infers, and more than it expects to see, which also errs on the side of width. 

As a schematic prototype for a better noise simulator, a first step would draw each physical noise source as a Gaussian process at the instrument level and propagate it through the instrument's measurement chain to the links before any \ac{TDI}, as instrument simulators do~\cite{bayle2023}; Eq.~\eqref{eq:noise_model} then becomes an effective description only, as would be the case during mission operation. A second, more realistic step would broaden the Poisson-distributed glitches of Sec.~\ref{sec:nonstationary} into overlapping artifacts with morphologies broad enough to contain \ac{LISA}'s, informed by \ac{LISA} Pathfinder~\cite{armano2022, baghi2022lpf} and by generators trained on ground-based glitches~\cite{dooney2024}, but not so broad that they absorb signals: this is the same trade-off of noise-model freedom discussed in Sec.~\ref{sec:noise}, but now reflected on the simulator rather than the parametrization used in inference. The simulator itself must then be tested with posterior predictive checks~\cite{gelman1996}, model criticism~\cite{ward2022, schmitt2023} and independently produced data such as the \ac{LISA} Data Challenges~\cite{baghi2022ldc}.

The hardest requirement is that the search treat instrumental artifacts as noise without explaining away as artifacts the \ac{GW} bursts that no block of current global fits models~\cite{littenberg2023, katz2025, deng2025}, from cosmic-string cusps~\cite{auclair2023} to unknown sources. Morphology alone may not separate the two; their interaction with the instrument may. However, this requires a more profound study of the instrument and spacecraft dynamics, which is outside the scope of this paper. In the current simulations, glitches are typically injected only in the \ac{TM} interferometers, which simplifies the task. Thus, a test-mass glitch reaches only two of the three Michelson variables (Sec.~\ref{sec:transients}), whereas a \ac{GW} enters all six links through a sky-dependent response. Using such patterns, Bayesian model selection could identify a burst as astrophysical above an \ac{SNR} of about five, with all channels, in known, stationary Gaussian noise~\cite{robson2019}, and a neural classifier could separate bursts from glitches of the same shape~\cite{houba2024glitch}; the per-link glitch templates of our embedder (Sec.~\ref{sec:embedding}) exploit the same handle. Learning the joint signature of response, transfer functions and time variation (drifting noise, a modulated foreground, a background assumed stationary and isotropic) suits machine learning, since a likelihood captures it only if every component has a tractable model, whereas a network needs each only simulated.

\subsection{Monitoring the instrument}\label{sec:disc_monitor}

Taken to its extreme, a noise estimate that assumes nothing about the signals is what detector characterization requires: the active estimate of the instrumental noise. This would constitute the most drastic deviation from the work here, because it would require a pipeline built around the inference network, operating online.

Like the stochastic search, it rests on the noise simulator and on parametrizations that keep noise and stochastic signals apart. In contrast to a search, however, the checks would have to be aimed at target classification and continual updating. The decisions drawn from the posterior would need known false-alarm and missed-detection rates (is a stretch of data nominal, is a transient an artifact, or is it a potential signal?~\cite{houba2024glitch}). Detecting, over many segments, data that the simulator does not cover, a threat to parameter estimation, becomes a desirable product here~\cite{schmitt2023}, as the question posed is about the instrument. Updating is then continual because the noise cannot be faithfully modeled before flight. It may drift during the mission~\cite{muratore2023}: the noise model of the first hours of data should be maximally agnostic, whereas years in, the stochastic contributions modeled and found should leave little unexplained under the name of noise. Amortization makes each new $8$~h chunk one posterior evaluation, while each change of the noise model would imply a retraining, fine-tuning or simply posterior guidance, as access to scores permits. Blind studies, on mock data whose noise content is withheld, can test the pipeline before launch.

A monitor can also potentially fuse information beyond the \ac{GW}-sensitive channels: glitch classes established at ground-based detectors~\cite{zevin2017}, null \ac{TDI} combinations, and weak monitors at low frequency~\cite{muratore2023}, as well as auxiliary channels insensitive to \ac{GW} could feed machine-learning data-quality monitors as is already done in ground-based detectors~\cite{essick2021}. Unlike classical techniques, the high degeneracy of this input space is directly exploited by networks, especially attention-based ones.

Every role above rests, in the end, on a simulator producing realistic multi-source \ac{LISA} data (a Galactic population, time-varying noise of \ac{LISA}'s complexity, \ac{MBHBs}, \ac{EMRIs}, etc) at the rate \ac{SBI} consumes them, a direction we are pursuing.

\section{Conclusion}\label{sec:conclusion}

We have separated the stochastic components of simulated \ac{LISA} data with a single, calibrated \ac{SBI} framework, a Simformer fed by a physics-aware embedding of the full \ac{TDI} cross-spectral matrix, removing the usual assumptions of stochastic analyses one at a time. To this end, we asked in the introduction four main questions: (i) Can the Galactic foreground and an \ac{SGWB} be separated from instrumental noise in realistic \ac{LISA} data? (ii) What aspects of instrumental model freedom are actually constrained by the data? (iii) How do our architectural choices affect the posteriors? (iv) Can these posteriors even be trusted? We address these one by one.

First, both in stationary Gaussian, stationary non-Gaussian and nonstationary non-Gaussian data, we have managed to isolate the high-\ac{SNR} Galactic foreground contribution from the low-\ac{SNR} \ac{SGWB}, and we show this with one $11.6$ day observation and over the whole prior regime. In Sec.~\ref{sec:stationary}, with a stationary Gaussian covariance and per-instrumental-link noise model freedom, we find in line with previous work~\cite{hartwig2023}, that the diagonal $AET$ \ac{TDI} channels suffice for the signal parameters, but not for the noise  parameters. When a single transient is included, a glitch or a massive black hole binary, Sec.~\ref{sec:gaussianity},  the stochastic foreground and background remain separable. In Sec~\ref{sec:nonstationary},with nonstationary noise, gapped data, and a realistic glitch population, this remains true: we find the non-stationarity to help the reconstruction of the Galactic contribution with an information gain of $\sim0.3$ nats; the $25$ glitches present in the observation barely affect the recoverability of the signal parameters; and data gaps degrade the information content of the observation. These findings are corroborated by local \ac{MCMC}-driven comparisons, with our network posteriors achieving up to $20\%$ importance sampling efficiency on stationary Gaussian data, and our localized \ac{SBC} checks showing rank-divergence values below $10^{-2}$ nats for all signal and most noise parameters in all data settings.

Second, regarding the instrumental noise model, throughout this manuscript we infer from $12$ to $32$ noise parameters, allowing for spectral shape variation, per-link amplitude freedom and time-varying amplitude deviations. In Sec.~\ref{sec:stationary}, we find the full \ac{TDI} spectral matrix to be key in constraining the \ac{OMS} per-link components. In fact, all \ac{OMS} parameters, the splines and all amplitude deviations, are constrainable (i.e. not recovering the prior and with a clearly defined degeneracy structure), even with the short $11.6$ day observation durations considered here. The \ac{TM} contribution on the other hand, really benefits from arm sum or difference parametrizations, with the former also recoverable throughout, while the latter usually follows the prior distribution due to an uninformative likelihood surface. As before, all these findings are corroborated with \ac{MCMC} reference posterior distributions, and by our localized \ac{SBC} scheme, with the only significant deviations from perfect calibration in Sec.~\ref{sec:gaussianity}. We find that the least calibrated noise parameters to be consistently the high-frequency spline knot amplitudes of \ac{TM} and \ac{OMS}: the exact knot location is the primary driver of over-the-prior network calibration.

Third, this work introduced an \ac{SBI} framework, building on the Simformer~\cite{gloeckler2024} architecture, and included a novel physics-aware embedding, and presented all results twice, once from a network trained with a denoising score matching (\ac{DSM}) objective and one with the state-of-the-art in \ac{GW} data analysis, flow matching (\ac{FM}) objective. This framework, simply reconfigured, was successfully used in all the data settings presented, with a variable parameter dimension, variable data content, quantity, and dependence structure. As input, we considered the impact of adding auxiliary information: some derived from the spectral matrix and simulator (rotation invariants of the matrix, coarse-graining bin number, and frequency index); and some model-dependent, including template banks and component \ac{TDI} response projections. In an ablation study over the same training duration, Sec.~\ref{sec:MBHBs_physics_emb}, we find that these extra channels mildly improve network training through a lower distance-to-injection metric (the energy score of the posterior to truth). As for the objective, we find that the choice between \ac{DSM} and \ac{FM} mattered less than data completeness, the noise model, the embedder content, or the glitch and gap rates. The only significant difference in accuracy was measured through \ac{HPD} in Sec.~\ref{sec:pop_checks}, where \ac{DSM} performed worse than \ac{FM} over the full prior, but not over the target observation. We conclude both to be viable for \ac{GW} data analysis so long as a rigorous \ac{SBI} calibration framework is included, as was done in this work.

Fourth, through the utilization of multiple existing \ac{SBI} posterior diagnostic checks and a novel posterior-local \ac{SBC} test, we quantify the deviations in our networks' posteriors, on one observation and over the whole training prior. High-power statistical tests, with complementary sensitivity to deviation types are utilised: \ac{TARP} detects bias  while \ac{HPD} is sensitive to posterior width discrepancies over the joint high-dimensional posterior distribution; \ac{SBC} detects any miscalibration over the marginals of the full posterior distribution and when applied locally to the target and adding multiple test-quantities, can detect posterior higher-order moment discrepancies. Population tests on $\sim10^{4}$ simulations, all read as effect sizes against a $0.15\sigma$ bias ($10\%$ width) tolerance, found the networks calibrated. Deviations are confined to the \ac{SGWB} tilt at very high \ac{SGWB} \ac{SNR} and to a few high-frequency noise knots, mostly for \ac{DSM}. These provide the machinery required for building trust in \ac{SBI} posteriors, demonstrating a clear ability to identify systematics as well as per-case discrepancies. This property of amortization allowed us to isolate modeling effects from simulator errors, and from network mis-calibration while designing this framework, thus proving invaluable in building trust in the resulting posteriors.

Finally, these four answers leave out several findings and new elements of this work. We collect them below:
\begin{itemize}
    \item \emph{Transients, marginalized:} with a glitch, four routes to the posterior agree (networks that marginalize or infer it, the non-central Wishart likelihood and an analytically marginalized one), with the networks as close to the exact posterior as a perfectly de-glitched analysis, whereas ignoring the glitch strongly biases it. With glitches and \ac{MBHBs} alike, the networks' distance to the truth does not depend on the transient's \ac{SNR}, verified up to \ac{SNR} $1000$.
    \item \emph{Glitch populations and gaps:} marginalizing a realistic glitch population costs far less than excising the glitched chunks (up to $26$ of the $35$), and gaps widen the posteriors no more than the lost observation time implies.
    \item \emph{Non-stationarity and faint signals:} unlike the Galactic foreground, the faint, prior-dominated \ac{SGWB} loses a little information to non-stationary noise over $11.6$ days, as its exact posterior does, a loss we expect to become a gain over longer observations with an anisotropic Galactic response~\cite{alvey2024}.
    \item \emph{Time as tokens:} non-stationarity and gaps required no new architecture, as a shared trunk turns each $8$~h chunk into one data token, a missing chunk enters as zeros with a presence channel, and an attention mask derived from the simulator's dependence graph confines each block's noise levels to the data of their block.
    \item \emph{Exact weighted ranks:} the importance-weighted rank of Eq.~\eqref{eq:weighted_rank} is exactly uniform for a correct estimator, for any tilt, number of draws and posterior shape, with ordinary and posterior \ac{SBC} as special cases. The network trained over the prior is thus tested near an observation without retraining, refitting or a surrogate (Appendix~\ref{app:sbc}).
    \item \emph{Effect sizes with a scale:} the rank divergence $D_u$ is a lower bound on the dataset-averaged Kullback--Leibler divergence between the true and estimated posteriors. Like the sliced 2-Wasserstein distance and the energy score, it is translated into the shift or width error of an exact posterior giving the same value. These elements are critical in quantifying posterior accuracy discrepancies in an interpretable manner.

\end{itemize}

The cost of each relaxed assumption falls on the checks more than on the network: they move from exact reference posteriors, which the networks match closely, to tests built from the simulator alone. Whether these results hold with every \ac{LISA} source in the data, rather than an empirical foreground and at most one type of astrophysical transient, is still an open question. However, every assumption diffused here is one fewer standing between simulation-based inference and the data \ac{LISA} will actually record.

\begin{acknowledgments}
We would like to thank Matt Edwards, Giorgio Mentasti, Nikolaos Karnesis, Neil Cornish, and Malvina Bellotti for their helpful comments and discussion. The authors have received support from the French Agence Nationale de la Recherche. The authors acknowledge
support from the CNES for the exploration of LISA
science. This work has made use of computational resources provided by the Azzurra cluster.
\end{acknowledgments}

\appendix

\section{The LISA instrument} \label{App:LISAmeasurement}

This appendix collects the conventions behind the transfer functions of Eq.~\eqref{eq:noise_cov} and the response kernels of Eq.~\eqref{eq:signal_cov}, writes out the noise models of Sec.~\ref{sec:noise}, and lists the training priors in Table~\ref{tab:priors}. We refer to~\cite{tinto2021} for a review of \ac{TDI}, to~\cite{bayle2023, hartwig2022, quangnam2023} for the measurement model and the noise transfer functions, and to~\cite{romano2017, flauger2021, hartwig2023} for the response to a stochastic background. We set $c = 1$, and all spectra are one-sided and in fractional-frequency units.

\subsection{Time-delay interferometry} \label{app:tdi}

The six links $k = ij$ are the measurements $\eta_{ij}$ on spacecraft $i$ of the beams received from spacecraft $j$, with light travel times $L_{ij}$ ~\cite{otto2012, bayle2023}. With the delay operator $\mathcal{D}_{ij}\,x(t) = x(t - L_{ij})$ and the chained delays $\mathcal{D}_{i_1 i_2 \dots i_n} = \mathcal{D}_{i_1 i_2}\mathcal{D}_{i_2 i_3}\cdots\mathcal{D}_{i_{n-1} i_n}$, the Michelson variable of \ac{TDI} 1.5, which accounts for $L_{ij} \neq L_{ji}$, reads~\cite{armstrong1999, shaddock2003, tinto2021}
\begin{align}
X_{1.5} \;=\; &(1 - \mathcal{D}_{121})\,(\eta_{13} + \mathcal{D}_{13}\,\eta_{31}) \nonumber \\
&- (1 - \mathcal{D}_{131})\,(\eta_{12} + \mathcal{D}_{12}\,\eta_{21}) ,
\label{eq:x15}
\end{align}
with $Y$ and $Z$ following from the cyclic permutation $1 \to 2 \to 3 \to 1$. Second-generation TDI adds delayed copies of these terms, so as to also cancel the laser noise to first order in the arm-length rates~\cite{shaddock2003, tinto2004, vallisneri2005, muratore2020}\footnote{We evaluate the $L_{ij}$ once, at the epoch $t_0$ of Keplerian orbits that minimize the flexing of the arms~\cite{nayak2006}, the same epoch at which the response is frozen (Sec.~\ref{sec:models}). There the six $L_{ij}$ lie between $8.303$ and $8.332$~s, a spread of $0.36\%$, with $|L_{ij} - L_{ji}|$ of order $10^{-4}\,L_{ij}$ since the spacecraft move while the light propagates.}.  Each delay then acts in the frequency domain as the phase shift $\mathcal{D}_{ij}(f) = e^{-2\pi i f L_{ij}}$, with the second-generation variable $\tilde{X}_{2.0} = (1 - \mathcal{D}_{121}\mathcal{D}_{131})\,\tilde{X}_{1.5}$. We collect the three variables as
\begin{equation}
\big(\tilde{X}, \tilde{Y}, \tilde{Z}\big)^{\mathsf{T}} \;=\; \sum_{k} \mathbf{c}_{k}(f)\, \tilde{\eta}_{k}(f) ,
\label{eq:tdi_map}
\end{equation}
with $\mathbf{c}_{k} \in \mathbb{C}^{3}$ the coefficients of link $k$. Since $X$ is built from the four links of the two arms meeting at spacecraft $1$, every link enters exactly two of the three Michelson variables, and the component of $\mathbf{c}_{ij}$ belonging to the spacecraft off arm $ij$ vanishes. The $A$, $E$, $T$ channels are the fixed rotation $A = (Z - X)/\sqrt{2}$, $E = (X - 2Y + Z)/\sqrt{6}$ and $T = (X + Y + Z)/\sqrt{3}$~\cite{prince2002}.

\subsection{Noise transfer functions} \label{app:noise_tf}

Ignoring tilt-to-length and clock noise, after \ac{TDI}, we are left with \ac{OMS} noise and two test-mass noises, that of the local test mass and, delayed by the arm, that of the distant one~\cite{hartwig2023, bayle2023}:
\begin{equation}
\tilde{\eta}_{ij}(f) \;=\; \tilde{n}^{\rm OMS}_{ij}(f) \;+\; \tilde{n}^{\rm TM}_{ij}(f)
\;+\; \mathcal{D}_{ij}(f)\, \tilde{n}^{\rm TM}_{ji}(f) , 
\label{eq:link_measurement}
\end{equation}
where $\tilde{n}^{\rm TM}_{ij}$ is the noise of the test mass on spacecraft $i$ facing $j$, with one-sided spectrum $S^{\rm TM}_{ij}$ of the form of Eq.~\eqref{eq:analytic_noise_tm}, and likewise for \ac{OMS}. The twelve sources being mutually independent, the only correlation between links comes from the test mass shared by the two links of an arm,
\begin{align}
\mathsf{S}^{\eta}_{ij,\,ij}(f) &\;=\; S^{\rm OMS}_{ij}(f) + S^{\rm TM}_{ij}(f) + S^{\rm TM}_{ji}(f) ,
\label{eq:link_diag} \\
\mathsf{S}^{\eta}_{ij,\,ji}(f) &\;=\; \mathcal{D}^{*}_{ji}(f)\, S^{\rm TM}_{ij}(f)
+ \mathcal{D}_{ij}(f)\, S^{\rm TM}_{ji}(f) ,
\label{eq:link_offdiag}
\end{align}
with all other entries of the $6 \times 6$ link spectral matrix vanishing. Substituting Eq.~\eqref{eq:link_measurement} into Eq.~\eqref{eq:tdi_map}, each source reaches $X$,$Y$,$Z$ through a vector,
\begin{equation}
\mathbf{t}^{\rm OMS}_{ij} \;=\; \mathbf{c}_{ij} ,
\qquad
\mathbf{t}^{\rm TM}_{ij} \;=\; \mathbf{c}_{ij} + \mathcal{D}_{ji}\,\mathbf{c}_{ji} ,
\label{eq:t_vectors}
\end{equation}
the second collecting the two appearances of test mass $ij$, in $\eta_{ij}$ and, delayed, in $\eta_{ji}$. The \ac{TDI} noise covariance follows as
\begin{equation}
C_{\rm noise} = \sum_{k,k'} \mathbf{c}_{k}\, \mathsf{S}^{\eta}_{kk'}\, \mathbf{c}^{\dagger}_{k'} = \sum_{k}\sum_{n} S^{(n)}_{k}\, \mathbf{t}^{(n)}_{k}\, \mathbf{t}^{(n)\dagger}_{k} ,
\label{eq:transfer}
\end{equation}
which is Eq.~\eqref{eq:noise_cov} with $\mathcal{T}^{(n)}_{k} = \mathbf{t}^{(n)}_{k}\mathbf{t}^{(n)\dagger}_{k}$, a rank-one, positive semi-definite Hermitian matrix.

\subsection{Stochastic signal response} \label{app:response}

A plane gravitational wave propagating along $\hat{\mathbf{k}}$, with polarization amplitudes $\tilde{h}_{A\in \{+, \times\}}$, produces in link $ij$ the fractional-frequency response, per unit $\tilde{h}_{A}$~\cite{estabrook1975, cornish2003},
\begin{equation}
G^{A}_{ij} = \frac{\xi^{A}_{ij}}{2\big(1 - \hat{\mathbf{k}}\cdot\hat{\mathbf{n}}_{ij}\big)} \Big[ e^{-2\pi i f (L_{ij} + \hat{\mathbf{k}}\cdot\mathbf{x}_{j})} - e^{-2\pi i f\, \hat{\mathbf{k}}\cdot\mathbf{x}_{i}} \Big] ,
\label{eq:gw_single_link}
\end{equation}
with $\mathbf{x}_{i}$ the position of spacecraft $i$, $\hat{\mathbf{n}}_{ij}$ the unit vector from $j$ to $i$, and $\xi^{+}_{ij} = (\hat{\mathbf{n}}_{ij}\cdot\hat{\mathbf{u}})^{2} - (\hat{\mathbf{n}}_{ij}\cdot\hat{\mathbf{v}})^{2}$, $\xi^{\times}_{ij} = 2(\hat{\mathbf{n}}_{ij}\cdot\hat{\mathbf{u}})(\hat{\mathbf{n}}_{ij}\cdot\hat{\mathbf{v}})$ the antenna patterns in a polarization basis $(\hat{\mathbf{u}}, \hat{\mathbf{v}})$ transverse to $\hat{\mathbf{k}}$. Its \ac{TDI} projection $\mathbf{r}^{A}(f, \hat{\mathbf{k}}) = \sum_{k} \mathbf{c}_{k}\, G^{A}_{k}$, for a stationary, unpolarized background of strain \ac{PSD} $S(f)$ and normalized sky distribution $p(\hat{\mathbf{k}})$, the \ac{TDI} covariance is $S(f)\,\mathcal{R}(f)$ with the averaged response:
\begin{equation}
\mathcal{R}(f) \;=\; \frac{1}{2} \sum_{A} \int {\rm d}^{2}\hat{\mathbf{k}}\; p(\hat{\mathbf{k}})\; \mathbf{r}^{A}\, \mathbf{r}^{A\dagger} .
\label{eq:response}
\end{equation}

We evaluate Eq.~\eqref{eq:response} for an isotropic $p = 1/4\pi$ on a HEALPix grid with $N_{\rm side} = 8$~\cite{gorski2005}, at the same epoch $t_0$ as the noise transfer functions, and use the result for both stochastic signals, $\mathcal{R}_{\rm gal} = \mathcal{R}_{\rm sgwb}$ (Sec.~\ref{sec:models}). An isotropic kernel depends on the shape of the constellation only, and not on its orientation. A realistic foreground instead weights Eq.~\eqref{eq:response} with the sky distribution of the Galaxy.

\subsection{Noise models and priors} \label{app:noise_models}

The noise models of Sec.~\ref{sec:noise} are Eq.~\eqref{eq:noise_model} with some of its terms fixed. Their priors, together with those of the stochastic signals, are listed in Table~\ref{tab:priors}. Because our noise model has some intrinsic hard degeneracies explained below, and the other considered sources bring in additional degeneracies, we use a mixture of techniques unique to the inference step, all defined relative to the simulation step (which is just Eq.~\ref{eq:noise_model} with some terms turned to zero depending on the section considered): reparametrization, lower-dimensional projections orthogonal to the degenerate planes, and directly excluding noise model terms at inference time (``implicit marginalization''). Per section, these are:
\begin{itemize}
\item \emph{Per-link analytic model} (Sec.~\ref{sec:stationary}): one amplitude per link at fixed shape, following~\cite{hartwig2023},
\begin{equation}
S^{(n)}_{k}(f) \;=\; \left(\frac{a^{(n)}_{k}}{\bar{a}^{(n)}}\right)^{2} \bar{S}^{(n)}(f) ,
\label{eq:perlink}
\end{equation}
with $\bar{a}^{\rm TM} = \bar{A}$ and $\bar{a}^{\rm OMS} = \bar{P}$, and $\bar{S}^{(n)}(f)$ is the requirements PSD~\cite{colpi2024}. We adopt the priors of~\cite{hartwig2023}, Gaussian on each amplitude with a $20\%$ width about the specification value, truncated at $4\sigma$, to avoid zeros.
\item \emph{Spline with per-link offsets} (Sec.~\ref{sec:gaussianity}): $\delta^{(n)}$ is the Akima interpolant through the knots $(\ln f^{(n)}_{i}, c^{(n)}_{i})$, $i = 1, \dots, n_s$, placed by warping a logarithmic spacing,
\begin{equation}
f^{(n)}_{i} \;=\; f_{\min} \left(\frac{f_{\max}}{f_{\min}}\right)^{l_{i}^{p_{n}}}, \quad l_{i} \;=\; \frac{i - 1}{n_s - 1} ,
\label{eq:knots}
\end{equation}
so that the two outer knots sit at the band edges, while $p_{n} > 1$ ($p_{n} < 1$) pulls the interior ones towards low (high) frequency; we use $(p_{\rm OMS}, p_{\rm TM}) = (0.8, 2.0)$ for the glitch networks and $(0.2, 0.9)$ for the \ac{MBHB} networks. The interpolant is fully specified by its knot values and is translation-equivariant, $\delta^{(n)}(\,\cdot\,; c + \lambda) = \delta^{(n)}(\,\cdot\,; c) + \lambda$, so a common shift of the six $\epsilon^{(n)}_{k}$ is exactly degenerate with the opposite uniform shift of the knots. The network therefore infers the offsets in an orthonormal basis of the five-dimensional sum-zero subspace orthogonal to $\mathbf{1} \coloneqq (1,1,1,1,1,1)^{\mathsf{T}}\in \mathbb{R}^6$, so that $\boldsymbol{\epsilon}^{(n)} = \mathsf{B}\,\mathbf{u}^{(n)}$ with $\mathsf{B}$ a $6 \times 5$ matrix satisfying $\mathsf{B}^{\mathsf{T}}\mathsf{B} = \mathbb{I}_5$ and $\mathbf{1}^{\mathsf{T}}\mathsf{B} = \mathbf{0}^{\mathsf{T}}$ (hence $\mathbb{I}_6 - \mathbf{1}\mathbf{1}^{\mathsf{T}}/6 = \mathsf{B}\mathsf{B}^\mathsf{T}$), the common level being carried by the knots. The simulator draws the contrast coordinates $\mathbf{u}^{(n)} \sim \mathcal{N}_5(0, \sigma_{\epsilon}^{2}\,\mathbb{I}_5)$, i.e.\ the projected prior $\boldsymbol{\epsilon}^{(n)} \sim \mathcal{N}\big(0, \sigma_{\epsilon}^{2}\,[\mathbb{I}_6 - \mathbf{1}\mathbf{1}^{\mathsf{T}}/6]\big)$ (up to the negligible $5\sigma_{\epsilon}$ truncation), with $\sigma_{\epsilon} = 2\ln 1.2$ (the $20\%$ amplitude scatter of Eq.~\eqref{eq:perlink} carried into log-\ac{PSD}). For \ac{TM}, the columns of $\mathsf{B}$ are two contrasts of the three arm means and the three within-arm differences $(\epsilon^{\rm TM}_{ij} - \epsilon^{\rm TM}_{ji})/\sqrt{2}$, and only the former are inferred: the latter, barely constrained by the data, are still drawn by the simulator and marginalized over by the network. The inferred \ac{TM} levels are thus arm means of the log-offsets, i.e.\ geometric means of the two spectra of an arm. \ac{OMS} uses the same basis, but all five contrasts are inferred.
\item \emph{Spline with block levels} (Sec.~\ref{sec:nonstationary}): the same interpolant with $p = 1.4$ for both components, unequal links (but we infer the \ac{TM} arm differences too, instead of marginalizing over them as in the previous model, to make the problem harder for the network), on the $8$ hour chunk band with $f_{\min} = 6.9 \times 10^{-5}$~Hz. Additionally, $n_b = 5$ independent levels per component, $d^{(n)}_{b} \sim \mathcal{N}(0, \sigma_{d}^{2})$ with $\sigma_{d} = 0.2$, i.e.\ $20\%$ in PSD and $10\%$ in amplitude, about half of $\sigma_{\epsilon}$; see~\cite{edwards2020, alvey2024} for segment-to-segment level variations of this kind. By the same equivariance, the mean of the block levels is degenerate with a uniform shift of the knots, a direction constrained by the prior alone. We explicitly do not restrict the block sum to zero (as we do with the per-link offsets), because that would make them conditionally dependent: this would break the attention masking, which depends on the variables being conditionally independent. In other words, addressing the degeneracy would correlate the time-varying component of the noise and we explicitly wanted to showcase the performance of the attention masking mechanism of the Simformer. 

\end{itemize}

Lastly, the degenerate, partially prior-dominated, partially likelihood-dominated, and high-dimensional model utilized in Sec.~\ref{sec:nonstationary} is exactly the difficult inference task we wish to test our networks on. It is {\it not} meant to be a practical noise model that one would use in likelihood-based, traditional approaches. The framework we constructed is perfectly suited to a complete change of noise model. 

\begin{table}
\caption{Training priors of the inferred parameters, with $\mathcal{U}[a, b]$ a uniform and $\mathcal{N}_{m}(\mu, \sigma^{2})$ a Gaussian truncated at $\mu \pm m\sigma$. The last column lists the configurations using each parameter, for which $\theta$ has dimension $17$ (Sec.~\ref{sec:stationary}), $24$ (Sec.~\ref{sec:gaussianity}; $26$ for the networks that infer the glitch) and $37$ (Sec.~\ref{sec:nonstationary}). $a^{\rm TM}_{k}$ is in units of $10^{-15}\,{\rm m\,s^{-2}\,Hz^{-1/2}}$, $a^{\rm OMS}_{k}$ in units of $10^{-12}\,{\rm m\,Hz^{-1/2}}$, $\Delta\dot{v}$ in ${\rm m\,s^{-2}}$, $\sigma_{\epsilon} = 2\ln 1.2$ and $\sigma_{d} = 0.2$; the $\mathbf{u}^{(n)}$ are the sum-zero contrasts of Appendix~\ref{app:noise_models}.}
\label{tab:priors}
\begin{ruledtabular}
\begin{tabular}{l l l}
parameter & prior & used in \\
\colrule
$\log_{10} A_{\rm gal}$ & $\mathcal{U}[-44, -41]$ & all \\
$\log_{10} (f_{r1}/{\rm Hz})$ & $\mathcal{U}[-3.5, -2]$ & all \\
$\log_{10} (f_{rk}/{\rm Hz})$ & $\mathcal{U}[-3.5, -2]$ & all \\
$\log_{10} \Omega$ & $\mathcal{U}[-13, -8]$ & all \\
$n_t$ & $\mathcal{U}[-4, 4]$ & all \\
\colrule
$a^{\rm TM}_{k}$ ($\times 6$) & $\mathcal{N}_{4}(3, 0.6^{2})$ & \ref{sec:stationary} \\
$a^{\rm OMS}_{k}$ ($\times 6$) & $\mathcal{N}_{4}(15, 3^{2})$ & \ref{sec:stationary} \\
$c^{(n)}_{i}$ ($\times 12$) & $\mathcal{U}[-\ln 10, \ln 10]$ & \ref{sec:gaussianity}, \ref{sec:nonstationary} \\
$\mathbf{u}^{\rm OMS}$ ($\times 5$) & $\mathcal{N}_{5}(0, \sigma_{\epsilon}^{2})$ & \ref{sec:gaussianity}, \ref{sec:nonstationary} \\
$\mathbf{u}^{\rm TM}$ ($\times 2$, $\times 5$) & $\mathcal{N}_{5}(0, \sigma_{\epsilon}^{2})$ & \ref{sec:gaussianity}, \ref{sec:nonstationary} \\
$d^{(n)}_{b}$ ($\times 10$) & $\mathcal{N}_{5}(0, \sigma_{d}^{2})$ & \ref{sec:nonstationary} \\
\colrule
$\ln \Delta\dot{v}$ & $\mathcal{U}[-34, -26]$ & \ref{sec:gaussianity} (glitch inferred) \\
$\log_{10} (\beta/{\rm s})$ & $\mathcal{U}[0, 3]$ & \ref{sec:gaussianity} (glitch inferred) \\
\end{tabular}
\end{ruledtabular}
\end{table}

\section{Network and training configurations} \label{app:configs}

Each training run occupies two H100 GPUs, one producing simulations and one training, with every sample consumed exactly once, so that the product of training steps and batch size counts distinct simulations; wall-clock times range from about ten hours to about eight days by configuration. The two main costs are the simulator, limited by device memory, and the embedder trunk, limited by throughput. The inference network is not the bottleneck, because token count and token width both stay modest (see Table~\ref{tab:configs}), which is the practical payoff of a tokenization that compresses aggressively. The simulation budgets are nonetheless large (order hundreds of millions), and we cannot separate how much of that is a property of the architecture from how much is the difficulty of the inference problem. What can be said is that neither binding cost looks fundamental: a cheaper reduction ahead of the trunk, or a producer parallelized across more devices with its memory and throughput limits addressed separately, would move the wall clock substantially. We have not optimized for leanness or for speed.

\begin{table*}
\caption{Data representation, network, and training settings for each of the three studies. Every configuration is run as a pair differing only in the objective block of Sec.~\ref{sec:objectives}: three pairs in Sec.~\ref{sec:stationary} (one per channel set), five in Sec.~\ref{sec:gaussianity} (glitch marginalized or inferred, three \ac{MBHB} embedders) and one in Sec.~\ref{sec:nonstationary}. Wall clocks are training times.}
\label{tab:configs}
\begin{ruledtabular}
\begin{tabular}{l l l l}
 & Sec.~\ref{sec:stationary} & Sec.~\ref{sec:gaussianity} & Sec.~\ref{sec:nonstationary} \\
\colrule
segment duration                 & $1.0 \times 10^{6}$~s   & $1.0 \times 10^{6}$~s  & $2.88 \times 10^{4}$~s $\times\, 35$ \\
coarse bins $n_{\rm cg}$         & $500$                   & $512$                  & $162$ \\
stored data object               & $9 \times 500$          & $9 \times 512$         & $35 \times 22 \times 162$ \\
data tokens $n_{\rm tok}$        & $1$ / $3$ / $6$         & $9$                    & $35$ \\
coordinate rows                  & yes                     & yes                    & yes \\
rotation invariants              & $AET$-diag, $XYZ$         & yes                    & yes \\
template bank                    & none                    & $E = 20$ (ablated)     & $E = 10$ \\

trunk self-attention             & yes                     & yes                    & no \\
\colrule
$\dim\theta$                     & $17$                    & $24$ / $26$            & $37$ \\
token number $N_{\rm tok}$    & $18$ / $20$ / $23$      & $33$ / $35$            & $72$ \\
token width $d_{\rm tok}$       & $128$                   & $192$                  & $384$ \\
layers / heads                   & $4$ / $4$               & $6$ / $6$              & $6$ / $6$ \\
attention size, widening         & $32$, $4$               & $32$, $4$              & $32$, $4$ \\
conditioning mode                & posterior               & posterior              & posterior \\
attention mask                   & unrestricted            & unrestricted           & derived \\
sub-bands $K$                    & $20$                    & $20$                   & $20$ \\
\colrule
objectives                       & \ac{DSM}, \ac{FM}                 & \ac{DSM}, \ac{FM}                & \ac{DSM}, \ac{FM} \\
paths                            & Eqs.~\eqref{eq:path_ve},~\eqref{eq:path_ot} & Eqs.~\eqref{eq:path_ve},~\eqref{eq:path_ot} & Eqs.~\eqref{eq:path_ve},~\eqref{eq:path_ot} \\
path noise $\sigma_{\min}$, $\sigma_{\max}$, $\gamma_{\min}$ & $10^{-4}$, $10$, $10^{-5}$ & $10^{-4}$, $10$, $10^{-5}$ & $10^{-4}$, $10$, $10^{-5}$ \\
solver steps                     & $200$                   & $200$                  & $200$ \\
batch size                       & $512$                   & $512$                  & $256$ \\
training steps                   & $4.0$ / $4.4$ / $5.0 \times 10^{5}$  & $8 \times 10^{5}$      & $1.2 \times 10^{6}$ \\
peak learning rate               & $1.0 \times 10^{-4}$    & $1.0 \times 10^{-4}$   & $1.5 \times 10^{-4}$ \\
distinct simulations             & $2.0$--$2.6 \times 10^{8}$ & $4.1 \times 10^{8}$ & $3.1 \times 10^{8}$ \\
wall clock per run               & $10$--$15$~h            & $1.2$--$3.6$~d         & $7.7$--$7.9$~d \\
\end{tabular}
\end{ruledtabular}
\end{table*}

\section{Likelihood-based checks}\label{app:likelihoods}

When the data are Gaussian draws about a covariance the coarse-grained statistic is sufficient and its distribution is:
\begin{equation}
\ln \mathcal{L}_0 \;=\; -\sum_{j} N_{j} \left[ \mathrm{tr} \! \left( \tilde{C}_{j}^{-1}(\theta)\, \tilde{S}_{j} \right) + \ln \det \tilde{C}_{j}(\theta) \right],
\label{eq:wishart_like}
\end{equation}
the central complex Wishart log-likelihood. 
$C_j$ and its prewhitened version $\tilde{C}_j$ are defined in the same way as $S$ and $\tilde{S}_j$, see Sec.~\ref{sec:datagen}.
The prewhitening map is fixed and invertible, so it shifts $\ln \mathcal{L}_0$ by a $\theta$-independent constant (omitted, as are the others). A coherent transient added to the Fourier amplitudes makes the statistic non-central, and the resulting likelihood is the non-central Wishart log-likelihood~\cite{ratnarajah2005}:

\begin{align}
\ln \mathcal{L}_1 \;=\; \ln \mathcal{L}_0 \;+\; \sum_{j} \Big[\, & \ln {}_0F_1 \big( N_j;\, N_j\, \tilde{C}_j^{-1} \tilde{M}_j\, \tilde{C}_j^{-1} \tilde{S}_j \big) \nonumber \\
& -\, \mathrm{tr} \big( \tilde{C}_j^{-1} \tilde{M}_j \big) \Big] ,
\label{eq:non_central_wishart} \\
M_j(\theta_{\rm g}) \;=\; \sum_{f \in j} & \, \delta z(f\mid \theta_{\rm g})\, \delta z^{\dagger}(f\mid \theta_{\rm g}) , \nonumber
\end{align} 
where the non-centrality $\tilde{M}_j$ is the glitch's prewhitenend cross-power in bin $j$, $\theta_{\rm g} = (\ln \Delta\dot{v}, \log_{10}\beta)$, and ${}_0F_1$ is the hypergeometric function of matrix argument (complex case)\footnote{The trigger time cancels in $M_j$ and is not inferred. In our configuration $M_j$ is rank one to better than $10^{-4}$ in the other eigenvalues, so ${}_0F_1$ reduces to a modified Bessel function of the trace of its argument.
}. 

The glitch can instead be marginalized analytically, at the price of needing the Fourier amplitudes $z(f)$ themselves, so $\mathcal{L}_2$ conditions on more than the network sees. Linearizing $\delta z$ about the maximum-likelihood glitch $\theta_{\rm g}^{0}$, with the trigger time held at its fitted value, and integrating over $\theta_{\rm g}$ with a flat prior gives
\begin{align} 
\ln \mathcal{L}_2 \;=\; & -\sum_{j} N_j \Big[ \mathrm{tr} \big( \tilde{C}_j^{-1} \tilde{R}_j \big) + \ln \det \tilde{C}_j \Big] \nonumber \\
& + \tfrac{1}{2}\, b^{\mathsf{T}} \Gamma^{-1} b \;-\; \tfrac{1}{2} \ln \det \Gamma ,
\label{eq:ML_marg_glitch} \\
b_\alpha \;=\; & 2\, \mathrm{Re} \sum_{j} \sum_{f \in j} \partial_\alpha \delta z^{\dagger}(f)\, C_j^{-1}\, r(f) , \nonumber \\
\Gamma_{\alpha \beta} \;=\; & 2\, \mathrm{Re} \sum_{j} \sum_{f \in j} \partial_\alpha \delta z^{\dagger}(f)\, C_j^{-1}\, \partial_\beta \delta z(f) , \nonumber
\end{align}
where $R_j$ is built like $S_j$ from the residual $r(f) = z(f) - \delta z(f; \theta_{\rm g}^{0})$, and $b$ and $\Gamma$ are the score and Fisher matrix of $\theta_{\rm g}$, with derivatives $\alpha,\beta$, at $\theta_{\rm g}^{0}$. The first term is $\ln\mathcal{L}_0$ on the residual; the last two are the gain from refitting the glitch (done once at the injection's covariance) and the volume of its marginalized directions.

Reference posteriors are obtained from these likelihoods by parallel-tempered \ac{MCMC}~\cite{karnesis2023eryn}. Comparing a network posterior against such a reference is less straightforward than it looks because of the dimensionalities involved. We quantify agreement with the sliced 2-Wasserstein distance ($sW_2$)~\cite{bonneel2015} between the network and reference whitened posteriors and closeness to the injected parameters with the energy score (ES)~\cite{gneiting2007}; for interpretability, each is translated into the width inflation or rigid shift of an exact posterior that gives the same value (Secs.~\ref{sec:agreement_test} and~\ref{sec:MBHBs_physics_emb}).

Where a likelihood exists it also supports the standard consistency check on an amortized posterior, \ac{IS} with the network as proposed in~\cite{dax2023is}. With draws $\theta_{i} \sim q_{\phi}(\theta\mid x)$,
$w_{i} \;=\; \frac{p(x \mid \theta_{i})\, p(\theta_{i})}{q_{\phi}(\theta_{i} \mid x)}$,
and $\mathrm{ESS} \;=\; \frac{\left( \sum_{i} w_{i} \right)^{2}}{\sum_{i} w_{i}^{2}}$, and the reweighted sample is asymptotically correct whatever the proposal, while the mean weight estimates the evidence. This collapses every aspect of posterior quality into two interpretable numbers, the efficiency $\mathrm{ESS}/N$ and the evidence. It is also limited however: it needs the likelihood; furthermore, its efficiency degrades quickly with parameter dimension; and a low efficiency indicts the proposal and the likelihood jointly, so it diagnoses the pair rather than just the network.

\section{Simulation-based calibration, its localized form, and the rank divergence} \label{app:sbc}

This appendix gives the reasoning behind the calibration tests of Sec.~\ref{sec:checks_intro}: why uniform ranks are expected, why the test may be restricted to the neighborhood of an observation without refitting, why the weighted rank of Eq.~\eqref{eq:weighted_rank} is exact, and why Eq.~\eqref{eq:kl_rank} is a lower bound.

The \ac{SBC} workflow is: draw $\theta^{*} \sim p(\theta)$, $x \sim p(x \mid \theta^{*})$ and $\theta_{1}, \dots, \theta_{K} \sim q_{\phi}(\theta \mid x)$. Since $p(\theta^{*})\, p(x \mid \theta^{*}) = p(x)\, p(\theta^{*} \mid x)$, the parameter that generated a dataset is, given that dataset, a draw from its posterior. If $q_{\phi}(\theta \mid x) = p(\theta \mid x)$, the $K + 1$ values are therefore exchangeable, and for any test quantity $f(\theta, x)$ the randomized rank $u = (r + V) / (K + 1)$, with $r$ the number of draws for which $f(\theta_{k}, x) < f(\theta^{*}, x)$ and $V$ uniform on $[0, 1]$, is exactly uniform~\cite{talts2018, modrak2023}. Uniform ranks are necessary for a correct estimator, not sufficient, and the kind of errors that can be detected depends on the test quantity~\cite{modrak2023}.

Posterior \ac{SBC}~\cite{sailynoja2026} takes the posterior at an observation, $p(\theta \mid x_{\rm obs})$, as the prior of the procedure above. The assessment is thereby focused on the region of parameter space around the posterior, and passing it remains a necessary, not a sufficient, condition~\cite{sailynoja2026}. For a replicate that is independent of the observation given $\theta$, Bayes' theorem gives $p(\theta \mid x, x_{\rm obs}) \;\propto\; p(\theta \mid x)\, \ell(\theta)$ and $p(\theta \mid x_{\rm obs}) \;\propto\; p(\theta)\, \ell(\theta)$, with $\ell(\theta) = p(x_{\rm obs} \mid \theta)$. Both distributions that posterior \ac{SBC} needs are thus reweightings, by one fixed function of $\theta$. 

Nothing in this depends on the weight being a likelihood. For any $g(\theta) \ge 0$ with $0 < \int p(\theta)\, g(\theta)\, {\rm d}\theta < \infty$, define the reweighted, or tilted, prior $p_{g} \propto p\, g$. If $\theta^{*} \sim p_{g}$ and $x \sim p(x \mid \theta^{*})$, then given $x$
\begin{equation}
\theta^{*} \;\sim\; p_{g}(\theta \mid x) \;=\; \frac{p(\theta \mid x)\, g(\theta)}{\mathbb{E}_{p(\theta \mid x)} \left[ g \right]} ,
\label{eq:tilted_post}
\end{equation}
which follows from $p(x \mid \theta)\, p(\theta)\, g(\theta) = p(x)\, p(\theta \mid x)\, g(\theta)$. Hence, if $q_{\phi}(\theta \mid x) = p(\theta \mid x)$ on the datasets that $p_{g}$ generates, then $\theta^{*}$ and draws from $q_{\phi}(\theta \mid x)$ reweighted by $g$ are exchangeable given $x$, for every $g$. 

The choice $g = 1$ is \ac{SBC} and $g = \ell$ is posterior \ac{SBC}. The function $g$ decides where the estimator is tested. It needs neither a likelihood nor the density of the estimator, and it may be built from the estimator's own posterior at $x_{\rm obs}$: an error in that posterior moves the region tested, not the validity of the test. That calibration can be examined under any distribution of parameters and data whose conditional is the posterior was noted for approximate Bayesian computation in Ref.~\cite{prangle2014}, where the restriction is made in data space. Drawing the parameters from a Gaussian mixture fitted to the posterior at the observation, with inflated covariances, is the design of the localized \ac{SBC} of Ref.~\cite{picchini2026}, whose name we adopt; there it tests a surrogate fitted under that design, whereas here the weights let it test the network trained under $p(\theta)$, which is the one used for inference.

The network can be sampled, but its reweighted distribution cannot be sampled from directly. Resampling a finite number of weighted draws does not, in general, produce exact samples from that distribution. We therefore use the following weighted rank, which is exact at finite \(K\). Let \(\theta_0=\theta^*\), \(w_j=g(\theta_j)\), and \(f_j=f(\theta_j,x)\) for \(j=0,\ldots,K\), and write \(W=\sum_{j=0}^{K}w_j\).

\emph{Claim}: Suppose that \(q_\phi(\theta\mid x)=p(\theta\mid x)\) for the marginal of $x$ under $p_g$-almost every \(x\), and that the test quantity distribution $f(\theta,x)$ under $p(\theta \mid x)$ is atomless, so ties occur with probability zero. Then the weighted rank of Eq.~\eqref{eq:weighted_rank} is exactly uniform on \([0,1]\), for every admissible tilt \(g\), every \(K\geq1\), and any posterior shape.

\emph{Proof}: Fix \(x\). Under the null, the \(K+1\) points have joint density
\[
p(\theta_{0:K}\mid x)
=\frac{g(\theta_0)}{c_g(x)}
\prod_{j=0}^{K}p(\theta_j\mid x),
\qquad
c_g(x)=\mathbb E_{p(\theta\mid x)}[g(\theta)].
\]
Randomly permute the points, denoting the permuted values by \(z_0,\ldots,z_K\), and let \(J\) be the position occupied by \(\theta_0=\theta^*\). Since the product of posterior densities is symmetric in the points, the conditional probability that position \(j\) contains the pseudo-truth is
\[
\Pr(J=j\mid z_0,\ldots,z_K,x)
=\frac{g(z_j)}{\sum_{l=0}^{K}g(z_l)}.
\]

Conditional on the permuted points, the pseudo-truth therefore occupies position \(j\) with probability \(w_j/W\). Because ties occur with probability zero \footnote{The expression assumes that ties have probability zero, as holds for the continuous test quantities considered here. For test quantities that can have ties, the exact rank is obtained by replacing \(Vw_*\) in the numerator with \(V\big[w_*+\sum_{k=1}^{K}w_k\,\mathbf{1}[f(\theta_k)=f(\theta^*)]\big]\), leaving the strict-inequality term unchanged.}, ordering the points by their test quantities partitions \([0,1]\) into consecutive intervals whose lengths are the corresponding weights divided by \(W\). If the pseudo-truth occupies a given position, the randomized rank is uniform on that position's interval, owing to the independent uniform variable \(V\). Since the probability of occupying each position equals the length of its interval, these conditional distributions combine to give a uniform distribution on \([0,1]\). This holds for every set of points and every \(x\). \(\square\)

We draw a pool of posterior samples at $x_{\rm obs}$ from the estimators to be compared, with mean $\mu$ and covariance $\Sigma$, and set, inside the prior support,
\begin{equation}
    g(\theta) \;=\; \exp \! \left[ -\tfrac{1}{2} (\theta - \mu)^{\top} \Lambda_{g}\, (\theta - \mu) \right] ,\label{eq:tilt_gauss_mean} 
\end{equation}
\begin{equation}
    \Lambda_{g} \;=\; \left[ (k^{2} \Sigma)^{-1} - \Lambda_{0} \right]_{+} ,
\label{eq:tilt_gauss_cov}
\end{equation}
where $\Lambda_{0}$ is the precision of the Gaussian factor of the prior, zero for a uniform prior, and $[\,\cdot\,]_{+}$ keeps the non-negative eigenvalues. The tilted prior is then a Gaussian of covariance $k^{2} \Sigma$ in the directions the data constrain and the prior itself in those they do not, truncated to the prior support, and we use rejection sampling for it. The width $k$ trades locality against efficiency: at $k = 1$ the neighborhood is one posterior width across, as in posterior \ac{SBC}, but the weights then leave few effective draws. We use $k = 3$, $N = 1024$ datasets and $K = 256$ or $512$ draws, for which typically a quarter to a half of the draws remain effective. 

Lastly, a comment on the rank divergence as a lower bound: let $q_{g}(\theta \mid x) \propto q_{\phi}(\theta | x)\, g(\theta)$ be the reweighted estimator. For a fixed $x$, Eq.~\eqref{eq:weighted_rank} is a randomized function of $\theta^{*}$. By the claim above, applied with $q_{\phi}$ in the role of the posterior, it returns a uniform variable when $\theta^{*} \sim q_{g}(\theta | x)$. In the test, $\theta^{*} \sim p_{g}(\theta | x)$ by Eq.~\eqref{eq:tilted_post}. Passing two distributions through the same randomized function cannot increase their Kullback--Leibler divergence, and that divergence is convex under the average over datasets. With $P_{u}$ the distribution of $u$ over the whole test and $\mathcal{U}$ the uniform one,
\begin{equation}
{\rm KL} \! \left( P_{u} \,\|\, \mathcal{U} \right) \;\le\; \mathbb{E}_{x} \! \left[ {\rm KL} \! \left( p_{g}(\theta \mid x) \,\|\, q_{g}(\theta \mid x) \right) \right] ,
\label{eq:kl_chain}
\end{equation}
the expectation running over the datasets of the test. The left-hand side is unchanged by the invertible map $z = \Phi^{-1}(u)$, which sends $\mathcal{U}$ to the standard normal. Among all distributions of $z$ with mean $m$ and variance $s^{2}$ the Gaussian has the largest entropy, hence the smallest divergence from the standard normal, and that smallest value is $D_{u}$ of Eq.~\eqref{eq:kl_rank}. Therefore $D_{u} \le {\rm KL}(P_{u} \| \mathcal{U})$, and $D_{u}$ bounds the right-hand side of Eq.~\eqref{eq:kl_chain} from below for every test quantity.  That statistics built from ranks measure a divergence which the one between the posteriors themselves dominates is the viewpoint of Ref.~\cite{yao2023}. 

We remark three points of caution for an adopter of this summary: the bound uses two moments of the ranks only, so $D_{u}$ is a floor, and the distribution function remains the ideal display. Next, $D_{u}$ is computed from $N$ ranks: for an exactly calibrated estimator $2 N D_{u}$ is approximately $\chi^{2}$ distributed with two degrees of freedom, so $D_{u}$ has mean $1/N$ and exceeds $4.6/N$ once in a hundred tests, and the corresponding level for the largest of several test quantities is obtained by simulating uniform ranks. Finally, a test quantity that does not depend on the data bounds only what such a quantity can see, as noted above. We highly encourage prospective adopters to read~\cite{modrak2023} and the wider \ac{SBC}-as-calibration literature.

\section{Noise-parameter posteriors}\label{app:extras}

\begin{figure*}
    \centering
    \includegraphics[width=\textwidth]{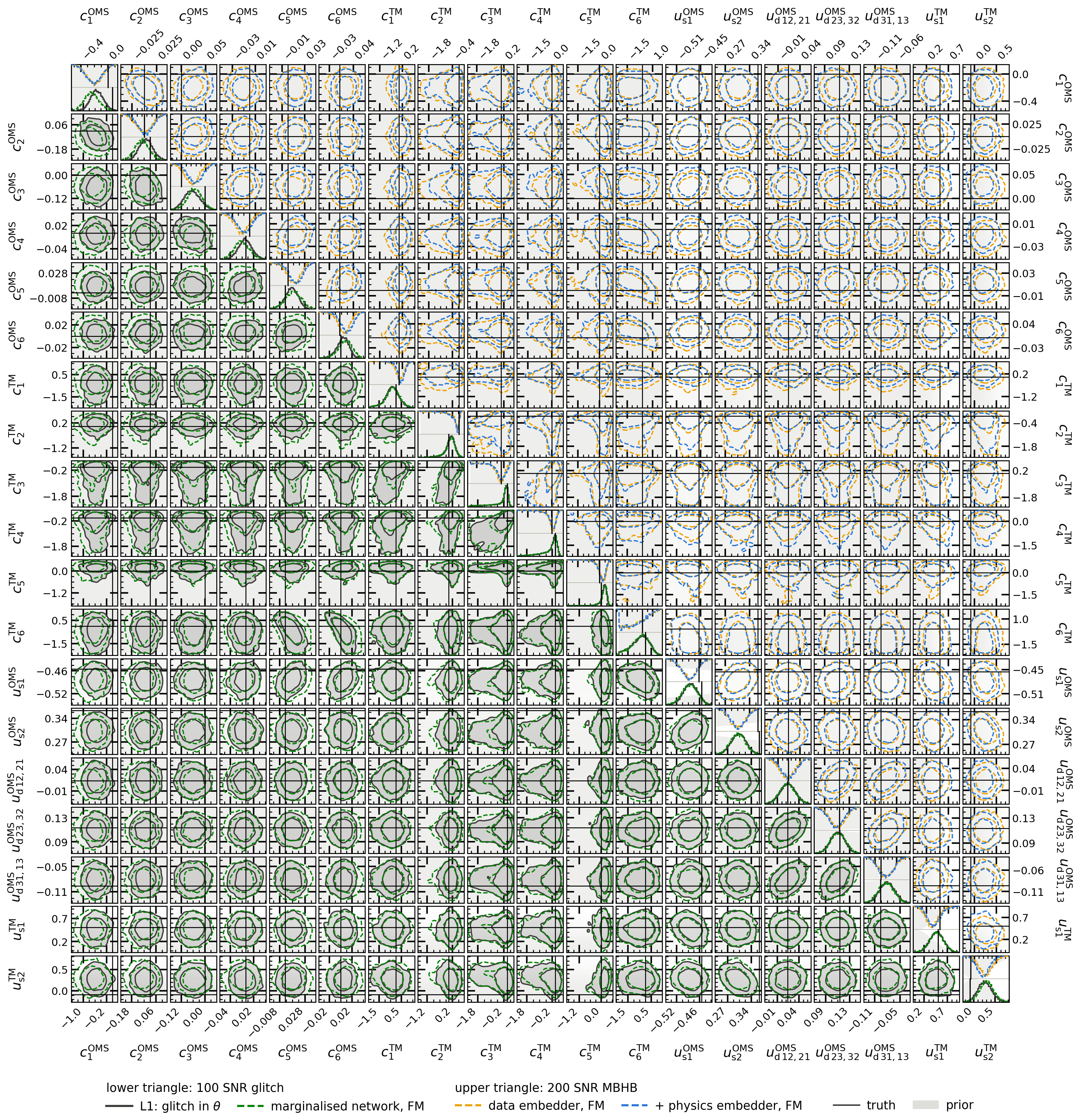}
\caption{Joint posteriors of the $19$ instrument-noise parameters for the two transient observations, one per triangle: the \ac{OMS} and \ac{TM} spline knots $c^{(n)}_{i}$ and the seven per-link contrasts $\mathbf{u}^{(n)}$ (five \ac{OMS}, two \ac{TM} arm means) of Appendix~\ref{app:noise_models}. Below the diagonal, the \ac{SNR} $100$ glitch observation of Fig.~\ref{fig:agreement}: the \ac{$L1$} likelihood, which carries the glitch in $\theta$ (filled grey), and the marginalized \ac{FM} network (green). Above the diagonal, the \ac{SNR} $200$ \ac{MBHB} observation of Fig.~\ref{fig:MBHB_sci}: the \ac{FM} networks with the data (gold) and the $+$physics (blue) embedders, with no likelihood reference. Network curves are dashed, the \ac{FM} line style used throughout. Each diagonal panel holds both observations: the glitch marginals rise from its bottom axis, on the scale of the column below, and the \ac{MBHB} marginals hang from its top axis, on the scale of the column above. Contours are the $68\%$ and $95\%$ highest-density regions, black lines mark the injected values and the gray background is the prior; both are the same for the two observations, which share one noise realization. The glitch and \ac{MBHB} networks place their interior knots at different frequencies (Eq.~\eqref{eq:knots}), so a knot index does not name the same parameter in the two triangles: each triangle has its own axis ranges, and only the contrasts or edge knots can be compared across the diagonal.}

    \label{fig:noise_monster}
\end{figure*}

\begin{figure*}
    \centering
    \includegraphics[width=0.9\textwidth]{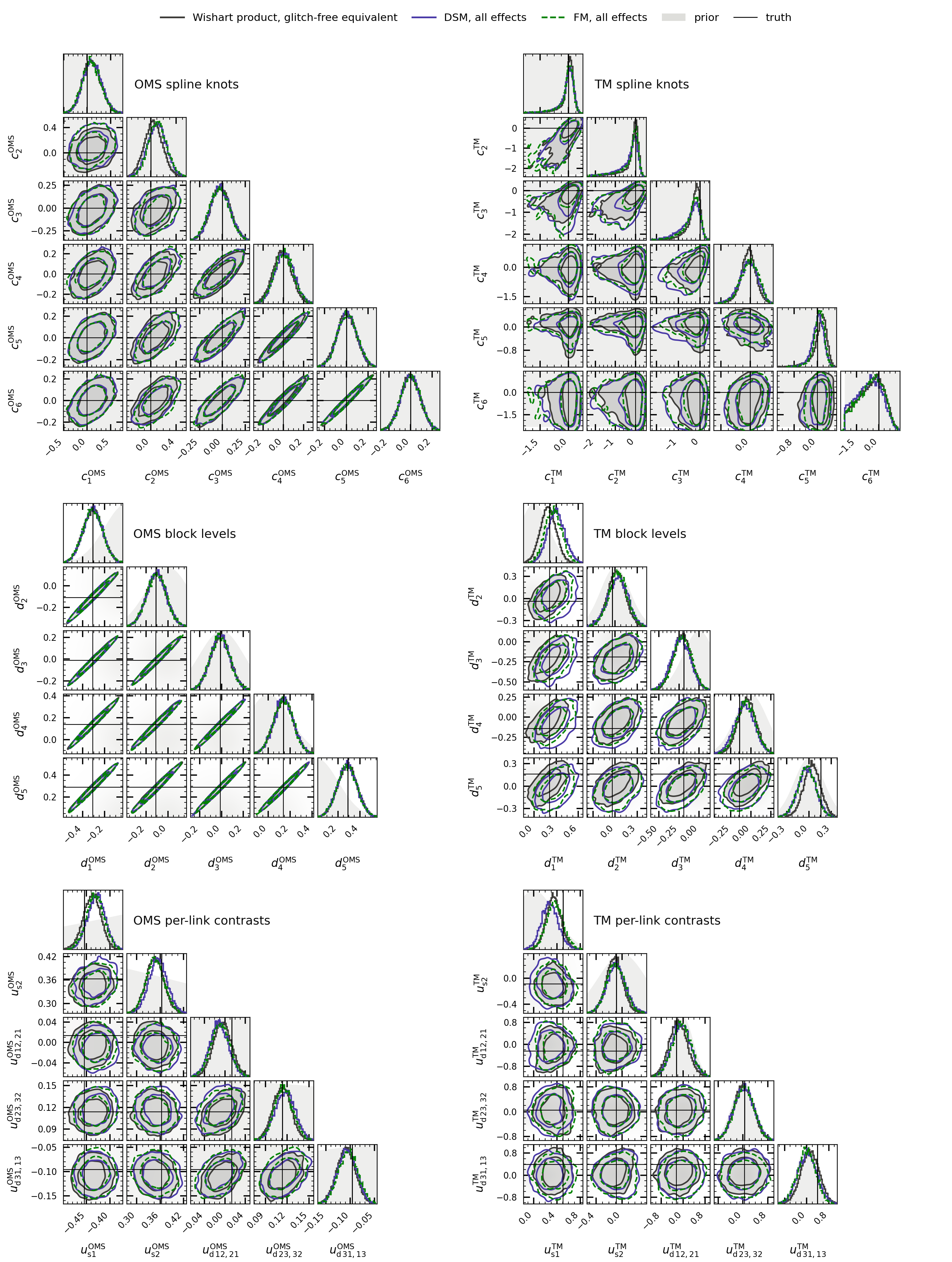}
\caption{The $32$ noise parameters of the all-effects observation of Fig.~\ref{fig:money_plot}: spline knots (top), block levels (middle) and per-link sum-zero contrasts (bottom), \ac{OMS} left and \ac{TM} right, with the posteriors of Fig.~\ref{fig:money_plot}: the exact posterior of the glitch-free data (filled), \ac{DSM} (solid) and \ac{FM} (dashed) on the glitched, gapped and nonstationary data. The narrow diagonals of the \ac{OMS} block levels are their common level, degenerate with a uniform shift of the \ac{OMS} knots and bounded by its prior alone.}

    \label{fig:nonstat_noise}
\end{figure*}

Fig.~\ref{fig:noise_monster} is the complementary noise posterior to the science parameters of Fig.~\ref{fig:agreement} (lower diagonal) and Fig.~\ref{fig:MBHB_sci} (upper diagonal). It includes $6$ spline parameters per noise (\ac{OMS} and \ac{TM}), and $5$ per-link amplitudes (parametrized as sums and differences, Appendix.~\ref{app:noise_models}) for the well-constrained \ac{OMS}, and only $2$ per-link amplitudes (only the sums) for the \ac{TM} component. Fig.~\ref{fig:nonstat_noise} is the noise companion to Fig.~\ref{fig:money_plot} and includes parameters for time-varying noise (block amplitudes), as well as the full per-link sum and differences parameters.

\section{Acronyms}

\begin{acronym}
  \acro{LISA}{Laser Interferometer Space Antenna}
  \acro{GW}{gravitational-wave}
  \acro{GBs}{Galactic binaries}
  \acro{MBHBs}{massive black hole binaries}
  \acro{MBHB}{massive black hole binary}
  \acro{EMRIs}{extreme mass-ratio inspirals}
  \acro{SOBBHs}{stellar-origin black hole binaries}
  \acro{SGWB}{stochastic gravitational wave backgrounds}
  \acro{TDI}{time-delay interferometry}
  \acro{DSM}{denoising score matching}
  \acro{FM}{flow matching}
  \acro{SBC}{simulation-based calibration}
  \acro{TARP}{tests of accuracy with random points}
  \acro{PE}{parameter estimation}
  \acro{SBI}{simulation-based inference}
  \acro{OMS}{optical metrology system}
  \acro{TM}{test-mass}
  \acro{HPD}{highest posterior density}
  \acro{MCMC}{Markov chain Monte Carlo}
  \acro{SNR}{signal-to-noise ratio}
  \acro{MOSA}{moving optical sub-assembly}
  \acro{PSD}{power spectral density}
  \acro{NPE}{neural posterior estimation}
  \acro{SI}{Stochastic Interpolants}
  \acro{SDE}{stochastic differential equation}
  \acro{ODE}{ordinary differential equation}
  \acro{IS}{importance sampling}
  \acro{ESS}{effective sample size}
  \acro{$L0$}{central Wishart likelihood}
  \acro{$L1$}{non-central Wishart likelihood}
  \acro{$L2$}{transient-marginalized likelihood}
  \acro{ECDF}{empirical cumulative distribution function}
\end{acronym}

\bibliography{apssamp}

@article{colpi2024,
  author  = {Colpi, Monica and Danzmann, Karsten and Hewitson, Martin and others},
  title   = {{LISA} Definition Study Report},
  journal = {arXiv e-prints},
  year    = {2024},
  eprint  = {2402.07571},
  archivePrefix = {arXiv},
  primaryClass  = {astro-ph.CO},
}

@article{tinto2021,
  author  = {Tinto, Massimo and Dhurandhar, Sanjeev V.},
  title   = {Time-delay interferometry},
  journal = {Living Reviews in Relativity},
  volume  = {24},
  pages   = {1},
  year    = {2021},
}

@article{prince2002,
  author  = {Prince, Thomas A. and Tinto, Massimo and Larson, Shane L. and Armstrong, J. W.},
  title   = {The {LISA} optimal sensitivity},
  journal = {Phys. Rev. D},
  volume  = {66},
  pages   = {122002},
  year    = {2002},
  eprint  = {gr-qc/0209039},
  archivePrefix = {arXiv},
}

@article{hartwig2023,
  author  = {Hartwig, Olaf and Lilley, Marc and Muratore, Martina and Pieroni, Mauro},
  title   = {Stochastic gravitational wave background reconstruction for a non-equilateral and unequal-noise {LISA} constellation},
  journal = {Phys. Rev. D},
  volume  = {107},
  pages   = {123531},
  year    = {2023},
  eprint  = {2303.15929},
  archivePrefix = {arXiv},
}

@article{kume2025,
  author  = {Kume, Jun'ya and Peloso, Marco and Pieroni, Mauro and Ricciardone, Angelo},
  title   = {Assessing the impact of unequal noises and foreground modeling on {SGWB} reconstruction with {LISA}},
  journal = {JCAP},
  volume  = {06},
  pages   = {030},
  year    = {2025},
  eprint  = {2410.10342},
  archivePrefix = {arXiv},
}

@article{vajpeyi2026,
  author  = {Vajpeyi, Avi and Meyer, Renate and Maturana-Russel, Patricio and Liu, Jianan},
  title   = {Multivariate {B}ayesian {P}-spline estimation of spectral density matrices, with application to {LISA} {TDI} noise},
  journal = {arXiv e-prints},
  year    = {2026},
  eprint  = {2607.04833},
  archivePrefix = {arXiv},
  primaryClass  = {stat.ME},
}

@article{aimen2026,
  author  = {Aimen, Nazeela and Maturana-Russel, Patricio and Vajpeyi, Avi and Christensen, Nelson and Meyer, Renate},
  title   = {{B}ayesian power spectral density estimation for {LISA} noise based on penalized splines with a parametric boost},
  journal = {Phys. Rev. D},
  volume  = {113},
  pages   = {024022},
  year    = {2026},
  eprint  = {2510.00533},
  archivePrefix = {arXiv},
}

@article{santini2025,
    author = "Santini, Alessandro and Muratore, Martina and Gair, Jonathan and Hartwig, Olaf",
    title = "{Flexible, GPU-accelerated approach for the joint characterization of LISA instrumental noise and stochastic gravitational wave backgrounds}",
    eprint = "2507.06300",
    archivePrefix = "arXiv",
    primaryClass = "gr-qc",
    doi = "10.1103/csx9-9trp",
    journal = "Phys. Rev. D",
    volume = "112",
    number = "8",
    pages = "084050",
    year = "2025"
}

@article{karnesis2021,
  author  = {Karnesis, Nikolaos and Babak, Stanislav and Pieroni, Mauro and Cornish, Neil and Littenberg, Tyson},
  title   = {Characterization of the stochastic signal originating from compact binary populations as measured by {LISA}},
  journal = {Phys. Rev. D},
  volume  = {104},
  pages   = {043019},
  year    = {2021},
  eprint  = {2103.14598},
  archivePrefix = {arXiv},
}

@article{cornish2017,
  author  = {Cornish, Neil J. and Robson, Travis},
  title   = {Galactic binary science with the new {LISA} design},
  journal = {J. Phys. Conf. Ser.},
  volume  = {840},
  pages   = {012024},
  year    = {2017},
  eprint  = {1703.09858},
  archivePrefix = {arXiv},
}

@article{boileau2021,
  author  = {Boileau, Guillaume and Lamberts, Astrid and Christensen, Nelson and Cornish, Neil J. and Meyer, Renate},
  title   = {Spectral separation of the stochastic gravitational-wave background for {LISA} in the context of a modulated {G}alactic foreground},
  journal = {Mon. Not. R. Astron. Soc.},
  volume  = {508},
  pages   = {803},
  year    = {2021},
  eprint  = {2105.04283},
  archivePrefix = {arXiv},
}

@article{muratore2025,
    author = "Muratore, Martina and Gair, Jonathan and Hartwig, Olaf and Katz, Michael L. and Toubiana, Alexandre",
    title = "{Pipeline for searching and fitting instrumental glitches in LISA data}",
    eprint = "2505.19870",
    archivePrefix = "arXiv",
    primaryClass = "gr-qc",
    doi = "10.1103/1sj2-219n",
    journal = "Phys. Rev. D",
    volume = "112",
    number = "6",
    pages = "063041",
    year = "2025"
}

@article{armano2022,
  author  = {Armano, M. and Audley, H. and Baird, J. and others},
  title   = {Transient acceleration events in {LISA} {P}athfinder data: Properties and possible physical origin},
  journal = {Phys. Rev. D},
  volume  = {106},
  pages   = {062001},
  year    = {2022},
  eprint  = {2205.11938},
  archivePrefix = {arXiv},
}

@article{edwards2020,
  author  = {Edwards, Matthew C. and Maturana-Russel, Patricio and Meyer, Renate and Gair, Jonathan and Korsakova, Natalia and Christensen, Nelson},
  title   = {Identifying and addressing nonstationary {LISA} noise},
  journal = {Phys. Rev. D},
  volume  = {102},
  pages   = {084062},
  year    = {2020},
  eprint  = {2004.07515},
  archivePrefix = {arXiv},
}

@article{alvey2024,
    author = "Alvey, James and Bhardwaj, Uddipta and Domcke, Valerie and Pieroni, Mauro and Weniger, Christoph",
    title = "{Leveraging time-dependent instrumental noise for the LISA stochastic gravitational wave background analysis}",
    eprint = "2408.00832",
    archivePrefix = "arXiv",
    primaryClass = "gr-qc",
    reportNumber = "CERN-TH-2024-127",
    doi = "10.1103/PhysRevD.111.102006",
    journal = "Phys. Rev. D",
    volume = "111",
    number = "10",
    pages = "102006",
    year = "2025"
}

@article{baghi2026,
  author  = {Baghi, Quentin and Karnesis, Nikolaos and Bayle, Jean-Baptiste},
  title   = {Statistics of time and frequency-averaged spectra in gravitational-wave background searches},
  journal = {arXiv e-prints},
  year    = {2026},
  eprint  = {2602.12781},
  archivePrefix = {arXiv},
  primaryClass  = {gr-qc},
}

@article{baghi2019,
  author  = {Baghi, Quentin and Thorpe, James Ira and Slutsky, Jacob and Baker, John and Dal Canton, Tito and Korsakova, Natalia and Karnesis, Nikolaos},
  title   = {Gravitational-wave parameter estimation with gaps in {LISA}: A {B}ayesian data augmentation method},
  journal = {Phys. Rev. D},
  volume  = {100},
  pages   = {022003},
  year    = {2019},
  eprint  = {1907.04747},
  archivePrefix = {arXiv},
}

@article{estelles2021,
  author  = {Estell\'es, H\'ector and Husa, Sascha and Colleoni, Marta and Keitel, David and Mateu-Lucena, Maite and Garc\'ia-Quir\'os, Cecilio and Ramos-Buades, Antoni and Borchers, Angela},
  title   = {Time-domain phenomenological model of gravitational-wave subdominant harmonics for quasicircular nonprecessing binary black hole coalescences},
  journal = {Phys. Rev. D},
  volume  = {105},
  pages   = {084039},
  year    = {2022},
  eprint  = {2012.11923},
  archivePrefix = {arXiv},
}

@article{Christensen:2018iqi,
    author = "Christensen, Nelson",
    title = "{Stochastic Gravitational Wave Backgrounds}",
    eprint = "1811.08797",
    archivePrefix = "arXiv",
    primaryClass = "gr-qc",
    doi = "10.1088/1361-6633/aae6b5",
    journal = "Rept. Prog. Phys.",
    volume = "82",
    number = "1",
    pages = "016903",
    year = "2019"
}

@misc{babak2021,
  title = {{{LISA Sensitivity}} and {{SNR Calculations}}},
  author = {Babak, Stanislav and Hewitson, Martin and Petiteau, Antoine},
  year = 2021,
  month = aug,
  number = {arXiv:2108.01167},
  eprint = {2108.01167},
  primaryclass = {astro-ph.IM},
  publisher = {arXiv},
  doi = {10.48550/arXiv.2108.01167},
  urldate = {2026-07-22},
  archiveprefix = {arXiv}
}

@misc{LISA:2017pwj,
    author = "Amaro-Seoane, Pau and others",
    collaboration = "LISA",
    title = "{Laser Interferometer Space Antenna}",
    eprint = "1702.00786",
    archivePrefix = "arXiv",
    primaryClass = "astro-ph.IM",
    month = "2",
    year = "2017"
}

@article{akima1970,
  author  = {Akima, Hiroshi},
  title   = {A new method of interpolation and smooth curve fitting based on local procedures},
  journal = {J. ACM}, volume = {17}, number = {4}, pages = {589--602}, year = {1970},
  doi     = {10.1145/321607.321609},
}

@article{baghi2023jcap,
  author  = {Baghi, Quentin and Karnesis, Nikolaos and Bayle, Jean-Baptiste and Besan\c{c}on, Marc and Inchausp\'e, Henri},
  title   = {Uncovering gravitational-wave backgrounds from noises of unknown shape with {LISA}},
  journal = {JCAP}, volume = {04}, pages = {066}, year = {2023},
  doi     = {10.1088/1475-7516/2023/04/066},
  eprint  = {2302.12573}, archivePrefix = {arXiv}, primaryClass = {gr-qc},
}

@article{karnesis2026,
  author  = {Karnesis, Nikolaos and Baghi, Quentin and Bayle, Jean-Baptiste and Galanis, Nikiforos},
  title   = {Effect of noise characterization on the detection of mHz stochastic gravitational waves},
  journal = {arXiv e-prints}, year = {2026},
  eprint  = {2601.19741}, archivePrefix = {arXiv}, primaryClass = {gr-qc},
}

@article{cranmer2020,   author={Cranmer, K. and Brehmer, J. and Louppe, G.}, title={The frontier of simulation-based inference}, journal={Proc. Natl. Acad. Sci.}, volume={117}, number={48}, pages={30055}, year={2020}}

@article{papamakarios2021normalizing,
  title={Normalizing flows for probabilistic modeling and inference},
  author={Papamakarios, George and Nalisnick, Eric and Rezende, Danilo Jimenez and Mohamed, Shakir and Lakshminarayanan, Balaji},
  journal={Journal of Machine Learning Research},
  volume={22},
  number={57},
  pages={1--64},
  year={2021}
}

@inproceedings{greenberg2019, author={Greenberg, D. S. and Nonnenmacher, M. and Macke, J. H.}, title={Automatic posterior transformation for likelihood-free inference}, booktitle={ICML}, year={2019}}

@article{papamakarios2016fast,
  title={Fast $\varepsilon$-free inference of simulation models with bayesian conditional density estimation},
  author={Papamakarios, George and Murray, Iain},
  journal={Advances in neural information processing systems},
  volume={29},
  year={2016}
}

@article{vincent2011,   author={Vincent, P.}, title={A connection between score matching and denoising autoencoders}, journal={Neural Comput.}, volume={23}, number={7}, pages={1661}, year={2011}}

@inproceedings{song2021, author={Song, Y. and Sohl-Dickstein, J. and Kingma, D. P. and Kumar, A. and Ermon, S. and Poole, B.}, title={Score-based generative modeling through stochastic differential equations}, booktitle={ICLR}, year={2021}}

@inproceedings{lipman2023, author={Lipman, Y. and Chen, R. T. Q. and Ben-Hamu, H. and Nickel, M. and Le, M.}, title={Flow matching for generative modeling}, booktitle={ICLR}, year={2023}}

@inproceedings{dax2023,  author={Dax, M. and Wildberger, J. and Buchholz, S. and Green, S. R. and Macke, J. H. and Sch\"olkopf, B.}, title={Flow matching for scalable simulation-based inference}, booktitle={NeurIPS}, year={2023}}

@article{dax2023is,     author={Dax, M. and Green, S. R. and Gair, J. and P\"urrer, M. and Wildberger, J. and Macke, J. H. and Buonanno, A. and Sch\"olkopf, B.}, title={Neural importance sampling for rapid and reliable gravitational-wave inference}, journal={Phys. Rev. Lett.}, volume={130}, pages={171403}, year={2023}}

@inproceedings{gloeckler2024, author={Gloeckler, M. and Deistler, M. and Weilbach, C. and Wood, F. and Macke, J. H.}, title={All-in-one simulation-based inference}, booktitle={ICML}, year={2024}}

@inproceedings{vaswani2017, author={Vaswani, A. and Shazeer, N. and Parmar, N. and Uszkoreit, J. and Jones, L. and Gomez, A. N. and Kaiser, L. and Polosukhin, I.}, title={Attention is all you need}, booktitle={NeurIPS}, year={2017}}

@inproceedings{ronneberger2015, author={Ronneberger, O. and Fischer, P. and Brox, T.}, title={U-Net: convolutional networks for biomedical image segmentation}, booktitle={MICCAI}, year={2015}}

@article{talts2018,  author={Talts, S. and Betancourt, M. and Simpson, D. and Vehtari, A. and Gelman, A.}, title={Validating Bayesian inference algorithms with simulation-based calibration}, journal={arXiv:1804.06788}, year={2018}}

@inproceedings{lemos2023, author={Lemos, P. and Coogan, A. and Hezaveh, Y. and Perreault-Levasseur, L.}, title={Sampling-based accuracy testing of posterior estimators for general inference}, booktitle={ICML}, year={2023}}

@article{hermans2021, author={Hermans, J. and Delaunoy, A. and Rozet, F. and Wehenkel, A. and Begy, V. and Louppe, G.}, title={A crisis in simulation-based inference? Beware, your posterior approximations can be unfaithful}, journal={Trans. Mach. Learn. Res.}, year={2022}, note={arXiv:2110.06581}}

@article{karnesis2023eryn, author={Karnesis, N. and Katz, M. L. and Korsakova, N. and Gair, J. R. and Stergioulas, N.}, title={Eryn: a multipurpose sampler for Bayesian inference}, journal={Mon. Not. R. Astron. Soc.}, volume={526}, pages={4814}, year={2023}}

@article{Caprini:2019pxz,
    author = "Caprini, Chiara and Figueroa, Daniel G. and Flauger, Raphael and Nardini, Germano and Peloso, Marco and Pieroni, Mauro and Ricciardone, Angelo and Tasinato, Gianmassimo",
    title = "{Reconstructing the spectral shape of a stochastic gravitational wave background with LISA}",
    eprint = "1906.09244",
    archivePrefix = "arXiv",
    primaryClass = "astro-ph.CO",
    reportNumber = "LISA-CosWG-19-02",
    doi = "10.1088/1475-7516/2019/11/017",
    journal = "JCAP",
    volume = "11",
    pages = "017",
    year = "2019"
}

@article{armstrong1999,
  author  = {Armstrong, J. W. and Estabrook, F. B. and Tinto, Massimo},
  title   = {Time-Delay Interferometry for Space-based Gravitational Wave Searches},
  journal = {Astrophys. J.},
  volume  = {527},
  pages   = {814},
  year    = {1999},
}

@article{shaddock2003,
  author  = {Shaddock, Daniel A. and Tinto, Massimo and Estabrook, Frank B. and Armstrong, J. W.},
  title   = {Data combinations accounting for {LISA} spacecraft motion},
  journal = {Phys. Rev. D},
  volume  = {68},
  pages   = {061303},
  year    = {2003},
}

@article{tinto2004,
  author  = {Tinto, Massimo and Estabrook, Frank B. and Armstrong, J. W.},
  title   = {Time delay interferometry with moving spacecraft arrays},
  journal = {Phys. Rev. D},
  volume  = {69},
  pages   = {082001},
  year    = {2004},
}

@article{vallisneri2005,
  author  = {Vallisneri, Michele},
  title   = {Geometric time delay interferometry},
  journal = {Phys. Rev. D},
  volume  = {72},
  pages   = {042003},
  year    = {2005},
  note    = {Erratum: Phys. Rev. D 76, 109903 (2007)},
}

@article{muratore2020,
  author  = {Muratore, Martina and Vetrugno, Daniele and Vitale, Stefano},
  title   = {Revisitation of time delay interferometry combinations that suppress laser noise in {LISA}},
  journal = {Class. Quantum Grav.},
  volume  = {37},
  pages   = {185019},
  year    = {2020},
}

@article{otto2012,
  author  = {Otto, Markus and Heinzel, Gerhard and Danzmann, Karsten},
  title   = {{TDI} and clock noise removal for the split interferometry configuration of {LISA}},
  journal = {Class. Quantum Grav.},
  volume  = {29},
  pages   = {205003},
  year    = {2012},
}

@article{bayle2023,
  author  = {Bayle, Jean-Baptiste and Hartwig, Olaf},
  title   = {Unified model for the {LISA} measurements and instrument simulations},
  journal = {Phys. Rev. D},
  volume  = {107},
  pages   = {083019},
  year    = {2023},
  eprint  = {2212.05351},
  archivePrefix = {arXiv},
}

@article{hartwig2022,
  author  = {Hartwig, Olaf and Muratore, Martina},
  title   = {Characterization of time delay interferometry combinations for the {LISA} instrument noise},
  journal = {Phys. Rev. D},
  volume  = {105},
  pages   = {062006},
  year    = {2022},
  eprint  = {2111.00975},
  archivePrefix = {arXiv},
}

@article{quangnam2023,
  author  = {Quang Nam, Dam and others},
  title   = {Time-delay interferometry noise transfer functions for {LISA}},
  journal = {Phys. Rev. D},
  volume  = {108},
  pages   = {082004},
  year    = {2023},
  eprint  = {2211.02539},
  archivePrefix = {arXiv},
}

@article{nayak2006,
  author  = {Nayak, K. Rajesh and Koshti, S. and Dhurandhar, S. V. and Vinet, J.-Y.},
  title   = {On the minimum flexing of {LISA}'s arms},
  journal = {Class. Quantum Grav.},
  volume  = {23},
  pages   = {1763},
  year    = {2006},
  eprint  = {gr-qc/0507105},
  archivePrefix = {arXiv},
}

@article{romano2017,
  author  = {Romano, Joseph D. and Cornish, Neil J.},
  title   = {Detection methods for stochastic gravitational-wave backgrounds: a unified treatment},
  journal = {Living Rev. Relativ.},
  volume  = {20},
  pages   = {2},
  year    = {2017},
  eprint  = {1608.06889},
  archivePrefix = {arXiv},
}

@article{flauger2021,
  author  = {Flauger, Raphael and Karnesis, Nikolaos and Nardini, Germano and Pieroni, Mauro and Ricciardone, Angelo and Torrado, Jes\'us},
  title   = {Improved reconstruction of a stochastic gravitational wave background with {LISA}},
  journal = {JCAP},
  volume  = {01},
  pages   = {059},
  year    = {2021},
  eprint  = {2009.11845},
  archivePrefix = {arXiv},
}

@article{estabrook1975,
  author  = {Estabrook, F. B. and Wahlquist, H. D.},
  title   = {Response of {D}oppler spacecraft tracking to gravitational radiation},
  journal = {Gen. Relativ. Gravit.},
  volume  = {6},
  pages   = {439},
  year    = {1975},
}

@article{cornish2003,
  author  = {Cornish, Neil J. and Rubbo, Louis J.},
  title   = {{LISA} response function},
  journal = {Phys. Rev. D},
  volume  = {67},
  pages   = {022001},
  year    = {2003},
  eprint  = {gr-qc/0209011},
  archivePrefix = {arXiv},
  note    = {Publisher's note: Phys. Rev. D 67, 029905 (2003)},
}

@article{adams2014,
  author  = {Adams, Matthew R. and Cornish, Neil J.},
  title   = {Detecting a stochastic gravitational wave background in the presence of a galactic foreground and instrument noise},
  journal = {Phys. Rev. D},
  volume  = {89},
  pages   = {022001},
  year    = {2014},
  eprint  = {1307.4116},
  archivePrefix = {arXiv},
}

@article{gorski2005,
  author  = {G\'orski, K. M. and others},
  title   = {{HEALPix}: A Framework for High-Resolution Discretization and Fast Analysis of Data Distributed on the Sphere},
  journal = {Astrophys. J.},
  volume  = {622},
  pages   = {759},
  year    = {2005},
  eprint  = {astro-ph/0409513},
  archivePrefix = {arXiv},
}

@article{modrak2023,
  author  = {Modr{\'a}k, Martin and Moon, Angie H. and Kim, Shinyoung and B{\"u}rkner, Paul and Huurre, Niko and Faltejskov{\'a}, Kate{\v r}ina and Gelman, Andrew and Vehtari, Aki},
  title   = {Simulation-Based Calibration Checking for {B}ayesian Computation: The Choice of Test Quantities Shapes Sensitivity},
  journal = {Bayesian Analysis},
  volume  = {20},
  number  = {2},
  pages   = {461--488},
  year    = {2025},
  doi     = {10.1214/23-BA1404},
  eprint  = {2211.02383},
  archivePrefix = {arXiv},
}

@article{sailynoja2026,
  author  = {S{\"a}ilynoja, Teemu and Schmitt, Marvin and B{\"u}rkner, Paul-Christian and Vehtari, Aki},
  title   = {Posterior {SBC}: simulation-based calibration checking conditional on data},
  journal = {Statistics and Computing},
  volume  = {36},
  pages   = {78},
  year    = {2026},
  doi     = {10.1007/s11222-026-10825-9},
  eprint  = {2502.03279},
  archivePrefix = {arXiv},
}

@article{sailynoja2022,
  author  = {S{\"a}ilynoja, Teemu and B{\"u}rkner, Paul-Christian and Vehtari, Aki},
  title   = {Graphical test for discrete uniformity and its applications in goodness-of-fit evaluation and multiple sample comparison},
  journal = {Statistics and Computing},
  volume  = {32},
  pages   = {32},
  year    = {2022},
  doi     = {10.1007/s11222-022-10090-6},
  eprint  = {2103.10522},
  archivePrefix = {arXiv},
}

@article{picchini2026,
  author  = {Picchini, Umberto},
  title   = {Amortized ratio-estimation importance sampling and localized simulation-based calibration for intractable likelihoods},
  journal = {arXiv e-prints},
  year    = {2026},
  eprint  = {2609.39712},
  archivePrefix = {arXiv},
}

@article{prangle2014,
  author  = {Prangle, Dennis and Blum, Michael G. B. and Popovic, Gordana and Sisson, Scott A.},
  title   = {Diagnostic tools for approximate {B}ayesian computation using the coverage property},
  journal = {Australian \& New Zealand Journal of Statistics},
  volume  = {56},
  pages   = {309--329},
  year    = {2014},
  eprint  = {1301.3166},
  archivePrefix = {arXiv},
}

@inproceedings{yao2023,
  author  = {Yao, Yuling and Domke, Justin},
  title   = {Discriminative Calibration: Check {B}ayesian Computation from Simulations and Flexible Classifier},
  booktitle = {Advances in Neural Information Processing Systems},
  volume  = {36},
  year    = {2023},
  eprint  = {2305.14593},
  archivePrefix = {arXiv},
}

@article{ratnarajah2005,
  author  = {Ratnarajah, T. and Vaillancourt, R. and Alvo, M.},
  title   = {Complex random matrices and {Rician} channel capacity},
  journal = {Problems of Information Transmission},
  volume  = {41},
  number  = {1},
  pages   = {1--22},
  year    = {2005},
  doi     = {10.1007/s11122-005-0006-6},
}

@article{alvey2024saqqara,
  author  = {Alvey, James and Bhardwaj, Uddipta and Domcke, Valerie and Pieroni, Mauro and Weniger, Christoph},
  title   = {Simulation-based inference for stochastic gravitational wave background data analysis},
  journal = {Phys. Rev. D}, volume = {109}, pages = {083008}, year = {2024},
  doi     = {10.1103/PhysRevD.109.083008},
  eprint  = {2309.07954}, archivePrefix = {arXiv},
}

@article{amaroseoane2023,
  author  = {Amaro-Seoane, Pau and Andrews, Jeff and Arca Sedda, Manuel and Askar, Abbas and Baghi, Quentin and others},
  title   = {Astrophysics with the {Laser Interferometer Space Antenna}},
  journal = {Living Rev. Relativ.}, volume = {26}, pages = {2}, year = {2023},
  doi     = {10.1007/s41114-022-00041-y},
  eprint  = {2203.06016}, archivePrefix = {arXiv},
}

@article{armano2016,
  author  = {Armano, M. and Audley, H. and Auger, G. and Baird, J. T. and Bassan, M. and others},
  title   = {Sub-Femto-g Free Fall for Space-Based Gravitational Wave Observatories: {LISA} {P}athfinder Results},
  journal = {Phys. Rev. Lett.}, volume = {116}, pages = {231101}, year = {2016},
  doi     = {10.1103/PhysRevLett.116.231101},
}

@article{aubin2021,
  author  = {Aubin, F. and Brighenti, F. and Chierici, R. and Estevez, D. and Greco, G. and others},
  title   = {The {MBTA} pipeline for detecting compact binary coalescences in the third {LIGO-Virgo} observing run},
  journal = {Class. Quantum Grav.}, volume = {38}, pages = {095004}, year = {2021},
  doi     = {10.1088/1361-6382/abe913},
  eprint  = {2012.11512}, archivePrefix = {arXiv},
}

@article{auclair2023,
  author  = {Auclair, Pierre and Bacon, David and Baker, Tessa and Barreiro, Tiago and Bartolo, Nicola and others},
  title   = {Cosmology with the {Laser Interferometer Space Antenna}},
  journal = {Living Rev. Relativ.}, volume = {26}, pages = {5}, year = {2023},
  doi     = {10.1007/s41114-023-00045-2},
  eprint  = {2204.05434}, archivePrefix = {arXiv},
}

@inproceedings{baghi2022ldc,
  author    = {Baghi, Quentin},
  title     = {The {LISA} Data Challenges},
  booktitle = {Proceedings of the 56th Rencontres de Moriond on Gravitation},
  pages     = {117--122}, year = {2022},
  doi       = {10.58027/1e1n-7973},
  eprint    = {2204.12142}, archivePrefix = {arXiv},
}

@article{baghi2022lpf,
  author  = {Baghi, Quentin and Korsakova, Natalia and Slutsky, Jacob and Castelli, Eleonora and Karnesis, Nikolaos and Bayle, Jean-Baptiste},
  title   = {Detection and characterization of instrumental transients in {LISA} {P}athfinder and their projection to {LISA}},
  journal = {Phys. Rev. D}, volume = {105}, pages = {042002}, year = {2022},
  doi     = {10.1103/PhysRevD.105.042002},
  eprint  = {2112.07490}, archivePrefix = {arXiv},
}

@article{burke2025,
  author  = {Burke, Ollie and Marsat, Sylvain and Gair, Jonathan R. and Katz, Michael L.},
  title   = {Addressing data gaps and assessing noise mismodeling in {LISA}},
  journal = {Phys. Rev. D}, volume = {111}, pages = {124053}, year = {2025},
  doi     = {10.1103/5jr8-k2ss},
  eprint  = {2502.17426}, archivePrefix = {arXiv},
}

@misc{cannon2022,
  author  = {Cannon, Patrick and Ward, Daniel and Schmon, Sebastian M.},
  title   = {Investigating the Impact of Model Misspecification in Neural Simulation-based Inference},
  year    = {2022},
  eprint  = {2209.01845}, archivePrefix = {arXiv},
}

@article{castelli2025,
  author  = {Castelli, Eleonora and Baghi, Quentin and Baker, John G. and Slutsky, Jacob and Bobin, J\'er\^ome and others},
  title   = {Extracting gravitational wave signals from {LISA} data in the presence of artifacts},
  journal = {Class. Quantum Grav.}, volume = {42}, pages = {065018}, year = {2025},
  doi     = {10.1088/1361-6382/adb931},
  eprint  = {2411.13402}, archivePrefix = {arXiv},
}

@article{cole2026,
  author  = {Cole, Philippa S. and Alvey, James and Speri, Lorenzo and Weniger, Christoph and Bhardwaj, Uddipta and others},
  title   = {Sequential simulation-based inference for extreme mass ratio inspirals},
  journal = {Phys. Rev. D}, volume = {113}, pages = {063030}, year = {2026},
  doi     = {10.1103/4cd3-wfjr},
  eprint  = {2505.16795}, archivePrefix = {arXiv},
}

@article{cornish2005,
  author  = {Cornish, Neil J. and Crowder, Jeff},
  title   = {{LISA} data analysis using {M}arkov chain {M}onte {C}arlo methods},
  journal = {Phys. Rev. D}, volume = {72}, pages = {043005}, year = {2005},
  doi     = {10.1103/PhysRevD.72.043005},
  eprint  = {gr-qc/0506059}, archivePrefix = {arXiv},
}

@article{cornish2022,
  author  = {Cornish, Neil J.},
  title   = {Low latency detection of massive black hole binaries},
  journal = {Phys. Rev. D}, volume = {105}, pages = {044007}, year = {2022},
  doi     = {10.1103/PhysRevD.105.044007},
  eprint  = {2110.06238}, archivePrefix = {arXiv},
}

@article{dax2021,
  author  = {Dax, Maximilian and Green, Stephen R. and Gair, Jonathan and Macke, Jakob H. and Buonanno, Alessandra and Sch\"olkopf, Bernhard},
  title   = {Real-Time Gravitational Wave Science with Neural Posterior Estimation},
  journal = {Phys. Rev. Lett.}, volume = {127}, pages = {241103}, year = {2021},
  doi     = {10.1103/PhysRevLett.127.241103},
  eprint  = {2106.12594}, archivePrefix = {arXiv},
}

@article{dax2025,
  author  = {Dax, Maximilian and Green, Stephen R. and Gair, Jonathan and Gupte, Nihar and P\"urrer, Michael and others},
  title   = {Real-time inference for binary neutron star mergers using machine learning},
  journal = {Nature}, volume = {639}, pages = {49--53}, year = {2025},
  doi     = {10.1038/s41586-025-08593-z},
  eprint  = {2407.09602}, archivePrefix = {arXiv},
}

@misc{delmond2026,
  author  = {Delmond, Tanguy and Korsakova, Natalia and Oberlin, Thomas and Marsat, Sylvain and Basset, Antoine and Dobigeon, Nicolas},
  title   = {Neural posterior estimation of {G}alactic {B}inary signals for the {LISA} mission},
  year    = {2026},
  eprint  = {2606.29039}, archivePrefix = {arXiv},
}

@article{deng2025,
  author  = {Deng, Senwen and Babak, Stanislav and Le Jeune, Maude and Marsat, Sylvain and Plagnol, \'Eric and Sartirana, Andrea},
  title   = {Modular global-fit pipeline for {LISA} data analysis},
  journal = {Phys. Rev. D}, volume = {111}, pages = {103014}, year = {2025},
  doi     = {10.1103/PhysRevD.111.103014},
  eprint  = {2501.10277}, archivePrefix = {arXiv},
}

@article{dimitriou2024,
  author  = {Dimitriou, Androniki and Figueroa, Daniel G. and Zald\'ivar, Bryan},
  title   = {Fast Likelihood-free Reconstruction of Gravitational Wave Backgrounds},
  journal = {JCAP}, volume = {09}, pages = {032}, year = {2024},
  doi     = {10.1088/1475-7516/2024/09/032},
  eprint  = {2309.08430}, archivePrefix = {arXiv},
}

@article{gelman1992,
  author  = {Gelman, Andrew and Rubin, Donald B.},
  title   = {Inference from iterative simulation using multiple sequences},
  journal = {Statist. Sci.}, volume = {7}, number = {4}, pages = {457--472}, year = {1992},
  doi     = {10.1214/ss/1177011136},
}

@article{gelman1996,
  author  = {Gelman, Andrew and Meng, Xiao-Li and Stern, Hal},
  title   = {Posterior predictive assessment of model fitness via realized discrepancies},
  journal = {Statist. Sinica}, volume = {6}, number = {4}, pages = {733--760}, year = {1996},
}

@article{gwtc4methods,
  author        = {Abac, A. G. and others},
  collaboration = {LIGO Scientific, Virgo, KAGRA},
  title         = {{GWTC-4.0}: Methods for Identifying and Characterizing Gravitational-wave Transients},
  journal       = {Astrophys. J. Lett.}, volume = {1004}, pages = {L21}, year = {2026},
  doi           = {10.3847/2041-8213/ae447b},
  eprint        = {2508.18081}, archivePrefix = {arXiv},
}

@article{houba2024mbhb,
  author  = {Houba, Niklas and Strub, Stefan H. and Ferraioli, Luigi and Giardini, Domenico},
  title   = {Detection and prediction of future massive black hole mergers with machine learning and truncated waveforms},
  journal = {Phys. Rev. D}, volume = {110}, pages = {062003}, year = {2024},
  doi     = {10.1103/PhysRevD.110.062003},
  eprint  = {2405.11340}, archivePrefix = {arXiv},
}

@article{houba2025,
  author  = {Houba, Niklas},
  title   = {Deep source separation of overlapping gravitational-wave signals and nonstationary noise artifacts},
  journal = {Phys. Rev. Res.}, volume = {7}, pages = {033115}, year = {2025},
  doi     = {10.1103/6bjw-xjj2},
  eprint  = {2503.10398}, archivePrefix = {arXiv},
}

@article{katz2025,
  author  = {Katz, Michael L. and Karnesis, Nikolaos and Korsakova, Natalia and Gair, Jonathan R. and Stergioulas, Nikolaos},
  title   = {Efficient {GPU}-accelerated multisource global fit pipeline for {LISA} data analysis},
  journal = {Phys. Rev. D}, volume = {111}, pages = {024060}, year = {2025},
  doi     = {10.1103/PhysRevD.111.024060},
  eprint  = {2405.04690}, archivePrefix = {arXiv},
}

@article{klimenko2016,
  author  = {Klimenko, S. and Vedovato, G. and Drago, M. and Salemi, F. and Tiwari, V. and others},
  title   = {Method for detection and reconstruction of gravitational wave transients with networks of advanced detectors},
  journal = {Phys. Rev. D}, volume = {93}, pages = {042004}, year = {2016},
  doi     = {10.1103/PhysRevD.93.042004},
  eprint  = {1511.05999}, archivePrefix = {arXiv},
}

@article{korsakova2024,
  author  = {Korsakova, Natalia and Babak, Stanislav and Katz, Michael L. and Karnesis, Nikolaos and Khukhlaev, Sviatoslav and Gair, Jonathan R.},
  title   = {Neural density estimation for {G}alactic binaries in the {LISA} data analysis},
  journal = {Phys. Rev. D}, volume = {110}, pages = {104069}, year = {2024},
  doi     = {10.1103/PhysRevD.110.104069},
  eprint  = {2402.13701}, archivePrefix = {arXiv},
}

@article{lamberts2019,
  author  = {Lamberts, Astrid and Blunt, Sarah and Littenberg, Tyson B. and Garrison-Kimmel, Shea and Kupfer, Thomas and Sanderson, Robyn E.},
  title   = {Predicting the {LISA} white dwarf binary population in the {M}ilky {W}ay with cosmological simulations},
  journal = {Mon. Not. R. Astron. Soc.}, volume = {490}, pages = {5888--5903}, year = {2019},
  doi     = {10.1093/mnras/stz2834},
  eprint  = {1907.00014}, archivePrefix = {arXiv},
}

@article{littenberg2023,
  author  = {Littenberg, Tyson B. and Cornish, Neil J.},
  title   = {Prototype global analysis of {LISA} data with multiple source types},
  journal = {Phys. Rev. D}, volume = {107}, pages = {063004}, year = {2023},
  doi     = {10.1103/PhysRevD.107.063004},
  eprint  = {2301.03673}, archivePrefix = {arXiv},
}

@article{lyu2026,
  author  = {Lyu, Huifang and Alvey, James and Anau Montel, Noemi and Pieroni, Mauro and Weniger, Christoph},
  title   = {Dynamic {SBI}: round-free sequential simulation-based inference with adaptive datasets},
  journal = {Mach. Learn.: Sci. Technol.}, volume = {7}, pages = {025036}, year = {2026},
  doi     = {10.1088/2632-2153/ae51e1},
  eprint  = {2510.13997}, archivePrefix = {arXiv},
}

@misc{mao2025,
  author  = {Mao, Ruiting and Lee, Jeong Eun and Edwards, Matthew C.},
  title   = {Robust and scalable simulation-based inference for gravitational wave signals with gaps},
  year    = {2025},
  eprint  = {2512.18290}, archivePrefix = {arXiv},
}

@article{messick2017,
  author  = {Messick, Cody and Blackburn, Kent and Brady, Patrick and Brockill, Patrick and Cannon, Kipp and others},
  title   = {Analysis framework for the prompt discovery of compact binary mergers in gravitational-wave data},
  journal = {Phys. Rev. D}, volume = {95}, pages = {042001}, year = {2017},
  doi     = {10.1103/PhysRevD.95.042001},
  eprint  = {1604.04324}, archivePrefix = {arXiv},
}

@article{muratore2023,
  author  = {Muratore, Martina and Hartwig, Olaf and Vetrugno, Daniele and Vitale, Stefano and Weber, William Joseph},
  title   = {Effectiveness of null time-delay interferometry channels as instrument noise monitors in {LISA}},
  journal = {Phys. Rev. D}, volume = {107}, pages = {082004}, year = {2023},
  doi     = {10.1103/PhysRevD.107.082004},
  eprint  = {2207.02138}, archivePrefix = {arXiv},
}

@inproceedings{schmitt2023,
  author    = {Schmitt, Marvin and B\"urkner, Paul-Christian and K\"othe, Ullrich and Radev, Stefan T.},
  title     = {Detecting Model Misspecification in Amortized {B}ayesian Inference with Neural Networks},
  booktitle = {Pattern Recognition (DAGM GCPR 2023)},
  series    = {Lecture Notes in Computer Science}, volume = {14264}, pages = {541--557},
  publisher = {Springer, Cham}, year = {2024},
  doi       = {10.1007/978-3-031-54605-1_35},
  eprint    = {2112.08866}, archivePrefix = {arXiv},
}

@article{spadaro2023,
  author  = {Spadaro, Alice and Buscicchio, Riccardo and Vetrugno, Daniele and Klein, Antoine and Vitale, Stefano and others},
  title   = {Glitch systematics on the observation of massive black-hole binaries with {LISA}},
  journal = {Phys. Rev. D}, volume = {108}, pages = {123029}, year = {2023},
  doi     = {10.1103/PhysRevD.108.123029},
  eprint  = {2306.03923}, archivePrefix = {arXiv},
}

@article{spadaro2026,
  author  = {Spadaro, Alice and Gair, Jonathan and Gerosa, Davide and Green, Stephen R. and Buscicchio, Riccardo and others},
  title   = {Accurate and efficient simulation-based inference for massive black-hole binaries with {LISA}},
  journal = {Phys. Rev. D}, volume = {114}, pages = {023038}, year = {2026},
  doi     = {10.1103/mmyl-wdgq},
  eprint  = {2603.20431}, archivePrefix = {arXiv},
}

@article{usman2016,
  author  = {Usman, Samantha A. and Nitz, Alexander H. and Harry, Ian W. and Biwer, Christopher M. and Brown, Duncan A. and others},
  title   = {The {PyCBC} search for gravitational waves from compact binary coalescence},
  journal = {Class. Quantum Grav.}, volume = {33}, pages = {215004}, year = {2016},
  doi     = {10.1088/0264-9381/33/21/215004},
  eprint  = {1508.02357}, archivePrefix = {arXiv},
}

@inproceedings{ward2022,
  author    = {Ward, Daniel and Cannon, Patrick and Beaumont, Mark and Fasiolo, Matteo and Schmon, Sebastian},
  title     = {Robust Neural Posterior Estimation and Statistical Model Criticism},
  booktitle = {Advances in Neural Information Processing Systems},
  volume    = {35}, pages = {33845--33859}, year = {2022},
  eprint    = {2210.06564}, archivePrefix = {arXiv},
}

@article{Hartig_2025,
   title={Postprocessing subtraction of tilt-to-length noise in LISA in the presence of gravitational wave signals},
   volume={111},
   ISSN={2470-0029},
   url={http://dx.doi.org/10.1103/PhysRevD.111.043048},
   DOI={10.1103/physrevd.111.043048},
   number={4},
   journal={Physical Review D},
   publisher={American Physical Society (APS)},
   author={Hartig, M.-S. and Paczkowski, S. and Hewitson, M. and Heinzel, G. and Wanner, G.},
   year={2025},
   month=Feb }

@article{hartwig2021clock,
  author        = {Hartwig, Olaf and Bayle, Jean-Baptiste},
  title         = {Clock-jitter reduction in {LISA} time-delay interferometry combinations},
  journal       = {Phys. Rev. D},
  volume        = {103},
  pages         = {123027},
  year          = {2021},
  doi           = {10.1103/PhysRevD.103.123027},
  eprint        = {2005.02430},
  archivePrefix = {arXiv}
}

@misc{phentax,
  author = {Santini, Alessandro},
  title  = {{phentax}: {JAX} implementation of {IMRPhenomTHM} waveforms},
  year   = {2026},
  url    = {https://github.com/asantini29/phentax},
  note   = {GitHub repository}
}

@article{albergo2025,
  title={Stochastic interpolants: A unifying framework for flows and diffusions},
  author={Albergo, Michael and Boffi, Nicholas M and Vanden-Eijnden, Eric},
  journal={Journal of Machine Learning Research},
doi = {10.48550/arXiv.2303.08797},
  volume={26},
  number={209},
  pages={1--80},
  year={2025}
}

@inproceedings{peebles2023,
  author        = {Peebles, William and Xie, Saining},
  title         = {Scalable Diffusion Models with Transformers},
  booktitle     = {Proceedings of the IEEE/CVF International Conference on Computer Vision (ICCV)},
  pages         = {4195--4205},
  year          = {2023},
  doi           = {10.1109/ICCV51070.2023.00387},
  eprint        = {2212.09748},
  archivePrefix = {arXiv}
}

@article{dooney2026,
  author  = {Dooney, T. and de Boer, M. and Narola, H. and Lopez, M. and Bromuri, S. and Tan, D. S. and {Van Den Broeck}, C.},
  title   = {Time-domain synthesis of gravitational-wave detector glitches using class-conditional derivative generative adversarial networks},
  journal = {Phys. Rev. D},
  volume  = {114},
  pages   = {082005},
  year    = {2026},
  doi     = {10.1103/d5yf-3jb4}
}

@article{dooney2024,
  author        = {Dooney, T. and Curier, R. L. and Tan, D. S. and Lopez, M. and {Van Den Broeck}, C. and Bromuri, S.},
  title         = {One flexible model for multiclass gravitational wave signal and glitch generation},
  journal       = {Phys. Rev. D},
  volume        = {110},
  number        = {2},
  pages         = {022004},
  year          = {2024},
  doi           = {10.1103/PhysRevD.110.022004},
  eprint        = {2401.16356},
  archivePrefix = {arXiv}
}

@article{gneiting2007,
  author  = {Gneiting, Tilmann and Raftery, Adrian E.},
  title   = {Strictly Proper Scoring Rules, Prediction, and Estimation},
  journal = {J. Am. Stat. Assoc.},
  volume  = {102},
  number  = {477},
  pages   = {359--378},
  year    = {2007},
  doi     = {10.1198/016214506000001437}
}

@article{bonneel2015,
  author  = {Bonneel, Nicolas and Rabin, Julien and Peyr{\'e}, Gabriel and Pfister, Hanspeter},
  title   = {Sliced and {Radon} {Wasserstein} Barycenters of Measures},
  journal = {J. Math. Imaging Vis.},
  volume  = {51},
  number  = {1},
  pages   = {22--45},
  year    = {2015},
  doi     = {10.1007/s10851-014-0506-3}
}

@article{sesana2016,
  author  = {Sesana, Alberto},
  title   = {Prospects for Multiband Gravitational-Wave Astronomy after {GW150914}},
  journal = {Phys. Rev. Lett.}, volume = {116}, pages = {231102}, year = {2016},
  doi     = {10.1103/PhysRevLett.116.231102},
  eprint  = {1602.06951}, archivePrefix = {arXiv},
}

@article{klein2016,
  author  = {Klein, Antoine and Barausse, Enrico and Sesana, Alberto and Petiteau, Antoine and Berti, Emanuele and others},
  title   = {Science with the space-based interferometer {eLISA}: Supermassive black hole binaries},
  journal = {Phys. Rev. D}, volume = {93}, pages = {024003}, year = {2016},
  doi     = {10.1103/PhysRevD.93.024003},
  eprint  = {1511.05581}, archivePrefix = {arXiv},
}

@article{babak2017,
  author  = {Babak, Stanislav and Gair, Jonathan and Sesana, Alberto and Barausse, Enrico and Sopuerta, Carlos F. and Berry, Christopher P. L. and Berti, Emanuele and Amaro-Seoane, Pau and Petiteau, Antoine and Klein, Antoine},
  title   = {Science with the space-based interferometer {LISA}. {V}: Extreme mass-ratio inspirals},
  journal = {Phys. Rev. D}, volume = {95}, pages = {103012}, year = {2017},
  doi     = {10.1103/PhysRevD.95.103012},
  eprint  = {1703.09722}, archivePrefix = {arXiv},
}

@article{digman2022,
  author  = {Digman, Matthew C. and Cornish, Neil J.},
  title   = {{LISA} Gravitational Wave Sources in a Time-varying Galactic Stochastic Background},
  journal = {Astrophys. J.}, volume = {940}, pages = {10}, year = {2022},
  doi     = {10.3847/1538-4357/ac9139},
  eprint  = {2206.14813}, archivePrefix = {arXiv},
}

@inproceedings{tran2026,
  author  = {Tran, Binh Thuan and Schreuder, Nicolas},
  title   = {Minimax-Optimal Two-Sample Test with Sliced {Wasserstein}},
  booktitle = {Proceedings of the 29th International Conference on Artificial Intelligence and Statistics},
  series  = {PMLR},
  volume  = {300},
  pages   = {1927--1935},
  year    = {2026},
  eprint  = {2510.27498},
  archivePrefix = {arXiv},
}

@article{robson2019,
  author        = {Robson, Travis and Cornish, Neil J.},
  title         = {Detecting gravitational wave bursts with {LISA} in the presence of instrumental glitches},
  journal       = {Phys. Rev. D},
  volume        = {99},
  pages         = {024019},
  year          = {2019},
  doi           = {10.1103/PhysRevD.99.024019},
  eprint        = {1811.04490},
  archivePrefix = {arXiv}
}

@article{houba2024glitch,
  author        = {Houba, Niklas and Ferraioli, Luigi and Giardini, Domenico},
  title         = {Detection and mitigation of glitches in {LISA} data: A machine learning approach},
  journal       = {Phys. Rev. D},
  volume        = {109},
  pages         = {083027},
  year          = {2024},
  doi           = {10.1103/PhysRevD.109.083027},
  eprint        = {2401.00846},
  archivePrefix = {arXiv}
}

@inproceedings{delaunoy2022,
  author        = {Delaunoy, Arnaud and Hermans, Joeri and Rozet, Fran{\c{c}}ois and Wehenkel, Antoine and Louppe, Gilles},
  title         = {Towards reliable simulation-based inference with balanced neural ratio estimation},
  booktitle     = {Advances in Neural Information Processing Systems},
  volume        = {35},
  year          = {2022},
  eprint        = {2208.13624},
  archivePrefix = {arXiv}
}

@article{delaunoy2023,
  author        = {Delaunoy, Arnaud and Miller, Benjamin Kurt and Forr{\'e}, Patrick and Weniger, Christoph and Louppe, Gilles},
  title         = {Balancing simulation-based inference for conservative posteriors},
  journal       = {arXiv e-prints},
  year          = {2023},
  eprint        = {2304.10978},
  archivePrefix = {arXiv}
}

@article{thrane2013,
  author        = {Thrane, Eric and Romano, Joseph D.},
  title         = {Sensitivity curves for searches for gravitational-wave backgrounds},
  journal       = {Phys. Rev. D},
  volume        = {88},
  pages         = {124032},
  year          = {2013},
  doi           = {10.1103/PhysRevD.88.124032},
  eprint        = {1310.5300},
  archivePrefix = {arXiv}
}

@article{zevin2017,
  author        = {Zevin, Michael and Coughlin, Scott and Bahaadini, Sara and Besler, Emre and Rohani, Neda and Allen, Sarah and Cabero, Miriam and Crowston, Kevin and Katsaggelos, Aggelos K. and Larson, Shane L. and Lee, Tae Kyoung and Lintott, Chris and Littenberg, Tyson B. and Lundgren, Andrew and Oesterlund, Carsten and Smith, Joshua R. and Trouille, Laura and Kalogera, Vicky},
  title         = {Gravity {Spy}: integrating {Advanced LIGO} detector characterization, machine learning, and citizen science},
  journal       = {Class. Quantum Grav.},
  volume        = {34},
  pages         = {064003},
  year          = {2017},
  doi           = {10.1088/1361-6382/aa5cea},
  eprint        = {1611.04596},
  archivePrefix = {arXiv}
}

@article{essick2021,
  author        = {Essick, Reed and Godwin, Patrick and Hanna, Chad and Blackburn, Lindy and Katsavounidis, Erik},
  title         = {{iDQ}: Statistical inference of non-{G}aussian noise with auxiliary degrees of freedom in gravitational-wave detectors},
  journal       = {Mach. Learn.: Sci. Technol.},
  volume        = {2},
  pages         = {015004},
  year          = {2021},
  eprint        = {2005.12761},
  archivePrefix = {arXiv}
}

@article{Berg__2019,
   title={Exponential shapelets: basis functions for data analysis of isolated features},
   volume={486},
   ISSN={1365-2966},
   url={http://dx.doi.org/10.1093/mnras/stz787},
   DOI={10.1093/mnras/stz787},
   number={1},
   journal={Monthly Notices of the Royal Astronomical Society},
   publisher={Oxford University Press (OUP)},
   author={Bergé, Joel and Massey, Richard and Baghi, Quentin and Touboul, Pierre},
   year={2019},
   month=Mar, pages={544–559} }

@article{auclair2023cosmic,
  title={Cosmic string bursts in LISA},
  author={Auclair, Pierre and Babak, Stanislav and Quelquejay Leclere, Hippolyte and Steer, Dani{\`e}le A},
  journal={Physical Review D},
  volume={108},
  number={4},
  pages={043519},
  year={2023},
  publisher={APS}
}

@article{rubbo2006event,
  title={Event rate for extreme mass ratio burst signals in the laser interferometer space antenna band},
  author={Rubbo, Louis J and Holley-Bockelmann, Kelly and Finn, Lee Samuel},
  journal={The Astrophysical Journal Letters},
  volume={649},
  number={1},
  pages={L25--L28},
  year={2006}
}

@misc{santini2026phentaxjaximplementationimrphenomthm,
      title={Phentax: A JAX implementation of the IMRPhenomTHM waveform model}, 
      author={Alessandro Santini and Diganta Bandopadhyay and Jonathan Gair},
      year={2026},
      eprint={2611.08596},
      archivePrefix={arXiv},
      primaryClass={gr-qc},
      url={https://arxiv.org/abs/2611.08596}, 
}

@misc{stochformer,
  author = {Cuceu, Iuliu},
  title = {Stochformer},
  year = {2025},
  publisher = {GitHub},
  journal = {GitHub repository},
  howpublished = {\url{https://github.com/iuliucuceu/stochformer}},
}

\end{document}